\documentclass[twocolumn,showpacs,floatfix,
superscriptaddress,
longbiblography,
amsmath,amssymb,
aps,
pre]{revtex4-2}
	
\usepackage{graphicx}
\usepackage{dcolumn}
\usepackage{amsmath}
\usepackage{bm}
\usepackage{appendix}
\usepackage{multirow}
\usepackage[bookmarks=true,colorlinks=true,urlcolor=blue,linkcolor=blue,citecolor=blue,breaklinks]{hyperref}
\usepackage[mathlines]{lineno}

\usepackage{inconsolata} 
\usepackage[usenames,dvipsnames]{color} 
\usepackage{listings}

\newcommand\underrel[2]{\mathrel{\mathop{#2}\limits_{#1}}}

\newcommand{\cwt}[1]{{\color{NavyBlue} #1}}

\newcommand{\typ}{\textrm{typ}}

\begin{document}

\preprint{APS/123-QED}

\title{ Describing the critical behavior of the Anderson transition in infinite dimension by random-matrix ensembles: logarithmic multifractality and critical localization }

\author{Weitao Chen}
\affiliation{Department of Physics, National University of Singapore, Singapore.}
\affiliation{MajuLab, CNRS-UCA-SU-NUS-NTU International Joint Research Unit, Singapore.}
\affiliation{Centre for Quantum Technologies, National University of Singapore, Singapore.}
\author{Olivier Giraud}
\affiliation{MajuLab, CNRS-UCA-SU-NUS-NTU International Joint Research Unit, Singapore.}
\affiliation{Centre for Quantum Technologies, National University of Singapore, Singapore.}
\affiliation{Universit\'e Paris-Saclay, CNRS, LPTMS, 91405 Orsay, France.}
\author{Jiangbin Gong}%
\email{phygj@nus.edu.sg}
\affiliation{Department of Physics, National University of Singapore, Singapore.}
\affiliation{MajuLab, CNRS-UCA-SU-NUS-NTU International Joint Research Unit, Singapore.}
\affiliation{Centre for Quantum Technologies, National University of Singapore, Singapore.}
\author{Gabriel Lemari\'e}%
\email{lemarie@irsamc.ups-tlse.fr}
\affiliation{MajuLab, CNRS-UCA-SU-NUS-NTU International Joint Research Unit, Singapore.}
\affiliation{Centre for Quantum Technologies, National University of Singapore, Singapore.}
\affiliation{Laboratoire de Physique Théorique, Université de Toulouse, CNRS, UPS, France.}

\date{\today}

\begin{abstract}

Due to their analytical tractability, random matrix ensembles serve as robust platforms for exploring exotic phenomena in systems that are computationally demanding. Building on a companion letter [\href{https://arxiv.org/abs/2312.17481}{arXiv:2312.17481}], this paper investigates two random matrix ensembles tailored to capture the critical behavior of the Anderson transition in infinite dimension, employing both analytical techniques and extensive numerical simulations. Our study unveils two types of critical behaviors: logarithmic multifractality and critical localization. In contrast to conventional multifractality, the novel logarithmic multifractality features eigenstate moments scaling algebraically with the logarithm of the system size. Critical localization, characterized by eigenstate moments of order $q>1/2$ converging to a finite value indicating localization, exhibits characteristic logarithmic finite-size or time effects, consistent with the critical behavior observed in random regular and Erdös-Rényi graphs of effective infinite dimensionality. Using perturbative methods, we establish the existence of logarithmic multifractality and critical localization in our models. Furthermore, we explore the emergence of novel scaling behaviors in the time dynamics and spatial correlation functions. Our models provide a valuable framework for studying infinite-dimensional quantum disordered systems, and the universality of our findings enables broad applicability to systems with pronounced finite-size effects and slow dynamics, including the contentious many-body localization transition, akin to the Anderson transition in infinite dimension.
\end{abstract}

\maketitle


\section{Introduction}

One of the most universally applicable theories for describing the statistical properties of complex quantum systems is Random Matrix Theory (RMT) \cite{mehta2004random}. This theory has found successful applications across a broad spectrum of domains, ranging from elucidating the dynamics of complex nuclei to quantum chaotic systems, mesoscopic physics to quantum information \cite{doi:10.1146/annurev.ns.38.120188.002225, doi:10.1080/00018732.2016.1198134, RevModPhys.69.731, 10.1063/1.4936880, doi:10.1146/annurev-conmatphys-031720-030658}. According to the fundamental Bohigas-Giannoni-Schmit (BGS) conjecture, which posits a quantum-to-classical correspondence, the statistical characteristics of the eigenspectrum and eigenvectors of quantum Hamiltonians that are fully chaotic in the classical limit should align with classical Wigner-Dyson random-matrix ensembles based on their global symmetries \cite{PhysRevLett.52.1}. Conversely, as suggested by the Berry-Tabor conjecture, the statistical properties of quantum systems with integrable classical dynamics are governed by random matrices within the Poisson universality class \cite{berry1977regular}. A substantial body of numerical evidence has consistently demonstrated the validity of these random matrix ensembles in describing the statistical properties of corresponding quantum systems, spanning both the fully chaotic and integrable limits \cite{GUHR1998189}.

Beyond the domains of these two extremes, intermediate systems garner significant attention, in part due to the notable Anderson localization transition \cite{RevModPhys.80.1355, abrahams201050,PhysRevLett.103.155703,PhysRevLett.101.255702, OhtsukiKawarabayashi,PhysRevLett.105.090601,PhysRevA.94.033615,PhysRevA.95.041602,martinez2022coherent,PhysRevA.100.043612,PhysRevE.108.054127}. This transition is driven by the nontrivial interplay between disorder (or chaotic classical dynamics) and interference effects and is characterized by a metallic phase at low disorder where states are delocalized and the spectrum follows random matrix statistics, to an insulating phase at large disorder, where states are localized with Poisson statistics. The Anderson transition manifests in sufficiently large dimensionalities (greater than 2D in the orthogonal class), and one of its most important characteristics is the emergence of quantum multifractality in the critical regime \cite{castellani1986multifractal, PhysRevLett.102.106406,PhysRevA.80.043626,PhysRevB.84.134209,PhysRevB.95.094204,PhysRevLett.76.1687,PhysRevB.75.174203,doi:10.7566/JPSJ.83.084711,PhysRevLett.124.186601}. This phenomenon arises from strong and scale-invariant spatial fluctuations at the critical point \cite{mandelbrot1974intermittent, mandelbrot1982fractal, falconer2004fractal} and is characterized by the moments $P_q$ of order $q$ of eigenstate amplitudes scaling algebraically with system size $N$:
\begin{equation}\label{eq:Pq}
\langle P_q\rangle =\langle \sum_i|\psi_\alpha(i)|^{2q}\rangle\sim N^{-D_q(q-1)}.
\end{equation}
Here, the sum is over the $N$ sites (indexed by $i$) of the system, with the normalized eigenstate amplitude at site $i$ denoted $|\psi_\alpha(i)|^{2}$. $\langle\dots \rangle$ denotes averaging over disorder configurations and eigenstates in a certain energy window.

Along this direction of investigation, several random-matrix ensembles have been introduced to capture the properties of the Anderson transition, and in particular quantum multifractality, thanks to analytical techniques for solving such ensembles. Wegner's $n$-orbital model served as a starting point, where replica tricks were employed to formulate a nonlinear $\sigma$ model \cite{PhysRevB.19.783}. Subsequently, a supersymmetric version of the nonlinear $\sigma$ model was introduced by Efetov \cite{Efetov1983}, and various random-matrix ensembles emerged, including the sparse random matrix ensemble \cite{PhysRevLett.67.2049}, random banded matrix ensemble \cite{PhysRevLett.67.2405}, and the power-law random banded matrix (PRBM) ensemble \cite{PhysRevE.54.3221, PhysRevB.62.7920}. The nonlinear $\sigma$ model approach provides valuable insights into the renormalization group flow across the transition, although its validity is confined to the weak disorder regime. Questions persist regarding its applicability in the strong disorder regime, where high-dimensional Anderson transitions occur \cite{PhysRevB.95.094204,baroni2023corrections,zirnbauer2023wegner}.

Among these models, the PRBM model stands out as particularly convenient for the analytical investigation of quantum multifractality. Several analogous models of the PRBM model have been introduced in the circular ensemble as variants of the quantum kicked rotor \cite{CHIRIKOV1979263,IZRAILEV1990299}, such as the Ruijsenaars-Schneider (RS) ensemble \cite{PhysRevLett.103.054103,PhysRevE.84.036212,PhysRevE.85.046208,Bogomolny_2011}. These models can be viewed as 1D disordered lattices with long-range algebraically decaying hopping, which drive a localization transition in 1D, exhibiting multifractal critical behavior when the algebraic exponent is $1$ (i.e., the dimension). They can be analytically studied, not only in the weak disorder regime through the mapping to a nonlinear sigma model but also in the strong disorder regime using nontrivial perturbative approaches such as Levitov renormalization \cite{L.S.Levitov_1989,PhysRevLett.64.547} and weighted L\'evy sums \cite{Monthus_2010}. Though the properties of localization and delocalization do not represent those of the standard localized or delocalized phases (for example, the states are algebraically localized due to long-range hopping, instead of exponentially localized) \cite{C.Yeung_1987,BORGONOVI1999317,CASATI1999293}, many critical properties of these models well represent the multifractal properties found at the usual finite-dimensional Anderson transition \cite{PhysRevE.54.3221, PhysRevB.62.7920,Kravtsov_2006,Kravtsov_2011,PhysRevLett.97.046803}. Describing these properties is usually particularly challenging, and the analytical predictions that these models allow make them particularly interesting.

Recently, there has been a significant surge in interest regarding the Anderson transition in infinite dimension (AT$^\infty$), particularly the Anderson transition on random graphs with an effective dimension of infinity.  This is because features of AT$^\infty$ are analogous to the many-body localization (MBL) transition \cite{sierant2024manybody,PhysRevB.34.6394,ZIRNBAUER1986375,PhysRevLett.67.2049,PhysRevLett.72.526,Monthus_2011,biroli2012difference,PhysRevLett.113.046806,PhysRevLett.117.156601,PhysRevB.94.220203,PhysRevB.94.184203,PhysRevB.96.214204,PhysRevLett.118.166801,PhysRevB.96.201114,PhysRevB.95.094204,KRAVTSOV2018148,PhysRevB.99.214202,PhysRevB.101.100201,Parisi_2020,PhysRevB.98.134205,PhysRevB.99.024202,PhysRevLett.125.250402,PhysRevB.105.094202,TIKHONOV2021168525,PhysRevResearch.2.012020,PhysRevB.105.174207,PhysRevB.106.214202,vanoni2023renormalization,baroni2023corrections,fyodorov1992novel,ADMirlin_1991,scoquart2024role}. The AT$^\infty$ exhibits exotic properties distinct from the transition in finite dimensions. The localized phase has important non-ergodic properties where wavefunctions lie on a few rare branches instead of exploring the exponentially many branches available \cite{biroli2012difference, PhysRevLett.118.166801, PhysRevResearch.2.012020, PhysRevResearch.2.012020, biroli2023largedeviation}. Their localization is characterized by two localization lengths; one, $\xi_\parallel$, describing localization along the few rare branches, is much larger than the other one, $\xi_\perp$, characterizing the decay perpendicular to these rare branches. $\xi_\parallel$ diverges at the transition, while $\xi_\perp$ reaches a finite universal value \cite{PhysRevLett.118.166801, PhysRevResearch.2.012020, PhysRevResearch.2.012020}.
Associated with these unusual properties is the strong multifractality characterizing the whole localized phase and also the critical behavior. Strong multifractality is associated with the existence of a $q^* > 0$ such that $D_q = 0$ for $q > q^*$ and $D_q > 0$ for $q < q^*$ \cite{RevModPhys.80.1355}. In the localized phase, $q^* = \xi_\perp \ln K < 1/2$, where $K$ is the branching number of the graph, while $q^* = \frac{1}{2}$ at the transition point, a universal value strongly constrained by a fundamental symmetry of the multifractal spectrum, \cite{PhysRevResearch.2.012020, PhysRevResearch.2.012020, PhysRevLett.97.046803, gruzberg2011symmetries, gruzberg2013classification, PhysRevResearch.3.L022023}.

The nature of the delocalized phase of the AT$^\infty$ has been the subject of an intense debate \cite{biroli2012difference, PhysRevLett.113.046806, PhysRevLett.117.156601, PhysRevB.94.220203,PhysRevB.96.214204,PhysRevLett.118.166801, KRAVTSOV2018148, PhysRevB.99.214202, PhysRevB.101.100201, Parisi_2020, PhysRevB.98.134205, TIKHONOV2021168525, biroli2018delocalization, PhysRevResearch.2.012020, PhysRevB.106.214202, PhysRevResearch.2.043346, 10.21468/SciPostPhys.15.2.045, pino2023correlated, pino2020scaling}, and is now understood to be ergodic at large scales in random regular graphs and smallworld networks, but extended non-ergodic, i.e.~multifractal, at least in the finite size Cayley tree. 
Interestingly, new random matrix ensembles such as generalized Rosenzweig-Porter ensembles have been introduced to capture the properties of this intriguing non-ergodic delocalized phase \cite{Kravtsov_2015, 10.21468/SciPostPhys.6.1.014, PhysRevResearch.2.043346, von2019non, Truong_2016, PhysRevE.98.032139, Monthus_2017, Amini_2017, PhysRevB.103.104205, kravtsov2020localization, PhysRevResearch.2.043346, 10.21468/SciPostPhys.11.2.045}. Such a phase shares a close analogy with the many-body localized phase, where states are multifractal in the Hilbert space \cite{PhysRevLett.123.180601}. In particular, similar slow dynamics have been found in many-body localized systems, Anderson model in random graphs and certain related random-matrix ensembles \cite{10.21468/SciPostPhys.11.2.045,10.21468/SciPostPhys.6.1.014,PhysRevLett.131.106301,PhysRevB.98.134205,PhysRevB.101.100201}.   

In this paper, which follows the companion letter Ref.~\cite{chen2023quantum}, we are interested in constructing random matrix models to describe analytically and numerically the strong multifractality emerging at the AT$^\infty$. Though the properties of the delocalized and localized phases have been extensively discussed, what makes the AT$^\infty$ highly nontrivial is the particularly unusual critical behavior emerging at the transition.

In fact, and very generally, the critical behavior of a transition characterizes its nature, allows for identifying the transition point, and plays a crucial role in the finite-size effects in the vicinity of a transition. For example, in the finite-dimensional Anderson transition, standard multifractality, i.e., the algebraic behavior of $\langle P_q \rangle$ with system size $N$, signals a second-order phase transition with scale invariance at the transition point. This property has been used to characterize the values of the critical disorder strength and the critical exponent of the transition \cite{castellani1986multifractal, PhysRevLett.102.106406,PhysRevA.80.043626,PhysRevB.84.134209,PhysRevB.95.094204,PhysRevLett.76.1687,PhysRevB.75.174203,doi:10.7566/JPSJ.83.084711,PhysRevLett.124.186601}. On the other hand, for the AT$^\infty$, knowing that the critical behavior should be strongly multifractal, i.e., $D_q=0$ for $q>1/2$, does not clearly allow for identifying neither the nature of the transition nor the value of the critical disorder. What is required is to know how $D_q$ vanishes with system size, or more precisely, how $P_q$ behaves with system size.

In recent years, it has become evident that the many-body localization transition is plagued by highly nontrivial and potent finite-size effects, rendering the study of this phenomenon through numerical simulations or experiments inconclusive \cite{PhysRevB.95.155129,PhysRevB.105.174205,PhysRevB.106.L020202,leonard2022signatures,PhysRevB.100.104204,PhysRevE.102.062144,PhysRevB.102.064207,PhysRevB.103.024203,PhysRevE.104.054105,PhysRevLett.127.230603,ABANIN2021168415,PhysRevLett.124.186601,Panda_2019,PhysRevB.102.100202}. We note that one of the reasons why it is challenging to characterize the critical disorder beyond which states are many-body localized lies in the fact that the nature of the transition and the critical behavior at the transition are not known. The AT$^\infty$ exhibits similar subtle finite-size effects and slow dynamics, but the existence of a transition and the precise value of the critical disorder can be clearly established \cite{PhysRevB.99.214202, Parisi_2020, 10.21468/SciPostPhys.15.2.045}. Nevertheless, a comprehensive understanding of the critical behavior is crucial for a thorough comprehension of finite-size effects in the vicinity of the transition. Thus, our random-matrix approach could, in the future, yield interesting consequences for many-body localization.

Analytical predictions have been made for random regular and Erdös-Rényi graphs: $\langle P_2\rangle \sim (\ln N)^{-1/2}+P_2^{N=\infty} $ where $P_2^{N=\infty}$ indicates that asymptotically the critical behavior is localized \cite{PhysRevB.34.6394, ZIRNBAUER1986375, PhysRevLett.72.526, PhysRevB.99.024202}. We will call this critical behavior "critical localization" in the following. We note that this type of critical behavior is reminiscent of that of a Kosterlitz-Thouless transition where $P_2^{N=\infty}$ would play the role of stiffness, which is finite in all the localized, analogous to quasi-ordered, phase up to the critical point, where it jumps to $0$ in the delocalized, analogous to disordered, phase \cite{doi:10.1142/8572}. Logarithmic finite-size corrections at criticality are characteristic of such a Kosterlitz-Thouless transition \cite{PhysRevB.37.5986, Hsieh_2013, PhysRevB.96.180202}. The analogy between AT$^{\infty}$ and the Kosterlitz-Thouless transition has been discussed recently in Ref.~\cite{PhysRevB.106.214202}.

As demonstrated in this paper (see also the companion letter \cite{chen2023quantum}), the aforementioned critical behavior is not the sole possibility for AT$^\infty$. Numerical simulations have hinted at an alternative scenario where $\langle P_2 \rangle$ follows an algebraic law in $\ln N$ (in contrast to $N$ for standard multifractality) \cite{PhysRevLett.118.166801,PhysRevResearch.2.012020,PhysRevB.106.214202, chen2023anderson}. This observation has motivated us to develop random-matrix ensembles featuring $\langle P_q \rangle \sim {(\ln N)}^{d_q}$, characterized by a power function of $\ln N$ with a $q$-dependent exponent $d_q$. This scaling behavior will be termed "logarithmic (or log-) multifractality" hereafter, representing a log-scale invariance that, to the best of our knowledge, has never been explored in the context of the Anderson transition or generally in critical phenomena. It is reminiscent of the multifractal behavior on a fractal support, see e.g. \cite{PhysRevA.35.4907}. Hence, one might view this behavior as the limiting case of a non-ergodic delocalized phase where it was shown that the support set of wavefunctions has a fractal dimension $D$ depending on the disorder strength, with $D$ vanishing at the localization transition, see e.g. \cite{kravtsov2018non}. 

In this paper, we introduce two random-matrix ensembles featuring log-multifractality or critical localization. The first one, an ensemble of Hermitian random matrices, is a variant of the PRBM ensemble \cite{PhysRevE.54.3221, PhysRevB.62.7920} with a specific decay pattern of the off-diagonal matrix elements. The second one is a unitary model related to the kicked rotor and RS models \cite{PhysRevLett.94.244102, PhysRevE.84.036212,PhysRevE.85.046208}.
We analytically demonstrate log-multifractality by establishing the algebraic behavior $P_q \sim {(\ln N)}^{d_q}$. Furthermore, we explore the slow decay of eigenstate spatial correlations, which serves as a crucial indicator of log-multifractality. Additionally, we derive the dynamics of the return probability analytically, revealing an algebraic decay with $\ln t$ rather than with time $t$. We also consider the expansion of a wave-packet, an observable accessible in different experimental platforms. Finally, we explain how to generalize our Hamiltonian model to describe critical localization. Our analytical predictions are thoroughly examined through numerical simulations, confirming their validity.

 The rest of the paper is organized as follows. In Sec.~\ref{sec2}, we present the two random-matrix models designed for log-multifractality. In Sec.~\ref{sec3}, we derive the emergence of the log-multifractality through a perturbation approach based on weighted L\'evy sums for the Hamiltonian model. In Sec.~\ref{sec4}, we derive the emergence of the log-multifractality through Levitov renormalization approach for the unitary model. In Sec.~\ref{sec5}, we show numerical evidence which support a new functional form of the spatial correlation functions subjected to log-multifractality. In Sec.~\ref{sec6}, we present the derivation of the dynamics of the return probabilty and describe in Sec.~\ref{sec6bis} the expansion of a wave packet in the presence of log-multifractality. In Sec.~\ref{sec7}, we extend the proposed ensembles to describe the critical localized behavior observed in the Anderson transitions in Random Regular Graphs and Erdős-Rényi graphs. We then conclude our study in Sec. \ref{sec8}.

\section{Critical random-matrix ensembles for the critical behavior of the Anderson transition in infinite dimension} \label{sec2}

\subsection{Recap of the power-law random matrix (PRBM) ensemble}

First, let us briefly review the two classes of random matrix models for quantum multifractality at the Anderson transition in finite dimensions and their main properties.
One extensively studied model is the PRBM ensemble \cite{PhysRevE.54.3221, PhysRevB.62.7920}. It is defined as an ensemble of $N\times N$ Hermitian matrices $\hat{H}$ whose entries $H_{ij}$ are independent Gaussian random variables with  mean $\langle H_{ij}\rangle=0$ and variance $\langle|H_{ij}|^2\rangle = \beta^{-1}$ for $i=j$ and  
\begin{equation}
\label{defPRBM}
\langle|H_{ij}|^2\rangle =\frac{1}{1+(|i-j|/b)^{2a}}
\end{equation}
for $i\neq j$. Here $\beta$ is the Dyson index for the orthogonal ($\beta=1$) and unitary ($\beta=2$) classes, and $a,b$ are two real parameters. The parameters $a$ and $b$ control the long-range algebraic decay of off-diagonal amplitudes and the eigenspectrum bandwidth, respectively. 

The PRBM model undergoes an Anderson transition at $a=1$, delineating a localized phase ($a>1$) and a delocalized phase ($a<1$). Although the properties of these phases differ from those observed in the Anderson transition in finite dimensions—such as the occurrence of algebraic localization for $a>1$ (see \cite{PhysRevE.54.3221, PhysRevLett.87.056601})—the critical properties at $a=1$ correspond to the quantum multifractal properties found at the Anderson transition in finite dimensions, where $b$ acts as an effective finite dimension. Specifically, as $b\rightarrow\infty$, weak multifractality emerges (e.g., $D_2 \approx 1$), akin to a low dimension $d\gtrsim 2$, where the critical disorder value is weak. Conversely, as $b\rightarrow0$, strong multifractal properties manifest (e.g., $D_2 \ll 1$), characteristic of large dimensionalities $d \gg 1$ with a large critical disorder.

The multifractal properties of the eigenstates can be analytically accessed in  both the weak ($b\gg 1$) and strong multifractal regimes ($b\ll 1$). In the weak multifractal regime, the model can be mapped to a supersymmetric nonlinear $\sigma $ model \cite{PhysRevE.54.3221, PhysRevB.62.7920}. In the strong multifractal regime, several techniques are utilized including real-space normalization \cite{PhysRevB.62.7920} and perturbation theory assisted by weighted L\'evy sums \cite{Monthus_2010}. In addition, the eigenspectrum and the dynamics can be also analytically studied by the technique of virial expansions \cite{Kravtsov_2006,Kravtsov_2011}. We will use these techniques in this paper to describe the effective infinite dimensional critical behavior we consider. 

Ensembles of unitary matrices with similar properties were constructed, based on the quantum kicked rotor, which itself serves as a paradigm for quantum chaos and dynamical localization, a generalization of Anderson localization in momentum space \cite{CHIRIKOV1979263,IZRAILEV1990299,PhysRevLett.75.4598}. The Hamiltonian of such kicked systems takes a general form
\begin{equation}
     \mathcal{H}=\frac{p^{2}}{2}+KV(q)\sum_{n}\delta(t-n),
\end{equation}
where $p,q$ stands for the momentum and real-space coordinate and $V(q+2\pi)=V(q)$, $q\in [0,2\pi)$. Such Hamiltonian yields a Floquet operator $U=\exp(-{\rm i} p^2/2\hbar)\exp(- {\rm i} KV(q)/\hbar)$, which can be quantized in a truncated Hilbert space with dimension $N$ with $p=i\hbar$, $i$ an integer between $-\frac{N}{2}$ and $\frac{N}{2}-1$, and $q=\frac{2\pi k}{N}$, $k$ an integer between $1$ and $N$ satisfying periodic boundary conditions in both $p$ and $q$. Explicitly,
\begin{equation}
\begin{split}
      U_{ij}=e^{-{\rm i} \Phi_i}\sum_{Q=1}^{N}F_{ik}e^{- {\rm i}KV(2\pi k/N)}F_{kj}^{-1},
\end{split}
\end{equation}
where $F_{jk}=\frac{1}{\sqrt{N}}e^{2 {\rm i}\pi jk/N}$. The phases corresponding to the kinetic energy $\Phi_{i}\equiv i^2\hbar/2$ are pseudo-random phases when $\hbar$ is irrational with $2 \pi$ \cite{PhysRevLett.49.509, birkhoff1931proof,PhysRevA.29.1639}, which are usually treated as fully-random phases for practical computation. If the real-space kicking potential of the rotor $V(q)$ is singular, this can induce long-range power-law decay in the off-diagonal entries of the corresponding unitary operator in momentum space  $|U_{ij}|\sim |i-j|^{-1}$. It was shown that 
systems with logarithmic singular potential \cite{PhysRevLett.94.244102}, or the RS model \cite{PhysRevLett.103.054103,PhysRevE.84.036212,PhysRevE.85.046208,PhysRevE.86.056215}, exhibit multifractal properties. These circular models offer advantages over their Hamiltonian counterpart, as they allow for efficient dynamical evolution through Fast Fourier Transforms \cite{santhanam2022quantum, PhysRevE.108.054127}; this enables to obtain eigenstates for large systems using sparse diagonalization with polynomial filters \cite{luitz2021polynomial}.

A remark is in order here. While it might seem that the PRBM and the unitary matrix models described above are already capable of describing strong multifractality, they access this regime asymptotically, e.g. by considering the limit of vanishing bandwidth $b \rightarrow 0$ in the PRBM case. However, this approach falls short of describing the critical properties at the AT$^\infty$ for the following reason. In the strongly multifractal regime $b \ll 1$ of the PRBM model, 
\begin{equation}
D_q\propto\frac{4b\rho(E)\sqrt{\pi}\Gamma(q-1/2)}{\Gamma(q)},
\end{equation}
for $q>q^*=1/2$, where $\rho(E)$ represents the density of diagonal elements, and the proportionality constant depends on the specific structure of the system \cite{PhysRevE.84.036212}. In other words, at finite $b$, $D_q$ does not tend to zero with system size but saturates to a small, finite value. 
To properly describe the critical behavior of the AT$^{\infty}$, one needs to work at infinite dimension, as effectively done by random graphs. The models we will introduce precisely address this requirement.


\subsection{Hermitian SRBM ensemble}
As explained above, to construct a random matrix ensemble describing the critical behavior of the Anderson transition in infinite dimensions (AT$^\infty$), characterized by a proper strong multifractal behavior where $D_q$ vanishes to $0$ as the system size $N$ tends to infinity for $q > q^* = 1/2$, we cannot rely on the asymptotic strong multifractal behavior of the PRBM model at small bandwidth $b\ll 1$. Alternatively, we observe that the localized phase ($a>1$) of the PRBM ensemble is characterized by algebraic localization and exhibits proper strong multifractal properties, where $D_q = 0$ for $q>q^*$, with $q^*<\frac12$ controlled by the algebraic localization "length", see \cite{PhysRevE.54.3221} and Fig.~\ref{figCorPRBM}. These properties resemble those of the localized phase of the AT$^{\infty}$. The critical behavior of the AT$^{\infty}$ can be regarded as the continuous limit of this localized-strong multifractal behavior as $q^* \rightarrow 1/2$ \cite{PhysRevResearch.2.012020, PhysRevB.106.214202}.

Hence, our approach is to consider the limit $a=1+0^{+}$ of the PRBM model, utilizing $|i-j|^{1+\epsilon}\simeq |i-j|(1+\epsilon \ln|i-j|)$ for $\epsilon \ll 1$. In our model, we retain only the term of order 1 in $\epsilon$.
We thus define the strongly multifractal random banded matrix (SRBM) ensemble as an ensemble of $N\times N$ Hermitian matrices $\hat{H}$ whose entries $H_{ij}$ are independent Gaussian random variables with  mean $\langle H_{ij}\rangle=0$ and variance $\langle|H_{ij}|^2\rangle = \beta^{-1}$ for $i=j$ and  
\begin{equation}
\label{defSRBM}
\langle|H_{ij}|^2\rangle =\frac{1}{1+[|i-j|\ln(1+|i-j|)/b]^2}
\end{equation}
for $i\neq j$. Again, $\beta$ is the Dyson index for the orthogonal ($\beta=1$) and unitary ($\beta=2$) classes, and $b>0$ is a real parameter. To reduce boundary effects in numerical simulations, as was done for the PRBM model \cite{PhysRevLett.84.3690, PhysRevB.62.7920}, we replace the term $|i-j|$ with $\sin(\pi|i-j|/N)/(\pi/N)$. 

In the limit $|i-j|\gg b$, the amplitude (standard deviation) of the off-diagonal elements decays as
\begin{equation}\label{eq3}
   \sqrt{ \langle|H_{ij}|^2\rangle}\simeq b \, (|i-j|\ln|i-j|)^{-1} \;.
\end{equation}
 The long-range decay described by Eq.~\eqref{eq3} with logarithmic dependence on $|i-j|$ will be seen to play an essential role in inducing log-multifractality in the following analytical derivations.
 
\begin{figure}
\includegraphics[width=0.48\textwidth]{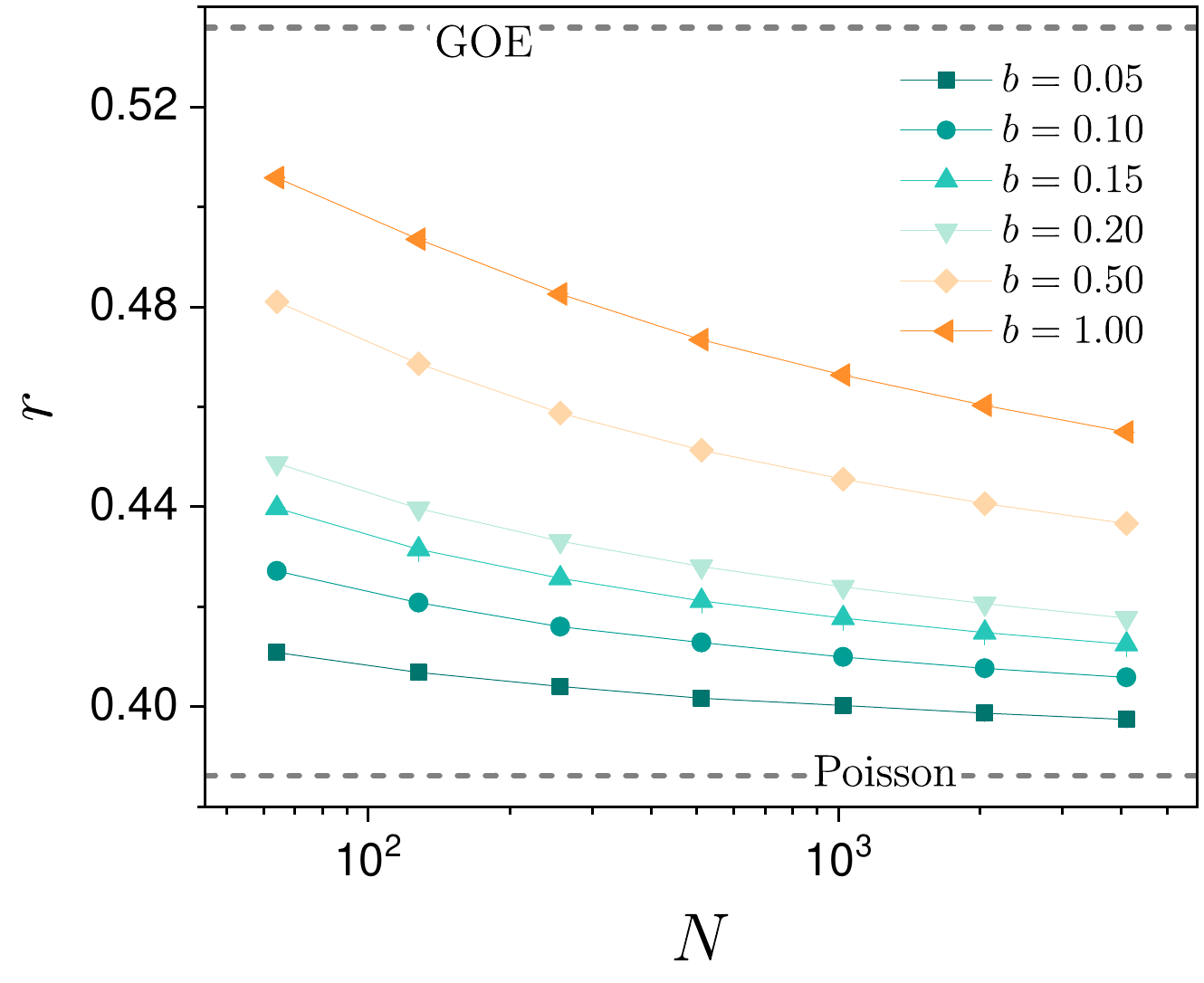}
    \caption{\label{figLevel} Average ratio of consecutive level spacings $r$ as a function of the system size $N$ for the SRBM model. The level statistics slowly converge to Poisson statistics. Disorder averaging ranges from $360,000$ realizations for $N=2^6$ to $18,000$ realizations for $N=2^{12}$. } 
\end{figure}

Before demonstrating in the next sections that this model exhibits logarithmic multifractality, we first compare its behavior with that of the PRBM model in terms of spectral statistics. In Fig.~\ref{figLevel}, we present numerical results for the mean level spacing ratio defined as
\begin{equation}
r=\Big\langle\frac{\text{min}(s_n,s_{n-1})}{\text{max}(s_n,s_{n-1})}\Big\rangle,
\end{equation}
where $s_n=e_{n+1}-e_n$ represents the nearest-neighbor spacings, with $e_n$ being an ordered sequence of eigenvalues \cite{PhysRevB.75.155111,PhysRevLett.110.084101,PhysRevX.12.011006}. The data indicate a slow convergence to Poisson statistics with increasing system size, in contrast to the intermediate statistics observed in the PRBM model for finite $b$ \cite{PhysRevB.61.R11859,Kravtsov_2006}. Poisson statistics are characteristic of the critical behavior at the AT$^\infty$ (see e.g. \cite{TIKHONOV2021168525, PhysRevResearch.2.012020, PhysRevB.106.214202}), consistent with the expectation $D_q=0$ for $q>1/2$.


\subsection{Unitary SRUM ensemble}
In the same way as circular ensembles were introduced with properties analogous to PRBM \cite{PhysRevLett.94.244102, PhysRevLett.103.054103,PhysRevE.84.036212,PhysRevE.85.046208,PhysRevE.86.056215}, we introduce a circular  ensemble as the counterpart to the SRBM, which we call the strongly multifractal random unitary matrix (SRUM) ensemble. Notice that the unitary matrix ensemble we are going to construct belongs to the orthogonal class ($\beta=1$) for real symmetric matrices since the time-reversal symmetry is preserved. The SRUM ensemble is comprised of random unitary matrices
\begin{equation}
\begin{split}
\label{defSRUM}
      U_{ij}=e^{{\rm i} \Phi_i}\sum_{k=1}^{N}F_{ik}e^{- {\rm i} KV\left[\frac{(2k+1)\pi}{N}\right]}F_{kj}^{-1},
\end{split}
\end{equation}
where $V(x) =\ln\left[-1/\ln(\lambda|\sin\frac{x}{2}|)\right]$ for $x\in [0,2\pi)$, $V(x+2\pi)=V(x)$ and the Fourier transform $F_{jk}=e^{2 {\rm i} \pi jk/N}/\sqrt{N}$. The parameter $\lambda$ is set to $\lambda=0.9$ to avoid the singularity of $V(x)$ at $x=\pi$. To avoid the singularity of $V(x)$ at $x=0\; (2\pi)$, the integer $k$ (from $1$ to $N$) is shifted as $k+1/2$ for numerical simulation. The $\Phi_{i}$ are random phases uniformly distributed over $[0,2\pi)$.  Due to the singular behavior of $V(x)$ when $x\rightarrow 0$ (or $2\pi$), the amplitudes of the matrix elements of $U_{ij}$ decay as
\begin{equation}\label{eq: Uij}
    |U_{ij}|\simeq \frac{K}{2|i-j|\ln |i-j|}
\end{equation}
for $|i-j|\gg1$, see App.~\ref{appendixA} for a detailed derivation of Eq.~\eqref{eq: Uij}.

\subsection{Generalized SRBM ensemble}
\label{secgenSRBM}
We shall also consider a generalization of the SRBM ensemble \eqref{defSRBM}, obtained by introducing a free parameter $\mu>0$ as an exponent controlling the logarithmic term in the decay of long-range hoppings $H_{ij}$. We define the generalized SRBM ensemble as the set of $N\times N$ Hermitian matrices $\hat{H}$ whose entries $H_{ij}$ are independent Gaussian random variables with mean $\langle H_{ij}\rangle=0$ and variance $\langle|H_{ij}|^2\rangle = \beta^{-1}$ for $i=j$ and  
\begin{equation}
\label{defGSRBM}
\langle|H_{ij}|^2\rangle =\frac{1}{1+\left[|i-j|\ln^{1+\mu}(1+|i-j|)/b\right]^2}
\end{equation}
for $i\neq j$. In this generalized ensemble, the amplitude (standard deviation) of the off-diagonal elements decays as
\begin{equation}
   \sqrt{ \langle|H_{ij}|^2\rangle}\simeq b \, \left(|i-j|\ln^{1+\mu}|i-j|\right)^{-1} \;,
\end{equation}
in the limit $|i-j|\gg b$. In section \ref{sec7}, we will show that this faster decay of the off-diagonal elements as compared to the SRBM ensemble \eqref{defSRBM} induces critical localization instead of logarithmic multifractality.

\subsection{A simple example}
\label{secinsight}
In the following sections we will calculate the multifractal dimensions of eigenvectors of the above models, either numerically or via perturbation theory. Before doing so, it is instructive to consider a case where multifractal dimensions can be calculated analytically exactly. Let $\psi$ be the state with components 
\begin{equation}
\label{defpsir}
    \psi_r=\frac{A}{r \ln r},\quad 2\leq r\leq N
\end{equation}
with $A$ a normalisation constant. This state corresponds to the vector of standard deviations \eqref{eq3} or to the vector of asymptotic matrix elements \eqref{eq: Uij}. The moments \eqref{eq:Pq} of $\psi$ are
\begin{equation}\label{eq:Pqexact}
P_q =A^{ 2q}\sum_{r=2}^N\frac{1}{(r \ln r)^{2q}}
\end{equation}
with
\begin{equation}
\label{a2q}
    A^{-2}=\sum_{r=2}^N\frac{1}{(r \ln r)^2}.
\end{equation}
The sum in \eqref{a2q} has a finite limit as $N\to\infty$. For $q>\frac12$, the sum in \eqref{eq:Pqexact} goes to a constant for $N\to\infty$, which implies from Eq.~\eqref{eq:Pq} that $D_q=0$. To obtain the leading-order expression of $P_q$ at large $N$ for $q<\frac12$, one can replace the sum by an integral
\begin{equation}
\sum_{r=2}^N\frac{1}{(r \ln r)^{2q}}\simeq \int_{2}^{N} \frac{dr}{(r\ln r)^{2q}}
\end{equation}
which, to leading order in $N$, behaves as
\begin{equation}
\label{sum2q}
    \int_{2}^{N} \frac{dr}{(r\ln r)^{2q}}\propto N^{1-2 q} (\ln N)^{-2 q}.
\end{equation}
Since we disregard the prefactors in \eqref{sum2q}, the same behavior remains valid for any fixed lower bound and any upper bound $\propto N$ in \eqref{sum2q}.
Therefore, for the state \eqref{defpsir} the multifractal dimensions defined in \eqref{eq:Pq} are \begin{equation}
D_q=\left \{\begin{array}{cc}0&\mathrm{for}\;q>1/2\\ \displaystyle{\frac{2q-1}{q-1}}&\mathrm{for}\;q<1/2\end{array}\right . \ ;
\label{final_D}
\end{equation}
note that for $q=\frac12$ we get  to leading order
\begin{equation}
\label{sum1}
\int_{2}^{N} \frac{dr}{r\ln r}=\ln \left(\frac{\ln N}{\ln (2)}\right)\simeq \ln(\ln N);
\end{equation}
for that integral, any fixed lower bound and any upper bound $\propto N$ yields the same behavior (including the prefactor).

\begin{figure}
\includegraphics[width=0.48\textwidth]{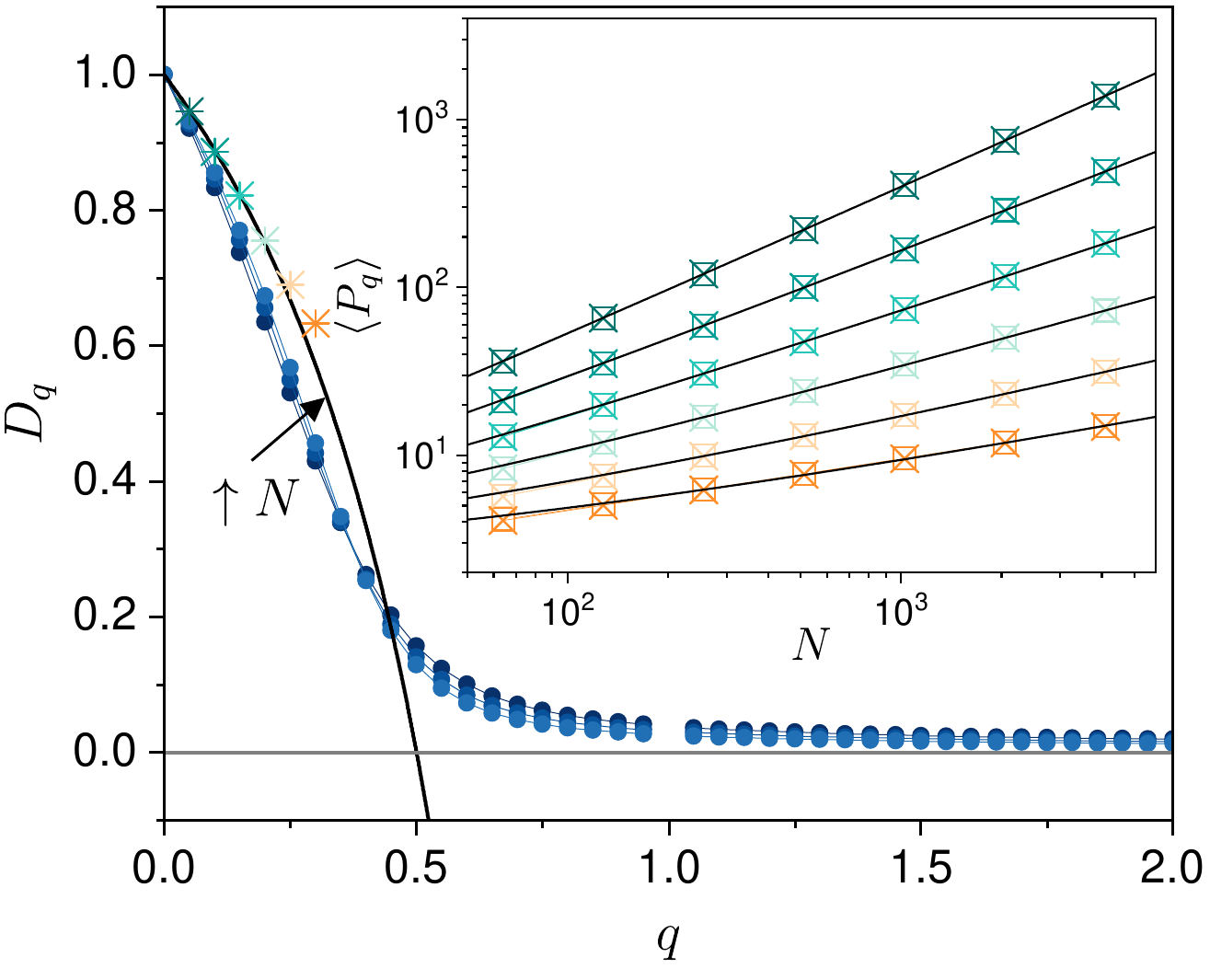}

    \caption{\label{fig1} Multifractal dimension $D_q$ vs $q$ for $b=0.05$ in the SRBM model with $\beta=1$. The finite-size estimate $D_q\equiv [\log_2 \langle P_q(N/2)\rangle-\log_2 \langle P_q(N) \rangle]/[q-1]$, represented by blue symbols (lines are an eyeguide) for system sizes $N=2^{8},2^{10},2^{12}$, converges slowly to the theoretical prediction $D_q=(2q-1)/(q-1)$ for $q<1/2$ and $D_q=0$ for $q>1/2$. The stars indicate the $D_q$ values obtained from the fits by Eq.~\eqref{eq:P2SRBM_} represented in the inset, which incorporate the log-corrections and agree well with $D_q=(2q-1)/(q-1)$. Inset: average (boxes) and typical (crosses) moments $P_q$ for $b=0.05$ and (from bottom to top) $q=0.05,0.10,\dots,0.30$; black dashed lines are fits by the conventional multifractality result Eq.~\eqref{eq:P2SRBM_}, with $A_{q}$ and $D_q$ two fitting parameters. Disorder averaging ranges from $360,000$ realizations for $N=2^6$ to $18,000$ realizations for $N=2^{12}$. }  
\end{figure}

\begin{figure*}
\includegraphics[width=0.46\textwidth]{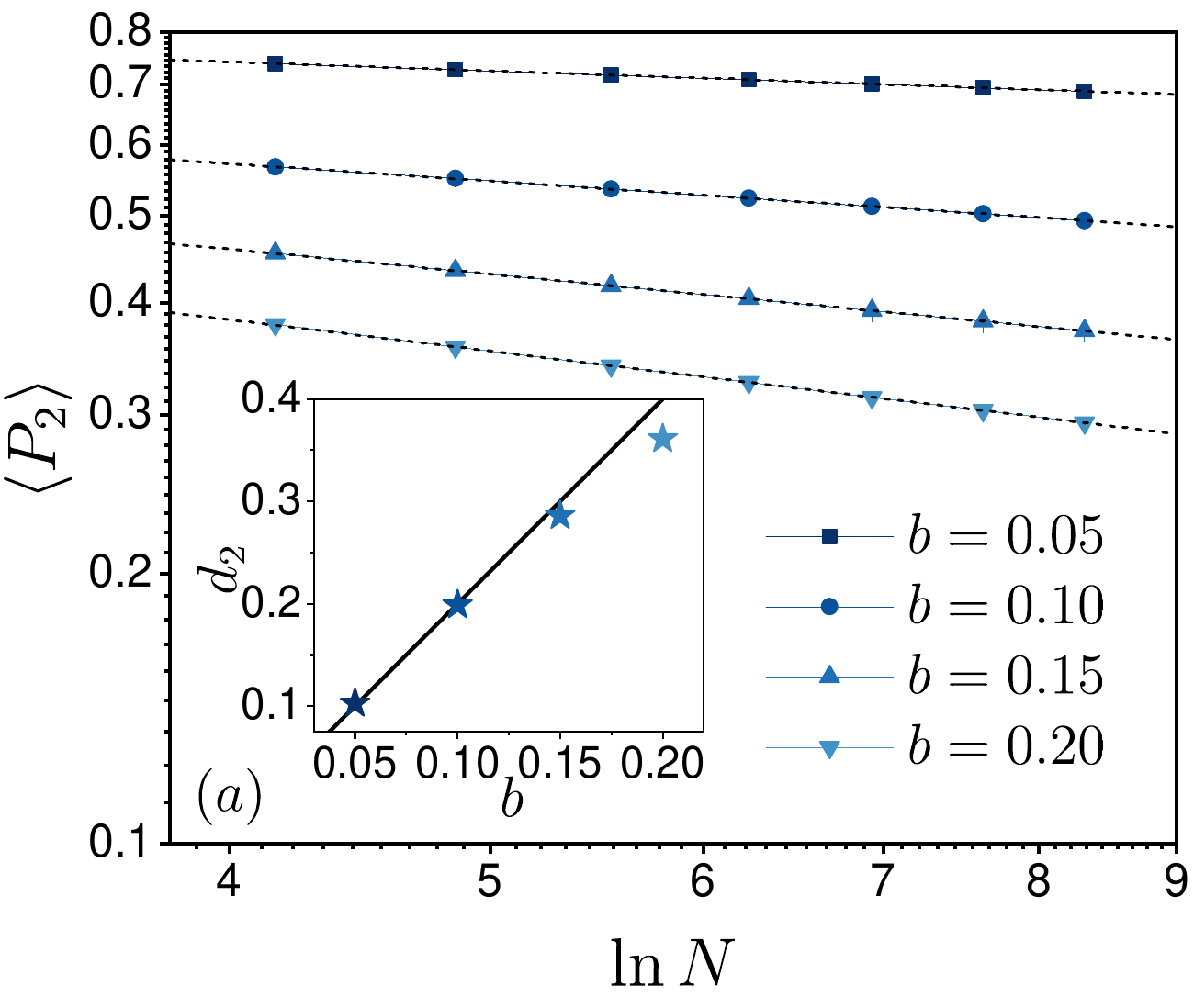}
\includegraphics[width=0.48\textwidth]{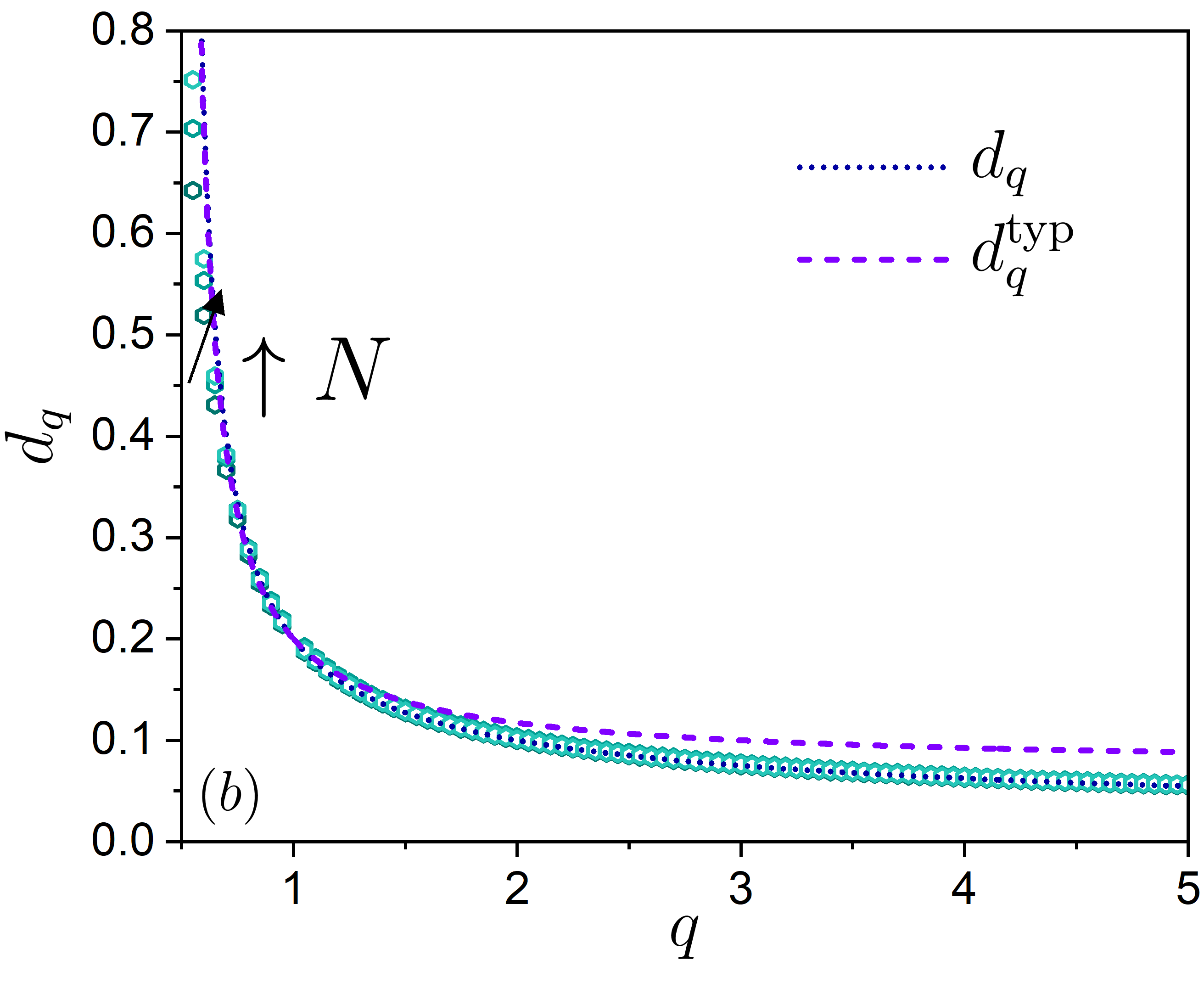}

    \caption{\label{fig2}   (a) Log-multifractality for the average moment $\langle P_2\rangle$ in the SRBM model with $\beta=1$, well described by Eq.~\eqref{eq:P2SRBM}. Different curves correspond to different $b$ values as indicated by the labels. The black dashed lines are power-law fits $\langle P_2\rangle = c (\ln N)^{-d_2}$ with $c$ and $d_2$ two fitting parameters. Inset: Comparison of the log-multifractal dimension $d_2$ obtained from fitting (symbols) and the analytical prediction Eq.~\eqref{eq:P2SRBM} (black solid line). (b) Average log-multifractal dimension $d_{q}$  (computed as $[\ln \langle P_{q}(N/2)\rangle-\ln \langle P_{q}(N) \rangle]/[(q-1)(\ln \ln N-\ln \ln \frac{N}{2})]$) as a function of $q$ for system sizes $N=2^{8},2^{10},2^{12}$. $d_{q}$ converges at large $N$ to the analytical law for the average $d_q$ \eqref{eq:dq} (blue dotted line). We also show the typical $d_q^{\text{typ}}$ for comparison (purple dashed line), see also Fig.~\ref{fig3}. Disorder averaging ranges from $360,000$ realizations for $N=2^6$ to $18,000$ realizations for $N=2^{12}$. } 
\end{figure*}

\begin{figure*}
\includegraphics[width=0.46\textwidth]{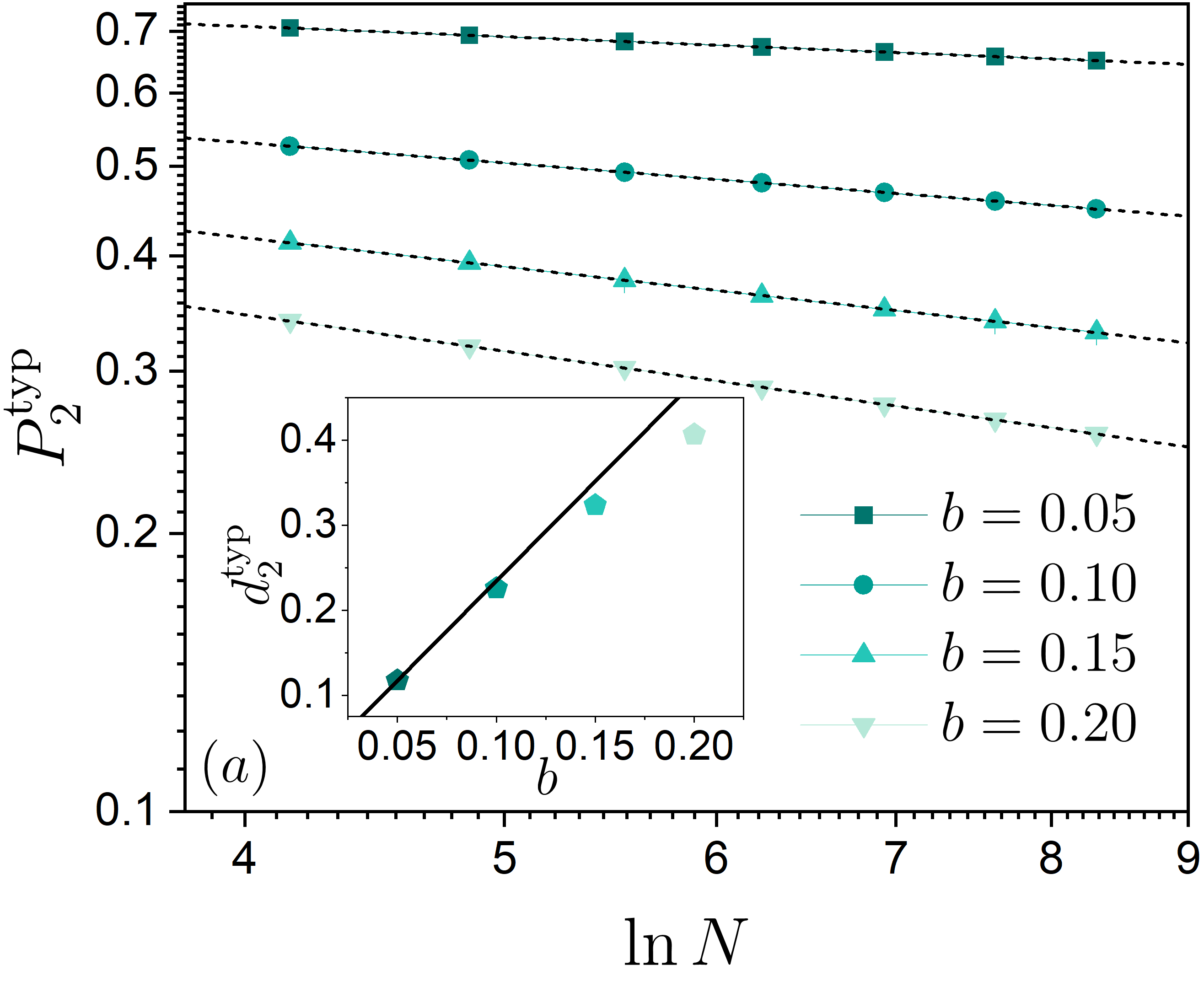}
\includegraphics[width=0.48\textwidth]{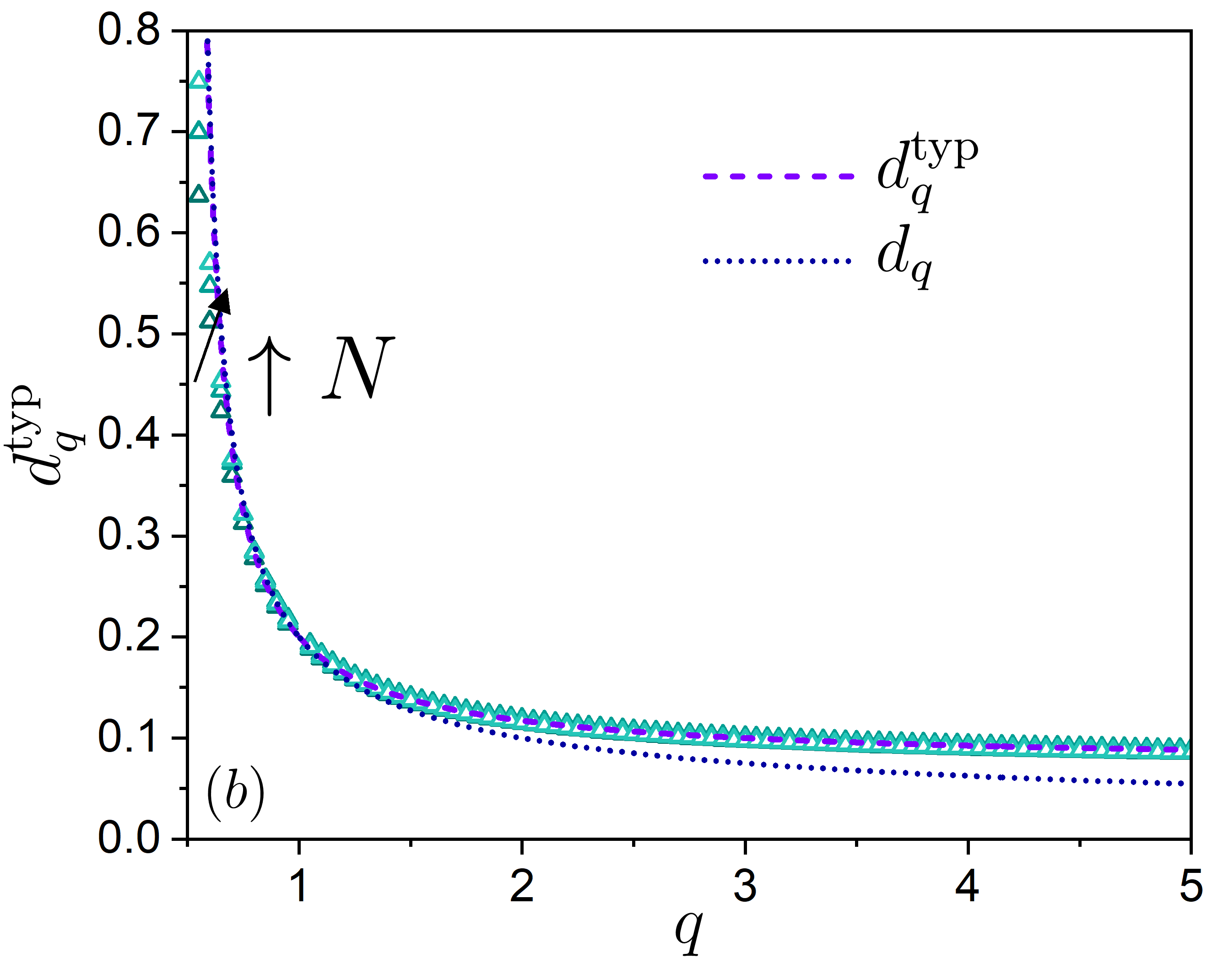}

    \caption{\label{fig3}   (a) Log-multifractality for the typical moment $ P_2^{\typ}$ in the SRBM model with $\beta=1$, well described by Eq.~\eqref{eq:P2SRBM}. Different curves correspond to different $b$ values as indicated by the labels. The black dashed lines are power-law fits $ P_2^{\typ} = c (\ln N)^{-d_2^{\typ}}$ with $c$ and $d_2^{\typ}$ two fitting parameters. Inset: Comparison of the log-multifractal dimension $d_2$ obtained from the fit (symbols) and the analytical prediction Eq.~\eqref{eq:P2SRBM} (black solid line). (b) Typical log-multifractal dimension $d_{q}^{\typ}$  (computed as $[\ln \langle P_{q}^{\typ}(N/2)\rangle-\ln \langle P_{q}^{\typ}(N) \rangle]/[(q-1)(\ln \ln N-\ln \ln \frac{N}{2})]$) as a function of $q$ for system sizes $N=2^{8},2^{10},2^{12}$. $d_{q}^{\typ}$ converges at large $N$ to the analytical law for the typical $d_q^\typ$ \eqref{eq:dq} (purple dashed line). We also show the average $d_q$ for comparison (blue dotted line), see also Fig.~\ref{fig3}. Disorder averaging ranges from $360,000$ realizations for $N=2^6$ to $18,000$ realizations for $N=2^{12}$. } 
\end{figure*}

\section{Logarithmic multifractality of SRBM through weighted L\'evy sums}\label{sec3}

In this Section we demonstrate the emergence of log-multifractality in the SRBM model, using a first-order perturbation approach aided by the technique of weighted L\'evy sums. The calculation essentially follows the derivation for the PRBM model described in detail in \cite{Monthus_2010}. In order for the present paper to be self-contained we recall the main steps of this derivation and adapt it to our case.

\subsection{Average moments $\langle P_q\rangle$}
We start from  the unperturbed order zero eigenstates  $|\psi_{i}^{0}\rangle=|i\rangle$ of the diagonal Hamiltonian $H_{ii}$. At first order in $b$, eigenfunctions are given by 
\begin{equation}
\label{psi1}
    |\psi_{i}^{1}\rangle=|i\rangle+\sum_{i\neq j} |j\rangle H_{ji}/(\varepsilon_i-\varepsilon_j)
\end{equation} 
with $\varepsilon_i=H_{ii}$ the unperturbed eigenenergy associated with state $|i\rangle$. 
States in \eqref{psi1} are not properly normalized. In order to calculate $P_q$ one introduces the normalization in the denominator, resulting in
\begin{equation}\label{eqpq}
  \langle P_{q}\rangle=\left\langle\frac{\sum_{j}{|\langle j|\psi_i^1\rangle|^{2q}}}{(\sum_{j}{|\langle j|\psi_i^1\rangle|^{2}})^{q}}\right\rangle=\left\langle\frac{1+\Sigma_{q}}{(1+\Sigma_{1})^{q}}\right\rangle,
\end{equation}
with 
\begin{equation}
\Sigma_{q}\equiv\sum_{\genfrac{}{}{0pt}{}{j=1}{j\neq i}}^N\frac{|H_{ij}|^{2q}}{|\varepsilon_i-\varepsilon_j|^{2q}}
\end{equation}
and the average in \eqref{eqpq} runs over labels $i$ and random realizations. 
Using the identity $ a^{-q}=\Gamma(q)^{-1}\int_{0}^{+\infty}t^{q-1}e^{-at}dt $, Eq.~\eqref{eqpq} yields\begin{equation}
\label{eq_pqex}
      \langle P_{q}\rangle=\frac{1}{\Gamma(q)}\int_{0}^{\infty} t^{q-1} e^{-t} \left(\langle e^{-t\Sigma_1}\rangle +\langle\Sigma_{q}e^{-t\Sigma_{1}}\rangle\right) dt.
\end{equation}
To 
simplify notation, we set $i-j\equiv\Vec{r}$, $r\equiv|\Vec{r}|$, and focus on the central site $i=N/2$ whose associated unperturbed eigenenergy is supposed to be at the band center, i.e., $\varepsilon_i \equiv 0$. Then, following Eq.~\eqref{eq3}, we have
\begin{equation}
\label{defSigmaq}
 \Sigma_{q}\simeq\sum_{\vec{r}\neq0}\left\vert\frac{V(\vec{r})}{\varepsilon_{\vec{r}}}\right\vert^{2q}, \qquad V(\vec{r})\underrel{r\gg b, b\rightarrow0}{\simeq}\frac{b}{r\ln r}u_{\vec{r}}.
\end{equation}
In Eq.~\eqref{defSigmaq}, the vector notation reminds that the sum runs over positive and negative values. For $\beta=1$, $u_{\vec{r}}$ is 
a standard normal Gaussian random variable such that 
\begin{equation}
    \label{defur}
\langle |u_{\Vec{r}}|^{2q}\rangle=\frac{2^q}{\sqrt{\pi}}\Gamma\left(q+\frac12\right).         
\end{equation}
For $\beta=2$, $u_{\vec{r}}$ is 
a standard complex normal Gaussian random variable with moments
\begin{equation}
    \label{defur_}
\langle |u_{\Vec{r}}|^{2q}\rangle=\Gamma\left(q+1\right).         
\end{equation}
Note that the moments $\langle|u_{\vec{r}}|^{2q}\rangle$ are independent of $\vec{r}$.

\begin{figure*}
\includegraphics[width=0.48\textwidth]{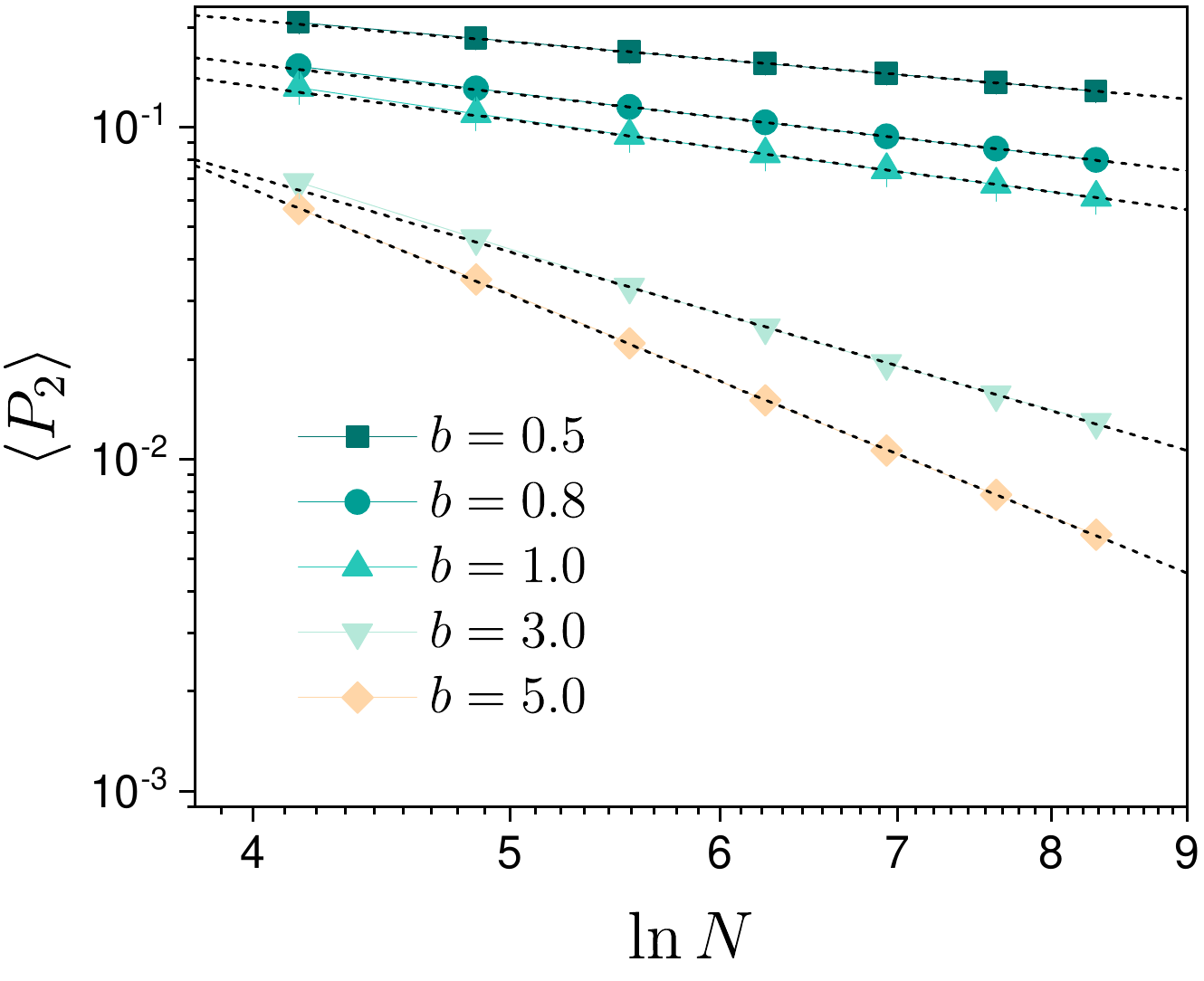}
\includegraphics[width=0.48\textwidth]{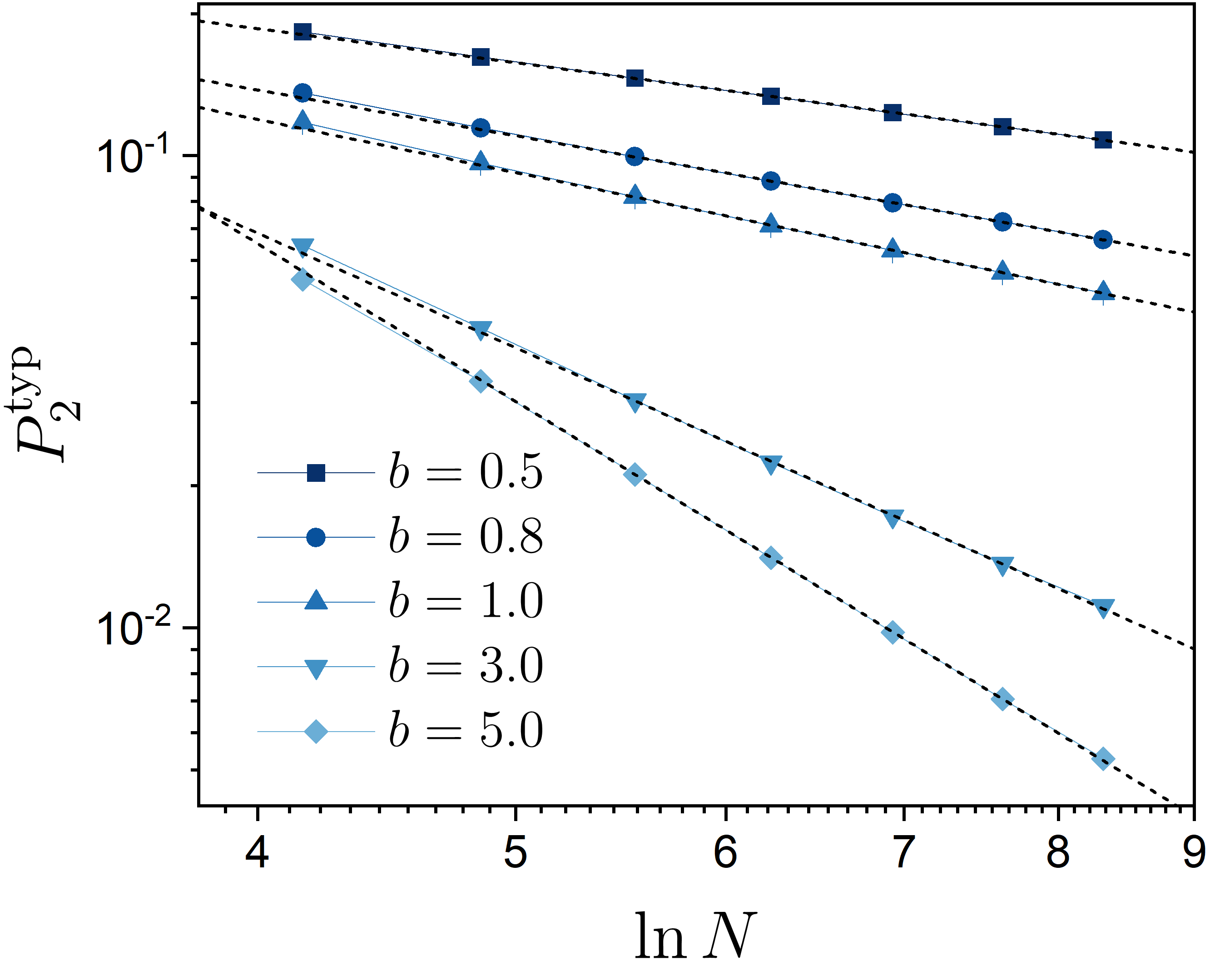}

    \caption{\label{figNP_SRBM} Log-multifractality for the average (left panel) and typical (right panel) moments $ P_2$ in the SRBM model with $\beta=1$, beyond the perturbative regime $b\ll 1$. Different curves correspond to different $b$ values as indicated by the labels. The black dashed lines are power-law fits $ \langle P_2\rangle = c (\ln N)^{-d_2}$ ($P_2^{\text{typ}} = c (\ln N)^{-d_2^{typ}} $, respectively) with $c$ and $d_2$ ($d_2^{typ}$, respectively) two fitting parameters. Disorder averaging ranges from $360,000$ realizations for $N=2^6$ to $18,000$ realizations for $N=2^{12}$.} 
\end{figure*}

The statistical average of $P_q$ in Eq.~\eqref{eq_pqex} can be studied by the statistics of the random variable $z_q=|\varepsilon_{\vec{r}}|^{-2q}$. Since $\varepsilon_{\vec{r}}$ is a Gaussian random variable with mean 0 and variance $1/\beta$, the probability distribution of $z_q$ reads
\begin{equation}
\label{pzqapprox}
    P(z_q)\underrel{z_q\rightarrow\infty}{\simeq}\sqrt{\frac{2\beta}{\pi}}\frac{\mu_q}{z_q^{1+\mu_q}}, \; \; \mu_q=\frac{1}{2q} \;.
\end{equation}
This distribution $P(z_q)$ has a fat tail analogous to L\'evy distributions, with exponent $-(1+\mu_q)$: the average of $ z_q $ diverges for $\mu_q<1$, i.e. $q>1/2$ and converges for $\mu_q>1$, i.e. $q<1/2$. 
In the following derivations, we will focus on the case of $\beta=1$.
Using \eqref{pzqapprox}, the moment generating function $\langle e^{-tz_q}\rangle$ can be evaluated in the limit $t\rightarrow0$ as
\begin{equation}
\label{etzqapprox}
 \begin{split}
       \langle e^{-tz_q} \rangle&\underrel{t\rightarrow0}{\simeq}1-\int_{0}^{\infty}\sqrt{\frac{2}{\pi}}\frac{\mu_q}{z_q^{1+\mu_q}}(1-e^{-tz_q})dz_q\\
      &\underrel{t\rightarrow0}{\simeq}1+\sqrt{\frac{2}{\pi}}t^{\mu_q}\mu_q\Gamma(-\mu_q)
\end{split}
\end{equation}
(see Eq.~(32) of \cite{Monthus_2010}). Similarly, since from \eqref{pzqapprox} we have $P(z_1)\sim z_1^{-3/2}$, the disorder average of $\langle z_1^q e^{-tz_1}\rangle$ can be evaluated in the same limit as
\begin{equation}\label{eqz1q}
\langle z_1^q e^{-tz_1}\rangle\underrel{t\rightarrow0}{\simeq}\langle z_1^q \rangle=\frac{\Gamma(1/2-q)}{2^q\sqrt{\pi}},\quad q<1/2,
\end{equation}
while for $q>1/2$ it yields
\begin{equation}
\label{z1qetz1}
\begin{split}
        \langle z_1^qe^{-tz_1} \rangle&\underrel{t\rightarrow0}{\simeq}\int_{0}^{\infty}\sqrt{\frac{2}{\pi}}\frac{1}{2z_1^{3/2}}z_1^qe^{-tz_1}dz_1\\
        &=\sqrt{\frac{1}{2\pi}}t^{\frac{1}{2}-q}\Gamma(q-\frac{1}{2}).
\end{split}
\end{equation}
Both averages in \eqref{eq_pqex} can be calculated in the same way as Eqs.~\eqref{etzqapprox}--\eqref{z1qetz1}.
First, if we assume independence of disorder at each site, the moment-generating function of $\Sigma_q$ expressed as \eqref{defSigmaq} reads
\begin{equation}
    \langle e^{-t\Sigma_q} \rangle=\prod_{\vec{r}}\left\langle e^{-t |V(\vec{r})|^{2q}|\varepsilon_{\vec{r}}|^{-2q}}\right\rangle\,.
\end{equation}
Since $V(\vec{r})$ and $\varepsilon_{\vec{r}}$ are independent random variables (the former corresponds to off-diagonal entries and the latter to diagonal ones), using Eq.~\eqref{etzqapprox} gives
\begin{equation}
\label{etSigmaq}
\begin{split}
        \langle e^{-t\Sigma_q} \rangle&\underrel{t\rightarrow0}{\simeq}\prod_{\vec{r}}\left(1+\sqrt{\frac{2}{\pi}}\langle |V(\vec{r})|\rangle t^{\mu_q}\mu_q\Gamma(-\mu_q)\right)\\
&\underrel{t\rightarrow0}{\simeq}\exp\left(\frac{4}{\pi}b\mu_q t^{\mu_q}\Gamma(-\mu_q)\ln \ln N\right)\; .
\end{split}
\end{equation}
In the last line we used
\begin{equation}
\label{eqsumv}
    \sum_{\vec{r}}\langle|V(\vec{r})|\rangle
\simeq 2b\langle |u_{\Vec{r}}|\rangle \ln\ln N\;
\end{equation}
(with a prefactor 2 coming from the sum over $\vec{r}$), 
which is a consequence of \eqref{sum1}, and $\langle |u_{\Vec{r}}|\rangle = \sqrt{2/\pi}$ for $\beta=1$, which comes from Eq.~\eqref{defur}.
Second, the average $\langle \Sigma_q e^{-t\Sigma_1}\rangle$ can be evaluated as
\begin{equation}\small\label{mom_gen_1}
\begin{split}
       \left\langle \Sigma_q e^{-t\Sigma_1}\right\rangle=\sum_{\vec{r}}&\left\langle|V(\vec{r})|^{2q} z_{\vec{r}}^q e^{-t|V(\vec{r})|^{2}z_{\vec{r}}}\right\rangle\\
       &\times  \prod_{\vec{r}^{\prime}\neq\vec{r}}\left\langle e^{-t|V(\vec{r^{\prime}})|^2z_{\vec{r}^{\prime}}}\right\rangle.
\end{split}
\end{equation}
For $q<1/2$ this yields
\begin{equation}\small\label{mom_gen_1b}
       \left\langle \Sigma_q e^{-t\Sigma_1}\right\rangle\underrel{t\rightarrow0}{\simeq}\sum_{\vec{r}}\left\langle|V(\vec{r})|^{2q}\right\rangle\langle z_1^q \rangle,
\end{equation}
and we have, as a consequence of Eq.~\eqref{sum2q},
\begin{equation}\label{eqsumv2q}
\sum_{\vec{r}}\langle|V(\vec{r})|^{2q}\rangle
\propto\langle |u_{\Vec{r}}|^{2q}\rangle b^{2q}N^{1-2q}(\ln N)^{-2q}\;.
\end{equation}
For $q>1/2$, the averages in \eqref{mom_gen_1} are obtained from \eqref{etzqapprox} (taken at $q=1$) and \eqref{z1qetz1}, with $t$ rescaled by the potentials $|V(\vec{r})|^2$ and $|V(\vec{r^{\prime}})|^2$; they give
\begin{equation}
\label{mom_gen_1b}
\begin{split}
           \left\langle \Sigma_q e^{-t\Sigma_1}\right \rangle\underrel{t\rightarrow0}{\simeq}\sum_{\vec{r}}&\langle|V(\vec{r})|\rangle 
       \sqrt{\frac{1}{2\pi}}t^{\frac{1}{2}-q}\Gamma\left(q-\frac{1}{2}\right)\\
     \times & \prod_{\vec{r}^{\prime}\neq\vec{r}}\left[1+\frac12\sqrt{\frac{2}{\pi}}\langle |V(\vec{r}^{\prime})|\rangle t^{\frac12}\Gamma\left(-\frac12\right)\right]\\
        \underrel{t\rightarrow0}{\simeq}t^{\frac{1}{2}-q}&\sum_{\vec{r}}\langle|V(\vec{r})|\rangle\frac{1}{\sqrt{2\pi}}\Gamma\left(q-\frac{1}{2}\right).
\end{split}
\end{equation}
Plugging all these expressions into Eq.~\eqref{eq_pqex} we finally obtain for $q<1/2$ 
\begin{equation}
      \langle P_{q}\rangle
     \underrel{N\rightarrow\infty}{\simeq}1+a_q b^{2q}N^{1-2q}(\ln N)^{-2q}
\label{pqavqpetit}
\end{equation}
with $a_q$ a prefactor, and for $q>1/2$ 
\begin{equation}\small
\begin{split}
      \langle P_{q}\rangle&=1-\frac{\Gamma(q+1/2)}{\Gamma(q)}\frac{4b}{\sqrt{\pi}}\ln\ln N+\frac{\Gamma(q-1/2)}{\Gamma(q)}\frac{2b}{\sqrt{\pi}}\ln\ln N\\
     &=1-\frac{4b}{\sqrt{\pi}}\frac{\Gamma(q-1/2)}{\Gamma(q-1)}\ln\ln N\\
     & \underrel{b\rightarrow0}{\simeq}(\ln N)^{-\frac{4b}{\sqrt{\pi}}\frac{\Gamma(q-1/2)}{\Gamma(q-1)}}.
\end{split}
\label{pqav}
\end{equation}

\subsection{Typical moments $P_q^{\typ}$}
Equations \eqref{pqavqpetit} and \eqref{pqav} give the average value of the moments $P_q$. The typical value of $P_q$, defined as $P_q^{\typ}\equiv\exp \langle \ln P_q\rangle$, can also be obtained through the same approach \cite{Monthus_2010}. Indeed, Eq.~\eqref{eqpq} can be written as 
\begin{equation}\label{eq: Pqtyp}
 \langle \ln P_{q}\rangle=\langle\ln (1+\Sigma_{q})\rangle-q\langle\ln (1+\Sigma_{1})\rangle,
\end{equation}
therefore the evaluation of $ P_q^{\typ}$ can be achieved by studying $\langle\ln (1+\Sigma_{q})\rangle$. For $q<1/2$, the disorder average of $\Sigma_{q}$ converges, yielding, from \eqref{mom_gen_1b}, 
\begin{equation}\label{logSigmaq}
\begin{split}
      \langle\ln (1+\Sigma_{q})\rangle&\simeq\langle \Sigma_{q}\rangle =\sum_{\vec{r}}\langle|V(\vec{r})|^{2q}\rangle\langle z_q \rangle;
\end{split}    
\end{equation}
for $q>1/2$, we represent  $\langle\ln (1+\Sigma_{q})\rangle$ as 
\begin{equation}
    \langle\ln (1+\Sigma_{q})\rangle=\int_{0}^{\infty}dt\frac{e^{-t}}{t}\left(1-\langle e^{-t \Sigma_{q}}\rangle\right);
\end{equation}
it is dominated by the region around $t=0$; taking into account Eq.~\eqref{etSigmaq}, this integral can be easily evaluated as
\begin{equation}\small\label{eq:lnSigmaq}
\begin{split}
    \langle\ln (1+\Sigma_{q})\rangle&\simeq\int_{0}^{\infty}dt\frac{e^{-t}}{t}\left(1-e^{\frac{4b\mu_q}{\pi}t^{\mu_q}\Gamma(-\mu_q)\ln \ln N}\right)\\
    &\simeq -\int_{0}^{\infty}dt\frac{e^{-t}}{t}\frac{4b\mu_q}{\pi}t^{\mu_q}\Gamma(-\mu_q)\ln \ln N\\
    &=\frac{4b}{\sin(\pi\mu_q)}\ln\ln N.
\end{split}
\end{equation}
 For $q<1/2$, the term $\langle\ln (1+\Sigma_{q})\rangle$ in Eq.~\eqref{eq: Pqtyp} dominates, hence
\begin{equation}
\begin{split}
        P_q^{\typ}&=\exp \langle \ln P_q\rangle\simeq \exp{\langle\Sigma_q\rangle}\\
        &\simeq 1+\frac{2b^{2q}\Gamma(1/2+q)\Gamma(1/2-q)N^{1-2q}(\ln N)^{-2q}}{2^{1-2q}\pi(1-2q)}.
\end{split}
\end{equation}
For $q>1/2$, substituting Eq.~\eqref{eq:lnSigmaq} into Eq.~\eqref{eq: Pqtyp}, the result reads
\begin{equation}
\begin{split}
     \langle \ln P_{q}\rangle&=\frac{4b}{\sin(\pi\mu_q)}\ln\ln N-q\frac{4b}{\sin(\pi\mu_1)}\ln\ln N\\
     &=4b\left[\frac{1}{\sin\left(\frac{\pi}{2q}\right)}-q \right]\ln \ln N,
\end{split}    
\end{equation}
i.e.,
\begin{equation}
     P_q^{\typ}=(\ln N)^{-4b\left[q-\frac{1}{\sin(\frac{\pi}{2q})} \right]}.
\end{equation}

\subsection{Log-multifractal dimensions}

In summary, two behaviors of the eigenstate moments are demonstrated: For $q<1/2$, both the average and typical eigenstate moments show conventional multifractality, but with logarithmic finite-size correction, i.e.,
\begin{equation}\label{eq:P2SRBM_}
      \langle P_{q}\rangle=P_q^{\typ} \simeq1+ A_{q} \, b^{2q} \,  (\ln N)^{-2q} N^{-D_q(q-1)}
\end{equation}
with $D_q=\frac{2q-1}{q-1}$ and $A_{q}$ a constant. For $q>1/2$, both the average and typical eigenstate moments show an algebraic behavior in the logarithm of the system size (which we call ``log-multifractality''), namely
\begin{equation}\label{eq:P2SRBM}
\begin{split}
      \langle P_{q}\rangle\sim  (\ln N)^{-d_q (q-1)}\,,\quad P_{q}^{\typ}\sim  (\ln N)^{-d_q^{\typ} (q-1)}\,,
\end{split}
\end{equation}
with two distinct log-multifractal dimensions
\begin{equation}\small\label{eq:dq}
    d_q = \frac{4b\Gamma(q-\frac{1}{2})}{\sqrt{\pi}\Gamma(q)}\;, \quad d_{q}^{\typ} = \frac{4b}{q-1}\left[q-\frac{1}{\sin(\frac{\pi}{2q})} \right]\;.
\end{equation}
The log-multifractal dimensions for $\beta=2$ can be obtained in the same way by replacing Eq.~\eqref{defur} with Eq.~\eqref{defur_} and taking into account the $\beta$-dependence in Eq.~\eqref{pzqapprox}, which leads to
\begin{equation}\small\label{eq:dq_}
    d_q = \frac{2\sqrt{\pi}b\Gamma(q-\frac{1}{2})}{\Gamma(q)}\;, \quad d_{q}^{\typ} = \frac{2\pi b}{q-1}\left[q-\frac{1}{\sin(\frac{\pi}{2q})} \right]\;.
\end{equation}
To validate the analytical predictions Eqs.~\eqref{eq:P2SRBM_}--\eqref{eq:dq}, we perform exact diagonalization of the SRBM model with high number of random realizations. Figure \ref{fig1} illustrates the conventional multifractal behavior of moments $P_q$ with $q<\frac12$, which agrees well with Eq.~\eqref{eq:P2SRBM_}. In Figs.~\ref{fig2} and \ref{fig3} we show the log-multifractality of both average moments $P_q$ and typical moments $P_q^{\textrm{typ}}$ with $q>\frac12$, fitting effectively with Eq.~\eqref{eq:P2SRBM}. The log-multifractal dimensions $d_q$ exhibit a nontrivial dependency on $q$ and $b$, well-accounted for by Eq.~\eqref{eq:dq}. In addition, we have verified the emergence of log-multifractality outside the perturbative regime for relatively large $b$ values, see Fig.~\ref{figNP_SRBM} for more details.

\section{Logarithmic multifractality of SRUM through Levitov renormalization}\label{sec4}

\begin{figure}
\includegraphics[width=0.48\textwidth]{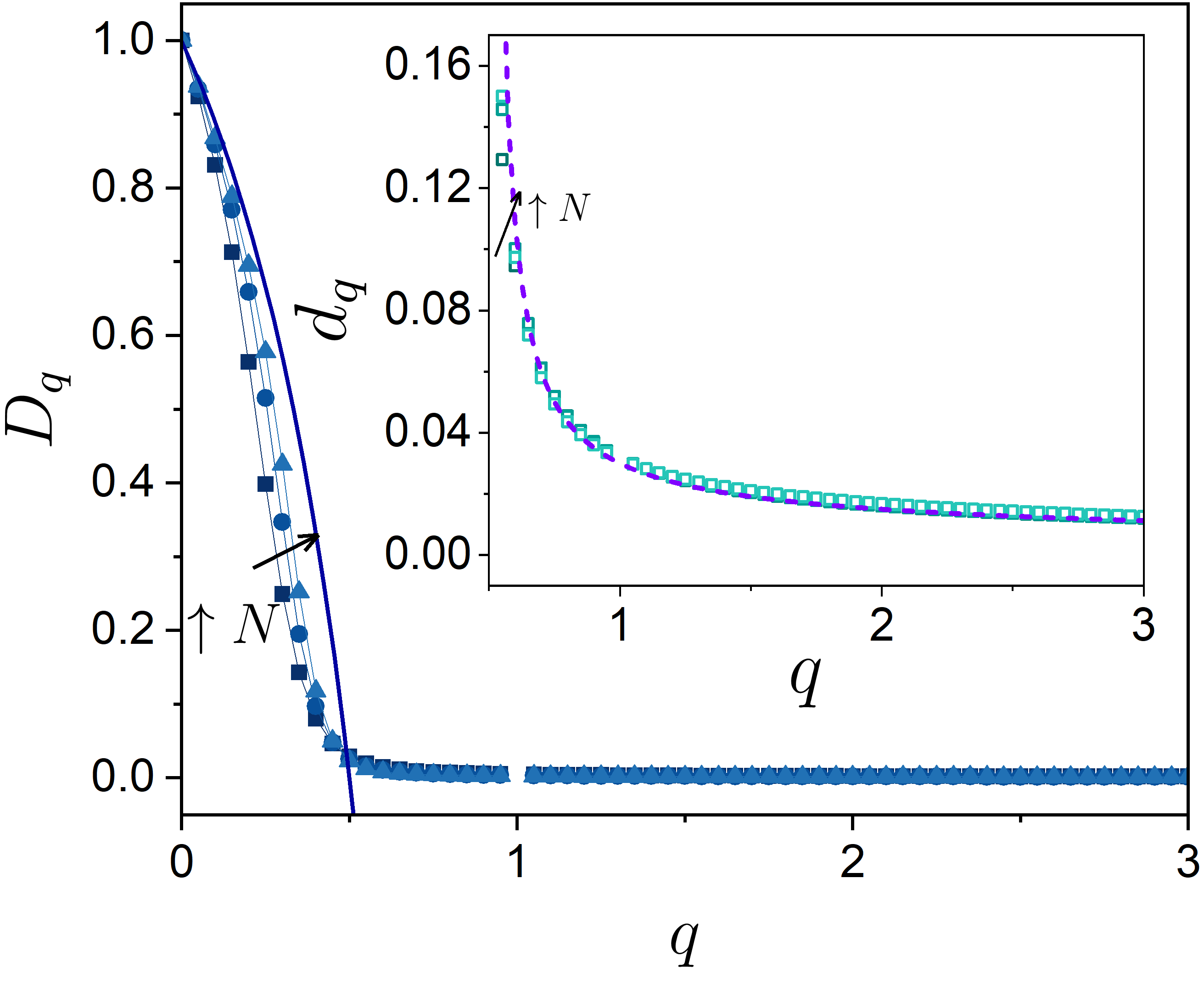}

    \caption{\label{figSRUM} Multifractal dimension $D_q$ vs $q$ for $K=0.03$ in the SRUM model. The finite-size estimate $D_q\equiv [\log_2 \langle P_q(N/2)\rangle-\log_2 \langle P_q(N) \rangle]/[q-1]$, represented by blue symbols (lines are eyeguide) for system sizes $N=2^{10},2^{14},2^{18}$, converges slowly to the theoretical prediction $D_q=(2q-1)/(q-1)$ for $q<1/2$. Inset:  Log-multifractal dimension $d_q$ (computed as $[\ln \langle P_q(N/2)\rangle-\ln \langle P_q(N) \rangle]/[(q-1)(\ln \ln N-\ln \ln \frac{N}{2})]$) as a function of $q$ for system sizes $N=2^{10},2^{14},2^{18}$. $d_q$ converges at large $N$ to the nontrivial analytical law \eqref{eq: P2SRUM} (violet line).  Disorder averaging ranges from $72,000$ realizations for $N=2^{10}$ to $1,800$ realizations for $N=2^{18}$. } 
\end{figure}

\begin{figure*}
\centering
\includegraphics[width=0.48\textwidth]{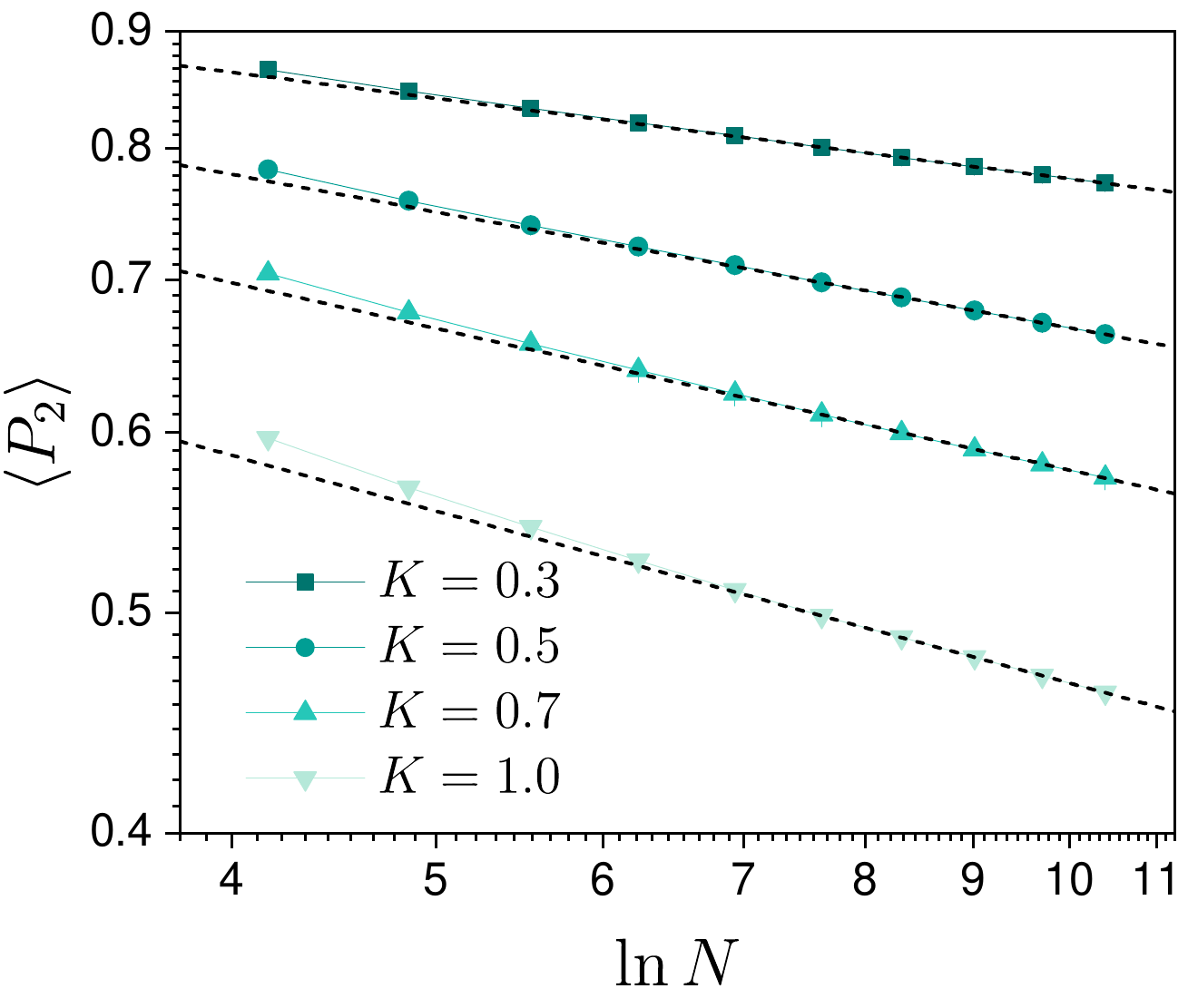}
\includegraphics[width=0.48\textwidth]{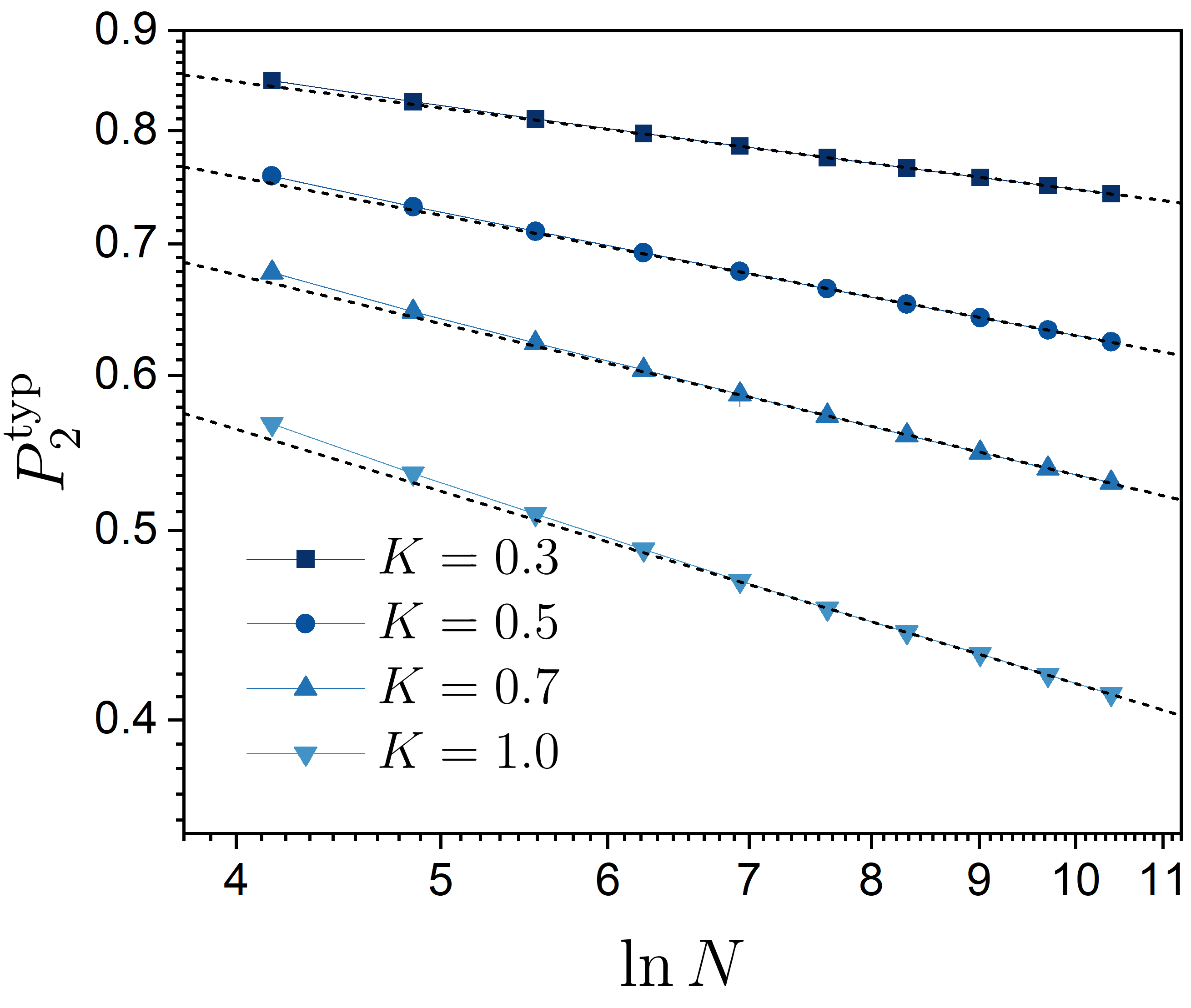}

    \caption{\label{figNP_SRUM}Log-multifractality for the average (left panel) and typical (right panel) moments $ P_2$ in the SRUM model, beyond the perturbative regime $K\ll 1$. Different curves correspond to different $K$ values as indicated by the labels. The black dashed lines are power-law fits $ \langle P_2\rangle = c (\ln N)^{-d_2}$ ($P_2^{\text{typ}} = c (\ln N)^{-d_2^{typ}} $, respectively) with $c$ and $d_2$ ($d_2^{typ}$, respectively) two fitting parameters. Disorder averaging ranges from $5,400,000$ realizations for $N=2^6$ to $36,000$ realizations for $N=2^{15}$. } 
\end{figure*}

In this Section, we employ another perturbation approach called Levitov renormalization \cite{L.S.Levitov_1989,PhysRevLett.64.547,PhysRevE.84.036212,PhysRevE.85.046208} to analytically compute $\langle P_q\rangle$ in the case of the SRUM model \eqref{defSRUM}. As for the previous calculation, the unperturbed matrix is diagonal and at order zero eigenstates are $|\psi_{i}^{0}\rangle=|i\rangle$. At first order in $K$, eigenfunctions are given by $ |\psi_{i}^{1}\rangle=|i\rangle+\sum_{i\neq j} |j\rangle U_{ji}/(e^{{\rm i}\Phi_i}-e^{{\rm i}\Phi_j})$. The moments $P_q$ averaged over all eigenvectors $i$ then read
\begin{equation}\small\label{eq9}
\begin{split}
      \langle P_q\rangle&\simeq 1+\frac{1}{N}\sum_{i\neq j}\langle|e^{{\rm i}\Phi_i}-e^{{\rm i}\Phi_j}|^{-2q}\rangle|U_{ij}|^{2q}
\end{split}
\end{equation}
(note that here, by contrast with the SRBM model, disorder only appears in the unperturbed eigenphases and $|U_{ij}|$ are nonrandom).
The disorder averaging
can be performed as 
\begin{equation}
    \langle|e^{{\rm i}\Phi_i}-e^{{\rm i}\Phi_j}|^{-2q}\rangle=\frac{1}{2\pi}\int_0^{2\pi}\frac{1}{|1-e^{{\rm i}\phi}|^{2q}}d\phi \; .
\end{equation}
For $q<\frac12$, this integral converges, and the result reads
\begin{equation}
      \langle|e^{{\rm i}\Phi_i}-e^{{\rm i}\Phi_j}|^{-2q}\rangle= \frac{\Gamma(\frac{1}{2}-q)}{2^{2q}\sqrt{\pi}\Gamma(1-q)}\;.
\end{equation}
The remaining sum is $\frac{1}{N}\sum_{i\neq j} |U_{ij}|$, where $|U_{ij}|$ only depends on $r=\vert i-j\vert$
and has an asymptotic behavior given by \eqref{eq: Uij}. This sum is again of the form of Eq.~\eqref{sum2q}, yielding for large $N$
\begin{equation}
   \frac{1}{N} \sum_{i\neq j} |U_{ij}|^{2q}\propto K^{2q}N^{1-2q}(\ln N)^{-2q}.
\end{equation}
Inserting this expression into Eq.~\eqref{eq9} finally yields for $q<1/2$
\begin{equation}\small\label{eq: P2SRUM_}
      \langle P_{q}\rangle \simeq1+ b_q K^{2q}N^{1-2q}(\ln N)^{-2q}
\end{equation}
with $b_q$ a constant.

When $q>\frac12$ the above disorder averaging diverges, as in the previous case of the SRBM model, necessitating more advanced treatments. In this context, we employ Levitov renormalization \cite{L.S.Levitov_1989,PhysRevLett.64.547}, known for its effectiveness in the PRBM and RS ensembles \cite{PhysRevB.62.7920, PhysRevE.84.036212}. We start with introducing a rescaled operator 
\begin{equation}
\begin{split}
      M_{ij}= U_{ij}\exp{\left[{\rm i}\frac{K}{N}\sum_{Q=1}^{N}V(2\pi Q/N)\right]},
\end{split}
\end{equation}
which has the same multifractal properties of its eigenstates as the original unitary operator $U$ since such a trivial global rescaling does not affect the eigenstates. The operator $M$ can be expanded at first order in the strongly-multifractal limit ($K\rightarrow0$) as
\begin{equation}
\begin{split}
M_{ij}=e^{{\rm i}\Phi_{i}}\delta_{ij}-iKe^{{\rm i}\Phi_{i}} W_{ij},\quad W_{ij}\equiv\sum_{k=1}^{N}F_{ik}V\!\!\left(\frac{2\pi k}{N}\right)\!F_{kj}^{-1}.
\end{split}
\end{equation}
The eigenstate moments $\langle P_q\rangle$ for states with eigenphases $\theta_{\alpha}$ around $\theta$ are defined as
\begin{equation}
    \langle P_q\rangle =\frac{1}{\rho(\theta)}\left\langle\frac{1}{N}\sum_{\alpha}\sum_{i}|\psi_i(\alpha)|^{2q}\delta(\theta-\theta_{\alpha})\right\rangle,
    \end{equation}
where $\rho(\theta)$ is the density of (eigen)states and the average is over random phases. We proceed by taking into account the contribution of all the $2\times2$ submatrices of $M$ to the eigenstate moments. These $2\times2$ matrices read, at first order, 
\begin{equation}
 \begin{pmatrix}
e^{{\rm i}\Phi_{i}}& -ie^{{\rm i}\Phi_{i}}K W_{ij}\\
-ie^{{\rm i}\Phi_{j}}K W_{ji} & e^{{\rm i}\Phi_{j}}
\end{pmatrix}.
\end{equation}
They can be diagonalized with eigenvalues
\begin{equation}
    \lambda_\sigma=e^{{\rm i}\gamma}\left(\cos\beta+{\rm i}\sigma\sqrt{h_{ij}^2+\sin^2\beta}\right),
\end{equation}
where $\sigma=\pm 1$, $\gamma=(\Phi_{i}+\Phi_{j})/2$, $\beta=(\Phi_{i}-\Phi_{j})/2$ and $h_{ij}\equiv K|W_{ij}|$. The corresponding eigenvectors are given by
\begin{equation}
\label{eigenvectorsuv}
\begin{split}
    |u_\sigma|^2&=\frac{h_{ij}^2}{\left|\sin\beta+{\rm i}\sigma\sqrt{h_{ij}^2+\sin^2\beta}\right|^2+h_{ij}^2}\\
    |v_\sigma|^2&=\frac{\left|\sin\beta+{\rm i}\sigma\sqrt{h_{ij}^2+\sin^2\beta}\right|^2}{\left|\sin\beta+{\rm i}\sigma\sqrt{h_{ij}^2+\sin^2\beta}\right|^2+h_{ij}^2}.
\end{split}    
\end{equation}
The contribution from eigenvectors of all $2\times2$ submatrices is then
\begin{equation}
    \langle P_q\rangle =\frac{1}{\rho(\theta)}\left\langle\frac{1}{N}\sum_{i<j}\sum_{\sigma=\pm}(|u_\sigma|^{2q}+|v_\sigma|^{2q})\delta(\theta-\theta_{\sigma})\right\rangle.
\end{equation}
The average over random phases, $\frac{1}{(2\pi)^2}\int_{0}^{2\pi}d\Phi_{i}\int_{0}^{2\pi}d\Phi_{j}$, becomes an average over $\beta$ and $\gamma$, that is, $2\frac{1}{(2\pi)^2}\int_{0}^{\pi}d\beta\int_{0}^{2\pi}d\gamma$. Since changing $\beta$ to $\beta+\pi$ exchanges the eigenvectors $\pm$, the average over $\beta$ and $\gamma$ can also be taken as $\frac{1}{2\pi^2}\int_{-\pi/2}^{\pi/2}d\beta\int_{0}^{2\pi}d\gamma$.
The average over $\gamma$ trivially cancels the density of states $\rho(\theta)$ with the $\delta$ function, and yields 
\begin{equation}\label{ordre1}
\langle P_q\rangle=  \frac{1}{N}\sum_{i<j}\sum_{\sigma=\pm}\left\langle|u_\sigma|^{2q}+|v_\sigma|^{2q}-1\right\rangle
\end{equation}
for the first-order contribution. The average over $\beta$ can be done by doing the change of variables with $|u_\sigma|^{2}=1/(1+e^{2\sigma t})$ and $|v_\sigma|^{2}=1/(1+e^{-2\sigma t})$. We obtain:
\begin{equation}\small
\begin{split}
    &\frac{1}{\pi}\int_{-\frac{\pi}{2}}^{\frac{\pi}{2}}\left(|u_\sigma|^{2q}+|v_\sigma|^{2q}-1\right)d\beta\\
    &=\frac{h_{ij}}{\pi}\int_{-\sinh^{-1}\frac{1}{h_{ij}}}^{\sinh^{-1}\frac{1}{h_{ij}}}\frac{\cosh t}{\sqrt{1-h_{ij}^2\sinh^2 t}}\left[\frac{2\cosh(qt)}{(2\cosh t)^q}-1\right]dt\\
    &\approx-\frac{h_{ij} \, \Gamma(q-\frac{1}{2})}{\sqrt{\pi}\Gamma(q-1)}
\end{split}    
\end{equation}
at lowest order. Substituting that expression into Eq.~\eqref{ordre1}, with $h_{ij}=K|W_{ij}|$, yields
\begin{equation}
\label{pqaux}
      \langle P_{q}\rangle=1-\frac{2}{N}\frac{\Gamma(q-\frac{1}{2})}{\sqrt{\pi}\Gamma(q-1)}\sum_{i<j}K|W_{ij}|
\end{equation}
(the coefficient 2 in \eqref{pqaux} comes from the sum over $\sigma$).
Using the asymptotic form \eqref{uijapp} for $K|W_{ij}|$, the double sum is of the form \eqref{sum1}, which yields
\begin{equation}\label{eq: P2SRUM}
\begin{split}
      \langle P_{q}\rangle&= 1-\frac{K\Gamma(q-\frac{1}{2})}{ \sqrt{\pi}\Gamma(q-1)}\ln(\ln N)\simeq  (\ln N)^{-\frac{K\Gamma(q-\frac{1}{2})}{\sqrt{\pi}\Gamma(q-1)}}.
\end{split}
\end{equation}
In Fig.~\ref{figSRUM} we present the results of large-scale numerical simulations of the SRUM model. These numerical results agree well with the analytical predictions Eqs.~\eqref{eq: P2SRUM_} and \eqref{eq: P2SRUM}. We refer the reader to \cite{chen2023quantum} for other numerical checks of these results. The emergence of log-multifractality is also demonstrated outside the perturbative regime for relatively large $K$ values in Fig.~\ref{figNP_SRUM}.

 \begin{figure*}[t]
\includegraphics[width=0.48\textwidth]{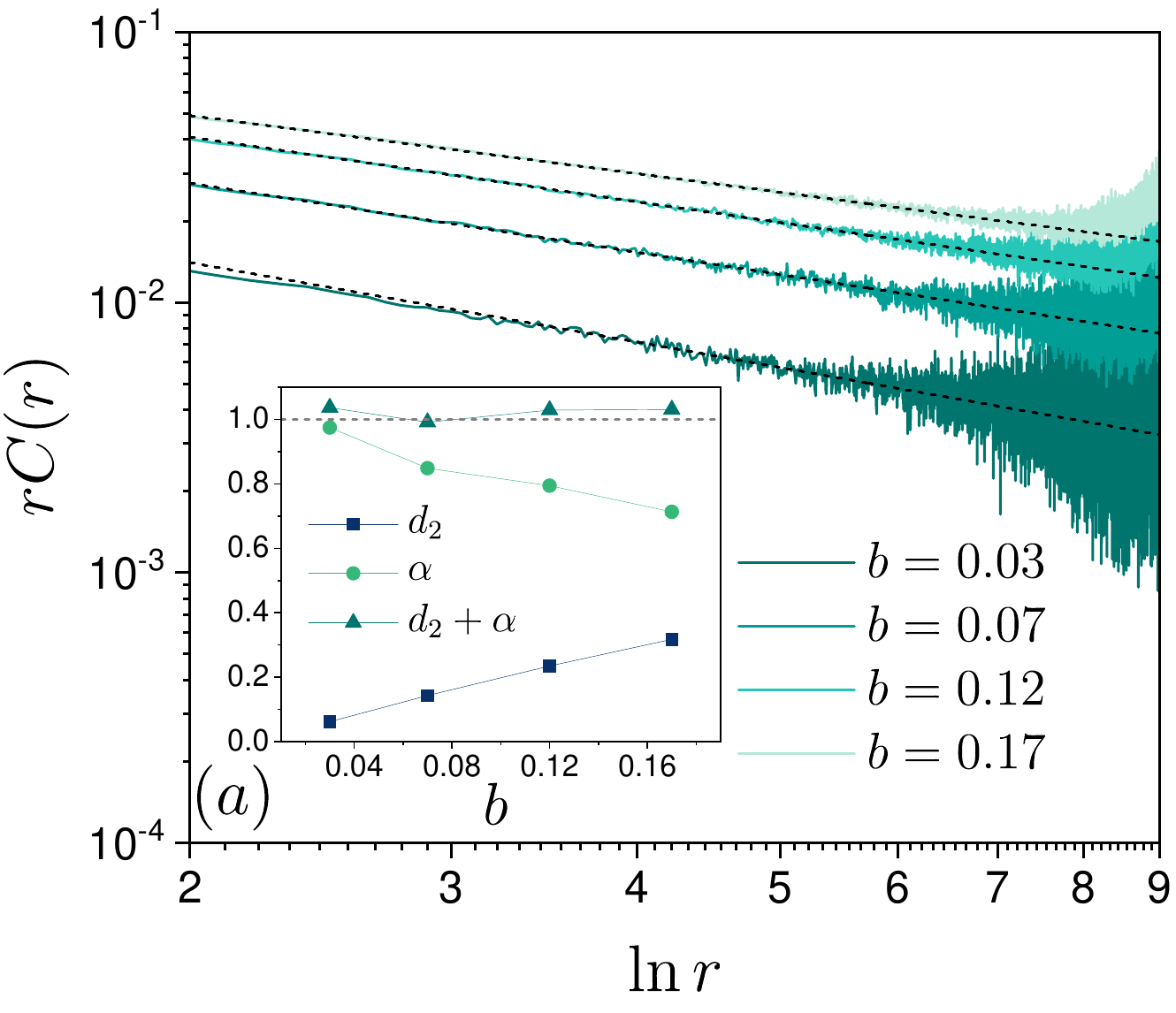}
\includegraphics[width=0.49\textwidth]{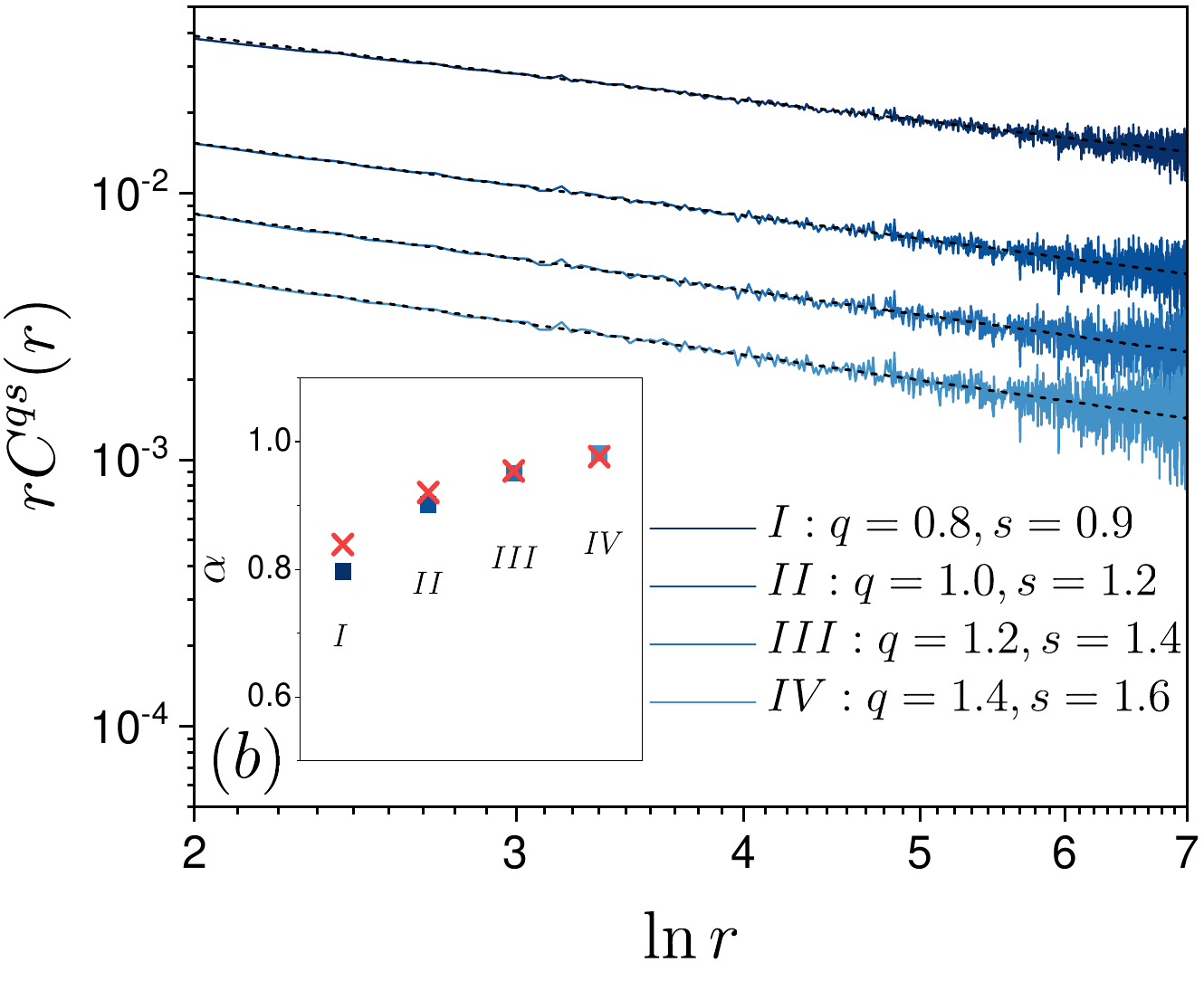}
 \caption{\label{figCorsRBM} (a) Spatial decay of the correlation function  $C(r)$ Eq.~\eqref{eq:cor11} in the SRBM model with $\beta=1$ for different $b$ values as indicated by the labels. The data is plotted as $rC(r)$ a function of $\ln r$ in log-log scale, the dash lines are power-law fits $rC(r)\sim(\ln r)^{-\alpha}$. Inset: Verification of the relationship $d_2+\alpha=1$. (b) Spatial decay of the generalized correlation function  $C^{qs}(r)$ Eq.~\eqref{eq:cor} in the SRBM model with $\beta=1$ for $q,s$ values as indicated by the labels with $b=0.05$. The data is plotted as $rC^{qs}(r)$ a function of $\ln r$ in log-log scale, the dash lines are power-law fits. Inset: The squares are fitting exponents $\alpha$ of power-law fits $rC^{qs}(r)\sim(\ln r)^{-\alpha}$ and the crosses are the corresponding predictions by Eq.~\eqref{eq:corqs}. Results have been averaged over $720$ realizations with $N=2^{14}$. }
 \end{figure*}

\section{Correlation functions}\label{sec5}
In this Section, we investigate the average correlation function of eigenstates, defined as
\begin{equation}\label{eq:cor}
C^{qs}(r) \equiv \left\langle|\psi(i)|^{2q} |\psi(i+r)|^{2s}\right\rangle,
\end{equation}
where the average is performed with respect to all sites $i$ and all sites at distance $r$ from $i$, over wave functions $\psi$ in an energy shell close to $E=0$, and over random matrix configurations. This correlation function serves as a crucial multifractality probe.

Let us first recall in Sec.~\ref{sec5A} the decay behavior of this average correlation function in the case of conventional multifractality in the $d$-dimensional Anderson model. 
In Sec.~\ref{sec5B} we will get back to our random matrix ensembles.

\subsection{Conventional multifractality}\label{sec5A}
In the context of the Anderson transition in dimension $d$, correlation functions behave as
\begin{equation}\label{eq:corqs}
    C^{qs}(r) \sim r^{\tau_{q+s}-\tau_{q}-\tau_s - d},
\end{equation}
with $\tau_q = D_q(q-1)$ and $0 \leq D_q \leq d$ being the conventional multifractal exponents and dimensions, respectively \cite{PhysRevE.54.3221, PhysRevB.62.7920}.  
However, wave functions may have a nontrivial multifractal support, in which case the definition of the correlation function should be modified to take the nontrivial support into account. 

Suppose that the set supporting the measure represented by wavefunction amplitudes $|\psi(i)|^{2}$ has fractal dimension $D < d$. In that case, as explained in \cite{PhysRevA.35.4907}, one should first redefine the spatial correlation function \textit{restricted} to the support of the wave function:
\begin{equation}\label{eq:corsup}
    C_\textrm{supp}^{qs}(r) \equiv \left\langle|\psi(i)|^{2q} |\psi(i+r)|^{2s}\right\rangle_{\textrm{supp}} \sim r^{\tau_{q+s}-\tau_{q}-\tau_s - D},
\end{equation}
with now $0 \leq D_q \leq D$ and $\langle \rangle_{\textrm{supp}}$ denotes an averaging over sites belonging to the support only.
The \textit{unrestricted} correlation function Eq.~\eqref{eq:cor} is related to the restricted one Eq.~\eqref{eq:corsup} by 
\begin{equation}
\label{relationUnresres}
    C^{qs}(r) \sim r^{D-d} C_\textrm{supp}^{qs}(r).
\end{equation}
The additional factor comes from the ratio between the number of sites at distance $r$ belonging to the support, which scales as $r^{D-1}$, and  sites at distance $r$, scaling as $r^{d-1}$.

Suppose we have a system of dimension $d$ where the support of wavefunctions is only 1D, with $D=1$. This is precisely what happens in the case of a tree or random graph of infinite effective dimension, 
since wavefunctions are delocalized along a single branch at the transition, see \cite{PhysRevLett.117.156601, PhysRevLett.118.166801, PhysRevResearch.2.012020, PhysRevB.106.214202, biroli2023largedeviation}.
Then from Eq.~\eqref{relationUnresres} we get for $C(r) \equiv C^{11}(r)$
\begin{equation}\label{eq:cor11}
   C(r) = \left\langle|\psi(i)|^{2} |\psi(i+r)|^{2}\right\rangle \sim r^{1-d} r^{D_2 - 1},
\end{equation}
with $0 \leq D_2 \leq D=1$. In dimension $d$, $r^{d-1}$ counts the number of terms at a distance $r$ from the origin; it can be interpreted as the surface of the volume $r^d$. In the infinite-dimensional case of a random graph with connectivity $K$, the number of terms at a distance $r$ grows exponentially as $K^r$, and Eq.~\eqref{eq:cor11} becomes 
 \begin{equation}
 \label{eqc11}
     C(r) \sim K^{-r} r^{D_2 - 1}. 
 \end{equation}
The prefactor $1/K^r$ can also be interpreted in the following way: the graph is locally tree-like and the wavefunction is supported on a single branch of that tree, which changes from realization to realization. In particular, for a \textit{given} fixed branch and at a distance $r$ from the root it takes $\sim K^r$ realizations of disorder to get a  wavefunction contributing to the correlation function \eqref{eqc11}.

Equation \eqref{eqc11} is reminiscent of what is predicted at the Anderson transition on RRG and the Bethe lattice, $C(r) \sim K^{-r} r^{-3/2}$, with the important difference that $0 < 1-D_2 < 1$ for a multifractal on a one-dimensional support, whereas the exponent of $r$ is $3/2 > 1$ for RRG and Bethe lattice (see section \ref{sec7}).

\subsection{Logarithmic multifractality}\label{sec5B}
The case of log-multifractality considered in the present paper is analogous to the above problem of tree/random graphs of infinite dimension, in the following sense. The support of wavefunctions scales logarithmically with system size, which is analogous to the graph case where wavefunctions lie on a single branch (with $\sim \ln N$ sites) of a graph of $N$ sites. The restricted correlation function $C_\textrm{supp}^{qs}(r)$ is now given by the analog of Eq.~\eqref{eq:corsup}, namely
\begin{equation}\label{eq:corsuplog}
    C_\textrm{supp}^{qs}(r)  \sim (\ln r)^{\tau_{q+s}-\tau_{q}-\tau_s - D},
\end{equation}
where $\tau_{q+s}-\tau_q-\tau_s-1 = d_{q+s}(q+s-1) - d_q(q-1)-d_s(s-1) -1$ with $0 \leq d_q \leq 1$. As for the prefactor needed to go from $C_\textrm{supp}^{qs}(r)$ to the unrestricted $C^{qs}(r)$, we can make the same reasoning as in Eq.~\eqref{eqc11}: the number of realizations of disorder required to get a  wavefunction contributing to the correlation function is exponential in the distance to the origin, thus for a given fixed site at distance $\ln r$ it takes $\sim r$ realizations. Therefore 
the factor $K^r$ in Eq.~\eqref{eqc11} should be replaced by $r$.

It should be noted that the transposition of the arguments described in Ref.~\cite{PhysRevA.35.4907} to the present case relies crucially on the assumption that coarse-graining at scale $r$ of log-multifractality results in moments $\langle P_q\rangle$ scaling as $\langle P_q\rangle \sim (\ln N/\ln r)^{-d_q(q-1)}$ for $q>1/2$, i.e., depends algebraically on the ratio of the logarithm of system size $N$ and the logarithm of coarse-graining size $r$. This assumption is well verified numerically, as we show in Fig.~\ref{figP2CGSRBM}.

 \begin{figure}
\includegraphics[width=0.48\textwidth]{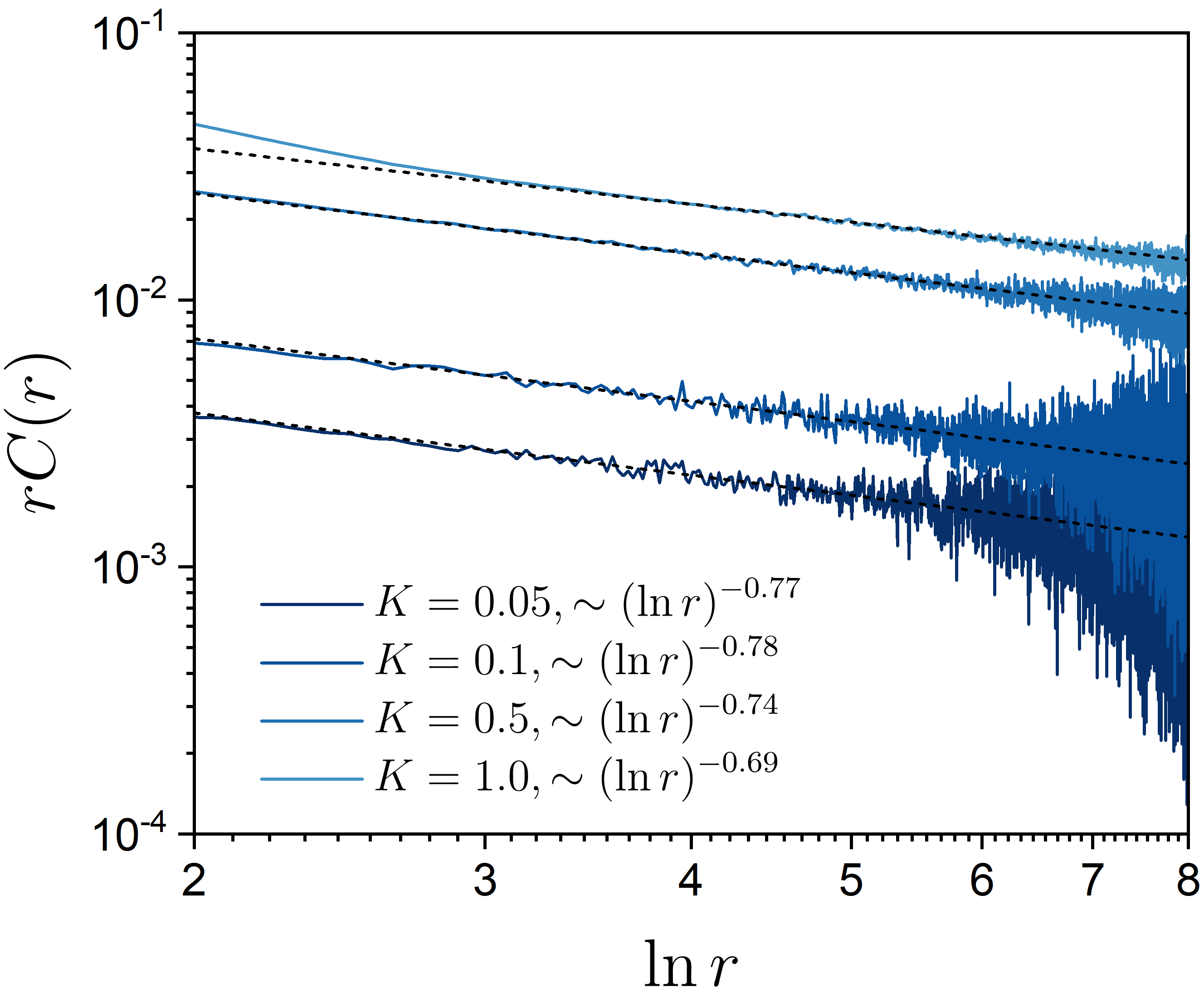}
 \caption{\label{figCorSRUM} Spatial decay of the correlation function  $C(r)$ Eq.~\eqref{eq:cor11} in the SRUM model for different $K$ values as indicated by the labels. The data is plotted as $rC(r)$ a function of $\ln r$ in log-log scale, the dash lines are power-law fits $rC(r)\sim(\ln r)^{-\alpha}$. Results have been averaged over $36,000$ realizations with $N=2^{16}$. }
 \end{figure}
 
\begin{figure}
\includegraphics[width=0.46\textwidth]{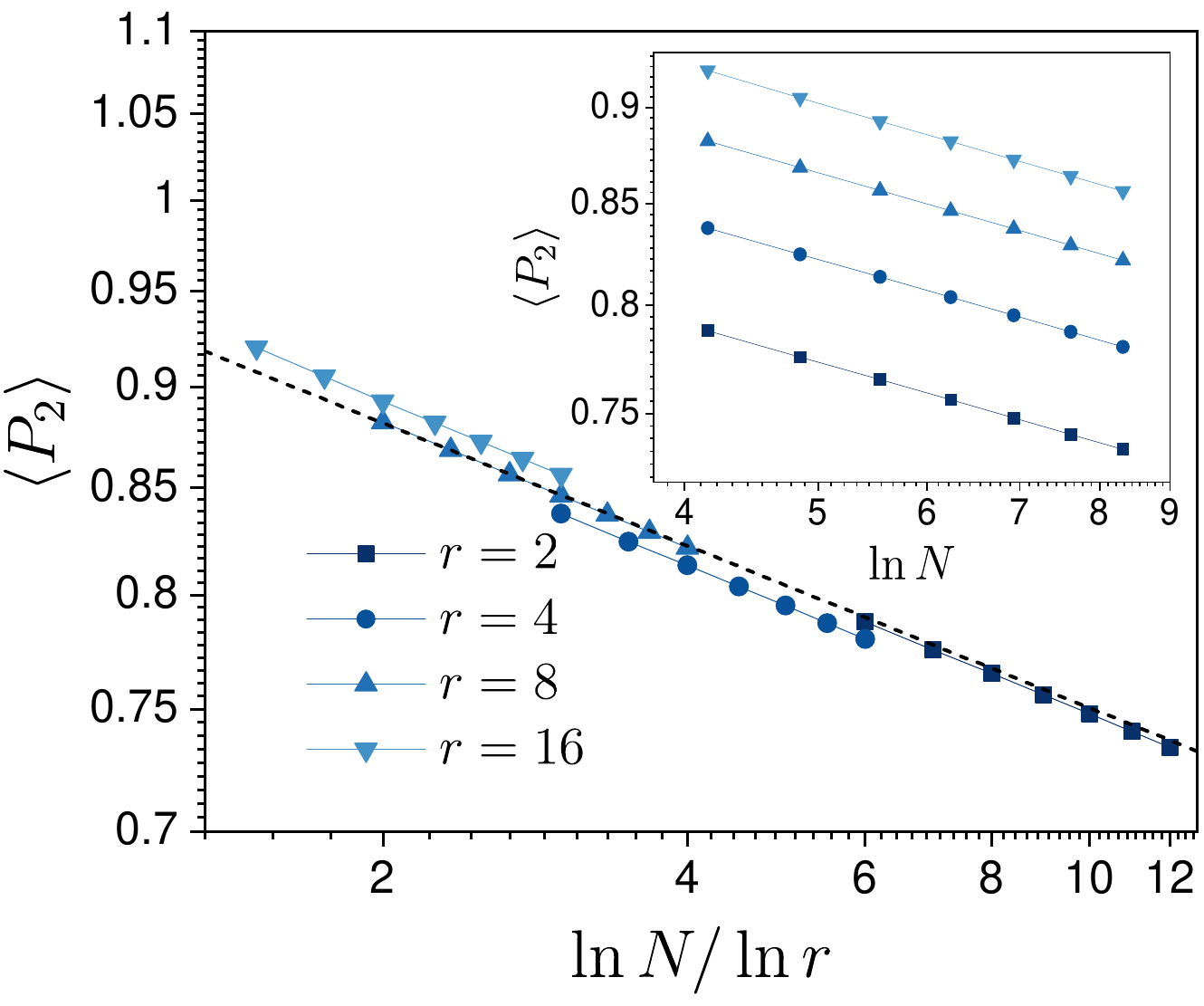}
 \caption{\label{figP2CGSRBM} Coarse-graining and log-multifractality in the SRBM model ($\beta=1$). Different curves correspond to $\langle P_2 \rangle$ obtained using different coarse-graining sizes $r$ indicated by the labels. The good data collapse of $\langle P_2 \rangle$ when plotted as a function of $\ln N/\ln r$, and the black dashed curve, demonstrate the scaling behavior $\langle P_2\rangle\sim (\ln N/\ln r)^{-d_2}$ associated with log-multifractality, distinct from $\langle P_2\rangle\sim (N/r)^{-D_2}$ for conventional multifractality. Inset: Raw data of $\langle P_2 \rangle$ before rescaling $\ln N$ by $\ln r$. Disorder averaging ranges from $360,000$ realizations for $N=2^6$ to $18,000$ realizations for $N=2^{12}$.}
 \end{figure}

The analogy with infinite-dimensional graphs featuring logarithmic support suggests a particular behavior for the unrestricted correlation function with log-multifractality:
\begin{equation}\label{eq:Cqs}
    C^{qs}(r) \sim r^{-1} (\ln r)^{\tau_{q+s}-\tau_q-\tau_s-1}.
\end{equation}
Our numerical results, which are presented in Fig.~\ref{figCorsRBM} for the SRBM model, confirm this prediction. In these figures, we plot $rC^{qs}(r)$ (with $C^{11} \equiv C$) as a function of $\ln r$ and observe behaviors that are consistent with algebraic laws, specifically $r C^{qs}(r) \sim (\ln r)^{-\alpha}$. The exponent $\alpha$ in Fig.~\ref{figCorsRBM} (a) for the case $q=s=1$ correlates well with $1-d_2$ as shown in the inset. We have further checked the exponents for different values of $q$ and $s$ shown in Fig.~\ref{figCorsRBM} (b), which also agree well with Eq.~\eqref{eq:Cqs}. Moreover, the numerical results presented in Fig.~\ref{figCorSRUM} for the SRUM model exhibit the same functional form as the prediction, albeit with deviations in the exponent $\alpha$ from $1-d_2$.

The non-trivial analytical prediction given by Equation~\eqref{eq:Cqs} for the correlation function $C^{qs}(r)$ is an important confirmation that the SRBM model accurately captures multifractality in infinite effective dimension.


\section{Quantum dynamics in presence of  logarithmic multifractality }

\subsection{Dynamics of return probability}\label{sec6}

The quantum dynamics of the return probability also encodes rich information on quantum multifractality \cite{PhysRevE.108.054127,PhysRevE.86.021136,PhysRevB.82.161102,PhysRevA.95.041602, PhysRevA.100.043612}. 
In the context of conventional multifractality, the return probability 
\begin{equation}
    \langle R_0(t) \rangle   \equiv \left \langle |\langle \psi(t)|\psi(0)\rangle|^2 \right \rangle \sim t^{-D_2}\;,
\end{equation}
exhibits a power-law decay with time $t$ with an exponent given by the multifractal dimension $D_2$ \cite{PhysRevLett.69.695,PhysRevB.82.161102, Kravtsov_2011}. In this Section, we show that $\langle R_0(t) \rangle$ behaves as 
\begin{equation}\label{eq:R0vslnt}
\langle R_0 \rangle\sim (\ln t)^{-d_2}\;,
\end{equation}
in the case of log-multifractality for the SRBM model in both orthogonal class and unitary class with Dyson index $\beta=1$ and $\beta=2$. The analytic expression for the return probability was obtained for the PRBM case in the limit $b\ll 1$ by means of a supersymmetric virial expansion \cite{Kravtsov_2011}, given as a series expansion
 $\langle R_0\rangle=\langle R_0^{(0)}\rangle+\langle R_0^{(1)}\rangle+\dots$ in terms of successive orders of the parameter $b$, with $\langle R_0^{(0)}\rangle=1$. For our SRBM model, we take as a starting point the expression obtained in \cite{Kravtsov_2011} for $\langle R_0^{(1)} \rangle$, with the only difference that the variance of the matrix entries $H_{ij}$, given by Eq.~\eqref{defSRBM}, has now the additional logarithmic term in $|i-j|$.

\begin{figure}
\includegraphics[width=0.5\textwidth]{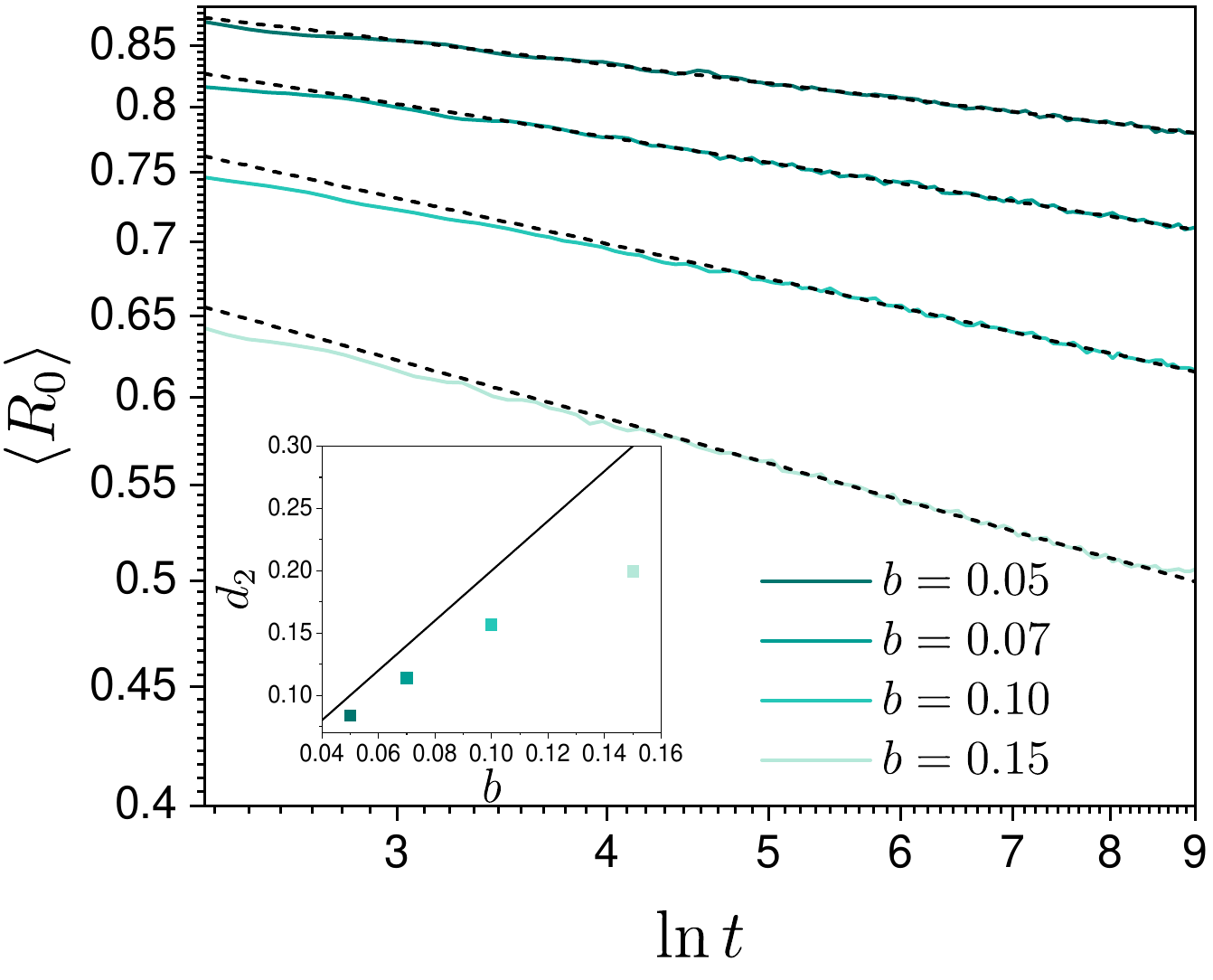}

    \caption{\label{figR0}  The return probability $\langle R_0 \rangle$ in the SRBM model with $\beta=1$ for different $b$ values indicated by the labels. The black dashed lines are power-law fits $\langle R_0 \rangle= c (\ln t)^{-d_2}$ with $c$ and $d_2$ two fitting parameters. Inset: Comparison of the exponent $d_2$ obtained from fitting (symbols) and the analytical prediction Eq.~\eqref{eq:R0vslnt} (black solid line). Results have been averaged over $14,400$ disorder configurations with the system size $N=2^{10}$.} 
\end{figure}

\begin{figure*}
\includegraphics[width=0.48\textwidth]{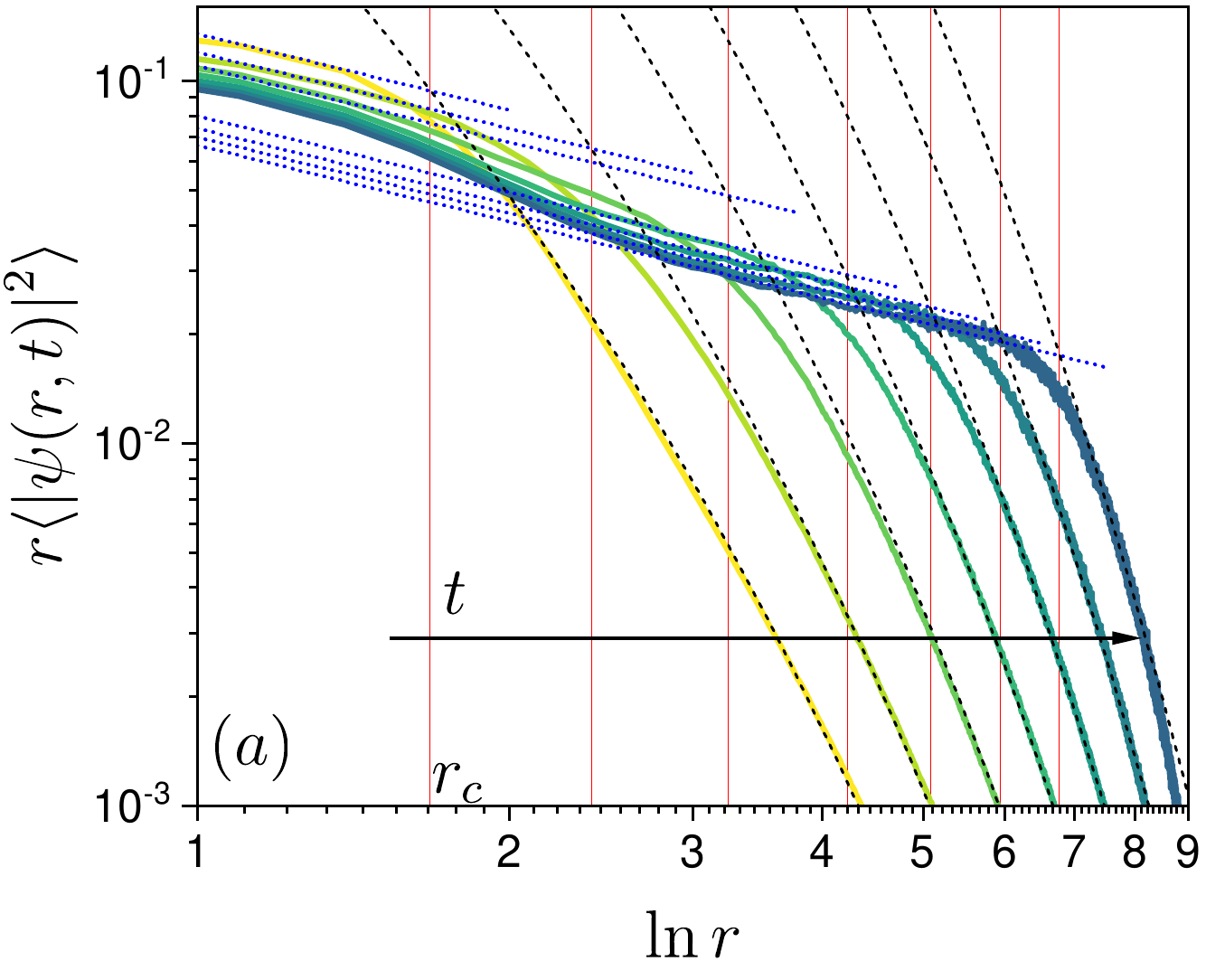}%
\includegraphics[width=0.47\textwidth]{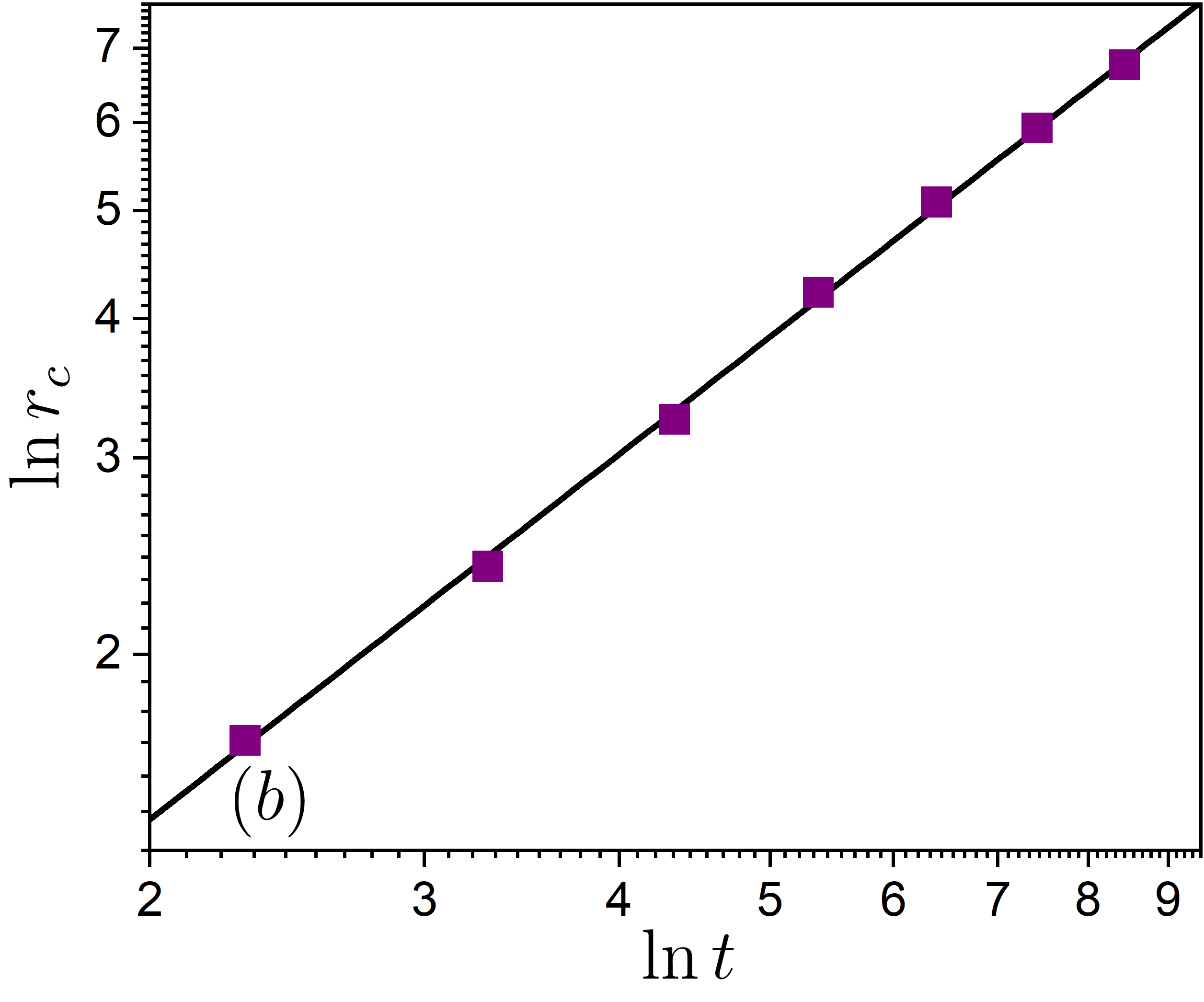}%
\caption{\label{FigDynamics}(a) Solid curves: Average probability distribution of wave packets in the SRUM model at different times plotted as $r\langle|\psi(r,t)|^2\rangle$ \textit{vs} $\ln r$, with initial condition $\psi(r,t=0) = \delta_{r,0}$ and $K=1.0$. The evolution time increases as indicated by the arrow and the data on the right panel. The violet dotted line fits the behavior corresponding to the correlation function $rC(r)\sim(\ln r)^{-\alpha}$ with $\alpha\approx0.7$ and the black dashed line fits the behavior corresponding to the typical decay of matrix element, i.e., $\langle|\psi(r)|^2\rangle\sim 1/(r\ln r)^2$, the red lines locates the crossover $r_c$ (the intersect) between the two decay behaviors of the wave packet. Results have been averaged over $720,000$ disorder configurations with system size $N=2^{15}$. (b) Symbols represent the crossover $r_c$ extracted from the wave packets, plotted as a function of time $t$. The solid line results from a power-law fit of the data for $r_c$, which yields $\ln r_c\sim (\ln t)^{1.09}$, in good agreement with Eq.~(\ref{eq:rc}). }
\end{figure*}

 For the SRBM in the orthogonal class $\beta=1$, Eq.~(B.1) of \cite{Kravtsov_2011} becomes
\begin{equation}
\label{R01}
\langle R_0^{(1)} \rangle=-\frac{\sqrt{2\pi}}{N}\sum_{i\neq j}^{N}2b_{ij} \, t \, g(2b_{ij}\,t^2),\quad g(x)\equiv e^{-x}I_0(x),
\end{equation}
where $b_{ij}=\frac{1}{2}\langle|H_{ij}|^2\rangle \simeq \frac{b^2}{2(|i-j|\ln|i-j|)^2}$ and $I_0(x) $ is the 0th order modified Bessel function. In the large-$N$ limit, we can replace the sum in Eq.~\eqref{R01} by an integral, 
\begin{equation}\small
\begin{split}
    \langle R_0^{(1)} \rangle &\underrel{N\rightarrow\infty}{\simeq}-2\sqrt{2\pi}\int_{1}^{\infty}dx2b_x \, t\, g(2b_xt^2)\\
    &=-2\sqrt{2\pi}\int_{1}^{\infty}dx\frac{b^2t}{(x\ln x)^2}\,g\left[\left(\frac{bt}{x\ln x}\right)^2\right].
\end{split}
\end{equation}
At large argument $\frac{bt}{x\ln x}\gg1$, or equivalently $x\ll e^{\mathcal W(bt)}$,
with $\mathcal W(x)$ the Lambert function defined by $\mathcal W(x)\exp\left[\mathcal W(x)\right]=x$, we have $g(z)\sim 1/\sqrt{2\pi z}$. 
Therefore, we replace the integral interval as $(1, e^{\mathcal{W}(x)})$, where the integrand dominates. The Lambert function has the asymptotic behavior $\mathcal W(x)\underrel{x\rightarrow+\infty}{\simeq}\ln x-\ln \ln x+\dots$. This gives, to lowest order in $b$,
\begin{equation}
\begin{split}
    \langle R_0^{(1)} \rangle\simeq -2b\ln [ \mathcal W(bt)] \simeq -2b \ln \ln (bt),
\end{split}
\end{equation}
and thus
\begin{equation}\label{eq:scaR0}
   \frac{\partial \ln \langle R_0\rangle}{\partial \ln \ln t}\simeq \frac{\partial \langle R_0^{(1)} \rangle}{\partial \ln \ln t}\underrel{t\rightarrow\infty}{\simeq}-2b=-d_2,
\end{equation}
where the last equality comes from Eq.~\eqref{eq:dq}. Therefore we finally get $\langle R_0\rangle\sim (\ln t)^{-d_2}$,
which indicates an algebraic decay of $\langle R_0\rangle$ in $\ln t$ controlled by the log-multifractal dimension $d_2$.
A similar approach applies to the unitary class $\beta=2$ but with more cumbersome derivations. Detail is given in Appendix \ref{returnprobbeta2}. The final result is 
\begin{equation}\label{eq:scaR0beta2}
   \frac{\partial \ln \langle R_0\rangle}{\partial \ln \ln t}\simeq -\pi b=-d_2,
\end{equation}
where the last equality comes from Eq.~\eqref{eq:dq_} and shows that in this case the algebraic decay of $\langle R_0\rangle$ is also governed by $d_2$.

We have performed numerical simulations for the SRBM model in the orthogonal class, and results presented in Fig.~\ref{figR0} confirm the predictions Eq.~\eqref{eq:R0vslnt} up to certain deviations due to strong finite-size effects. The same scaling behavior of the return probability $\langle R_0\rangle$ is expected for SRUM ensemble, as the amplitudes of its off-diagonal elements decay in the same way. The results presented in Ref.~\cite{chen2023quantum} for SRUM, confirm the validity of the scaling described by Eq.~\eqref{eq:R0vslnt}.

\subsection{Wave packet dynamics}\label{sec6bis}
In addition to examining the local observable $\langle R_0\rangle$, we propose a characterization of the wave packet shape in the context of log-multifractality, inspired by the approach in Ref.~\cite{PhysRevE.108.054127}. 

Generalizing the results of Refs. \cite{PhysRevE.108.054127, PhysRevLett.79.1959, CHALKER1990253, PhysRevA.100.043612, PhysRevE.86.021136} for conventional multifractality in the PRBM model leads us to the following predictions:
Initiated at a single site, a wave packet subject to long-range hoppings will exhibit a tail primarily influenced by these hopping elements. In the SRBM model where the long-range hoppings adhere to the behavior described in Eq.~\eqref{eq3}, then the tail of the wave packet should follow a scaling behavior $\langle|\psi(r)|^2\rangle\sim (r\ln r)^{-2}$. However, in the proximity of the initial site $r=0$ where the wave packet was initialized, a nontrivial power-law decay $\langle|\psi(r)|^2\rangle\sim C(r)$, with the correlation function $C(r)$ given by Eq.~\eqref{eq:Cqs}, dynamically emerges. Therefore, the functional form of the wave packet can be constructed as
\begin{equation}
\label{eqpsi2}
\begin{split}
\begin{aligned}
\langle|\psi(r,t)|^{2}\rangle =     \begin{cases}
       & \langle R_0\rangle\left[r(\ln r)^{\alpha}\right]^{-1} , \quad 1< r< r_c, \\
       &B\left[\frac{r}{r_c}\ln(\frac{r}{r_c})\right]^{-2}, \quad r_c< r\leq \frac{N}{2} ,
    \end{cases} 
    \end{aligned}
\end{split}
\end{equation}
with $B= \langle R_0\rangle\left[r_c(\ln r_c)^{\alpha}\right]^{-1}$,  $\alpha$ the decay exponent obtained from the correlation function $C(r)$ in Eq.~\eqref{eq:Cqs}. For the SRBM ensemble, Eq.~\eqref{eq:Cqs} gives $\alpha=1-d_2$, see Fig.~\ref{figCorsRBM}. 

Given that this argument does not assume any specific structure but rather relies on the existence of log-multifractality, it is applicable to the SRUM model as well. Numerical simulations, as presented in Ref.~\cite{chen2023quantum}, confirm the validity of this wave packet characterization in the SRUM model.

\begin{figure*}
\includegraphics[width=0.46\textwidth]{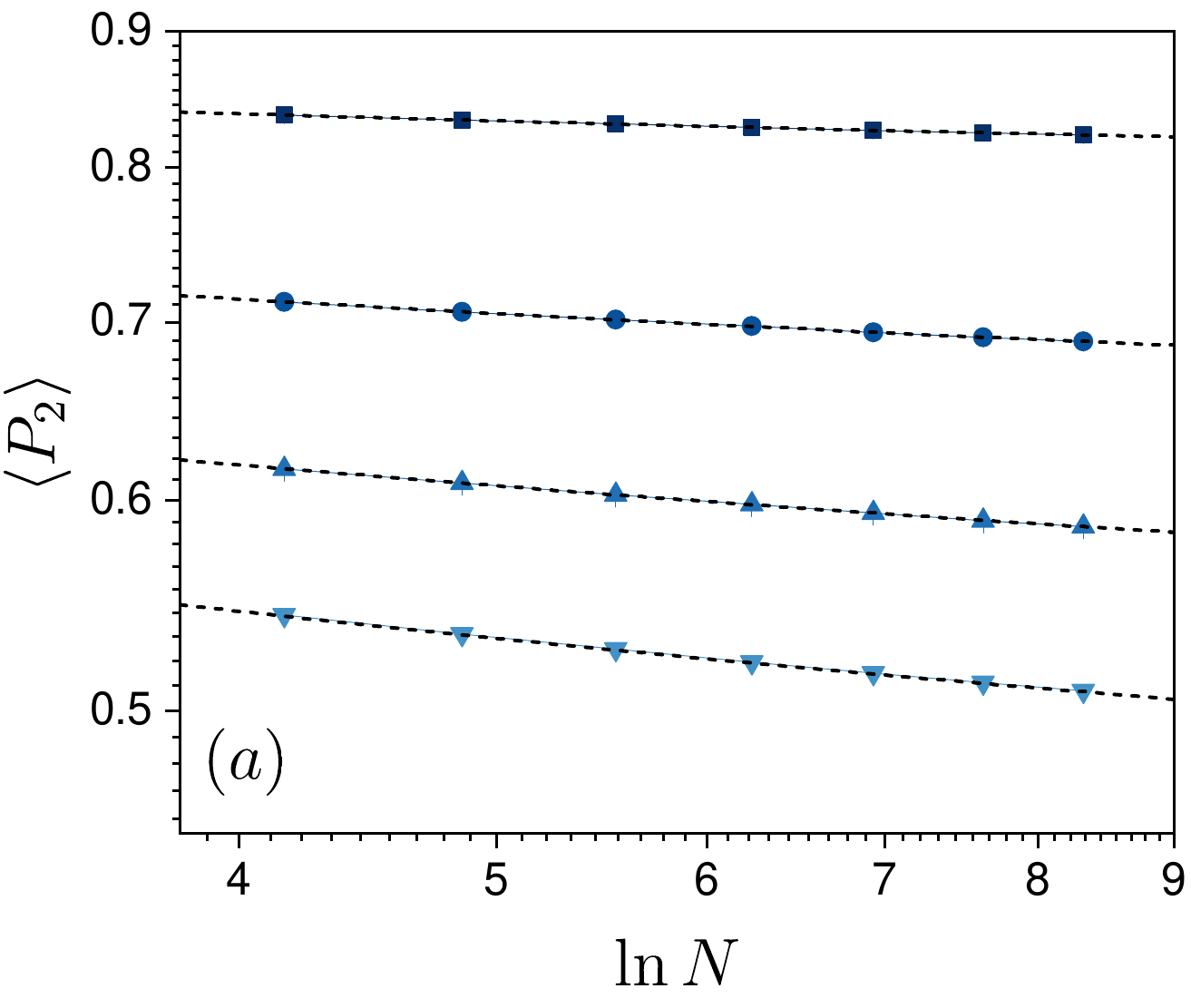}
\includegraphics[width=0.48\textwidth]{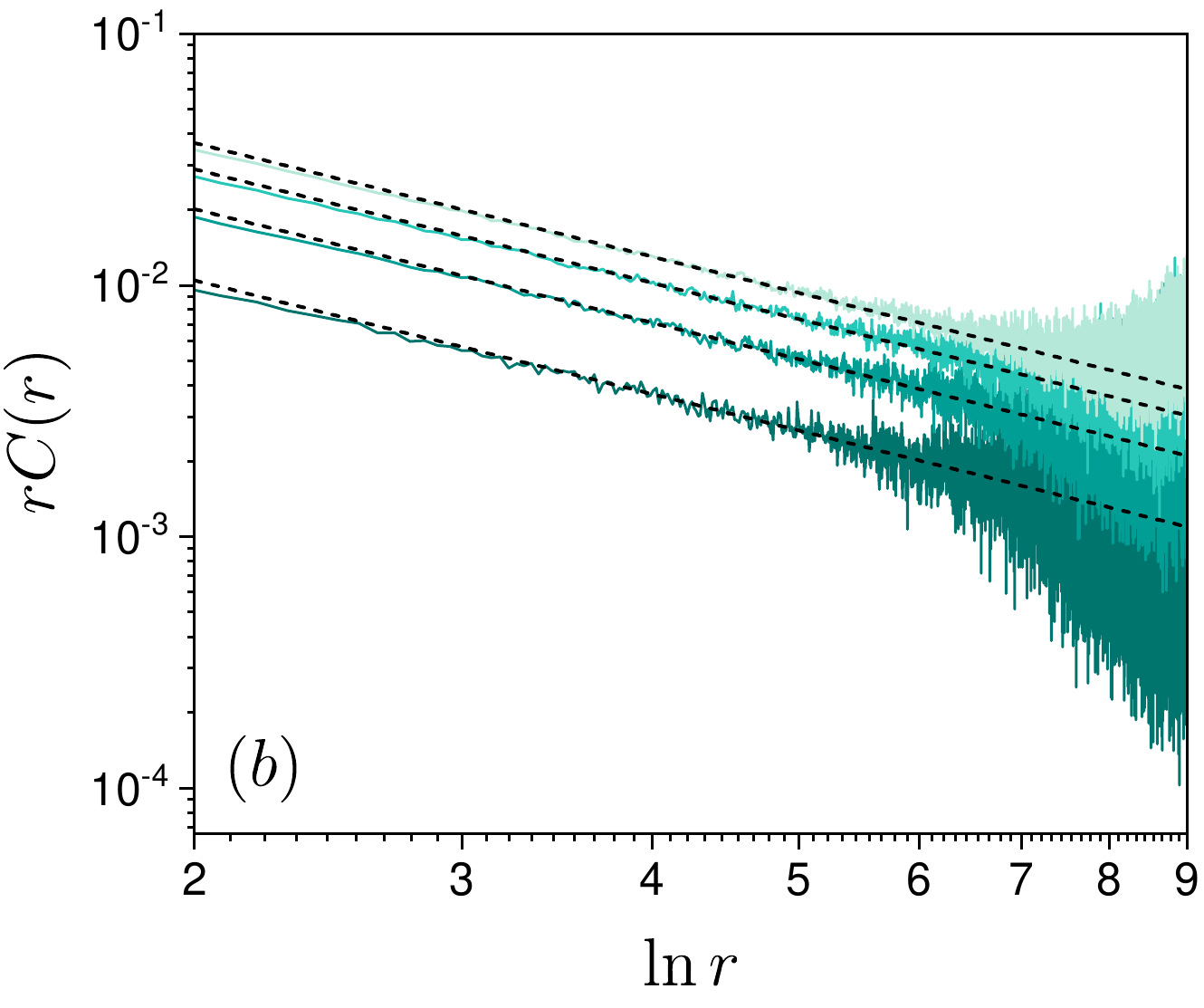}
 \caption{\label{figHOSRBM}(a) Critical localization behavior of $\langle P_2\rangle$ for the generalized SRBM model Eq.~\eqref{defGSRBM} with $\beta=1$ and $\mu=1/2$; different curves correspond to different $b$ values: $b=0.03, 0.06, 0.09, 0.12$ from top to bottom. The black dashed lines are fits by $\langle P_2\rangle =  c_0 (\ln N)^{-\mu}+P_2^\infty$ with $c_0$ and $P_2^\infty$ two fitting parameters, confirming Eq.~\eqref{eq:P2GSRBM}. Disorder averaging ranges from $360,000$ realizations for $N=2^6$ to $18,000$ realizations for $N=2^{12}$. (b) Spatial decay for the correlation function  $C(r)$ in the generalized SRBM model with $\beta=1$ and $\mu=1/2$ for different $b$ values: $b=0.03, 0.06, 0.09, 0.12$ from bottom to top. The data are plotted as $rC(r)$ a function of $\ln r$ in log-log scale, and the black dashed lines are fits by $rC(r)=A(\ln r)^{-(1+\mu)}$ with $A$ a fitting parameter, confirming Eq.~\eqref{eq:CorGSRBM}. These results for $\mu=1/2$ are in perfect agreement with the critical behavior predicted in random regular and Erdös-Rényi graphs of effective infinite dimensionality, see \cite{PhysRevB.34.6394, ZIRNBAUER1986375, PhysRevLett.72.526, PhysRevB.99.024202}. Results have been averaged over $720$ realizations with $N=2^{14}$.}
 \end{figure*}

Furthermore, through a normalization argument, one can deduce the dynamical dependence of the crossover $r_c$ as $\ln r_c \sim (\ln t)^{\alpha/(1-d_2)}$. The normalization of the wave packets requires

\begin{equation}\label{eq:norm}\small
\begin{split}
\begin{aligned}
   1=||\psi||^{2}\simeq&2\int_{1}^{r_{c}}\langle R_0\rangle\left[r(\ln r)^{\alpha}\right]^{-1}dr\\
   &+2\int_{r_c}^{\frac{N}{2}}B\left[\frac{r}{r_c}\ln(\frac{r}{r_c})\right]^{-2}dr\\
    \simeq&\frac{2\langle R_0\rangle}{1-\alpha}(\ln {r_c})^{1-\alpha}+2\int_{r_c}^{\frac{N}{2}}B\left[\frac{r}{r_c}\ln(\frac{r}{r_c})\right]^{-2}dr.
\end{aligned}
\end{split}
\end{equation}
The second term on the right-hand side is evaluated using Eq.~\eqref{sum2q} for $q=1$; it vanishes when $N\rightarrow\infty$. Therefore, in the thermodynamic limit, Eq.~\eqref{eq:norm} gives:
\begin{equation}
   \frac{2\langle R_0\rangle}{1-\alpha}(\ln {r_c})^{1-\alpha}  \simeq  1 \; ,
\end{equation}
i.e., $\ln {r_c}\sim \langle R_0\rangle^{-1/(1-\alpha)}$. Imposing $ \langle R_0\rangle \sim (\ln t)^{-d_2}$, we immediately obtain
\begin{equation}\label{eq:rc}
    \ln r_c\sim (\ln t)^{\frac{d_2}{1-\alpha}}.
\end{equation}
In Fig.~\ref{FigDynamics} we determine the crossover $r_c$ by locating $r_c$ as the intersect of the two behaviors in Eq.~(\ref{eqpsi2}) in the SRUM model. The results extracted from the wave packets at different times align with our prediction.


\begin{figure*}
\includegraphics[width=0.48\textwidth]{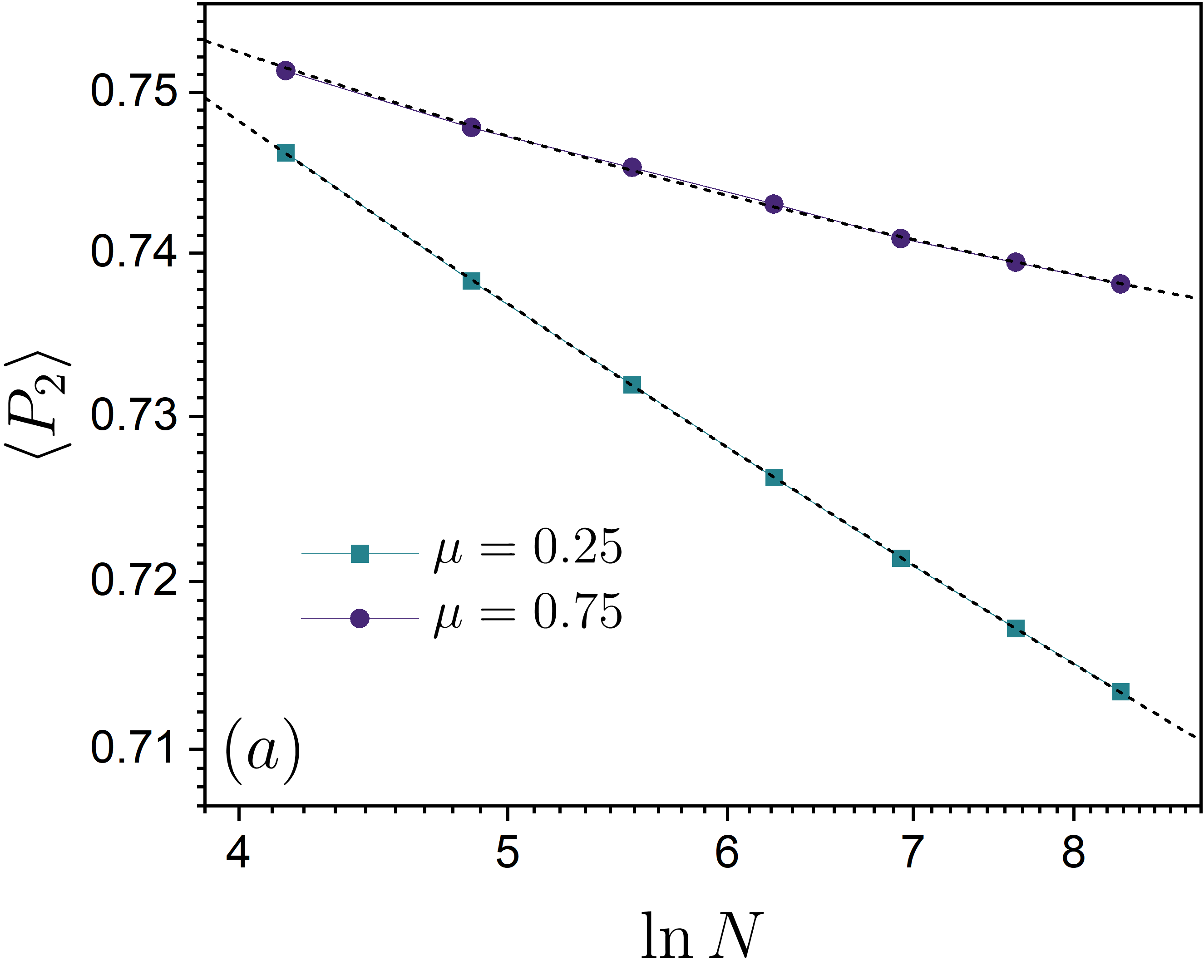}
\includegraphics[width=0.48\textwidth]{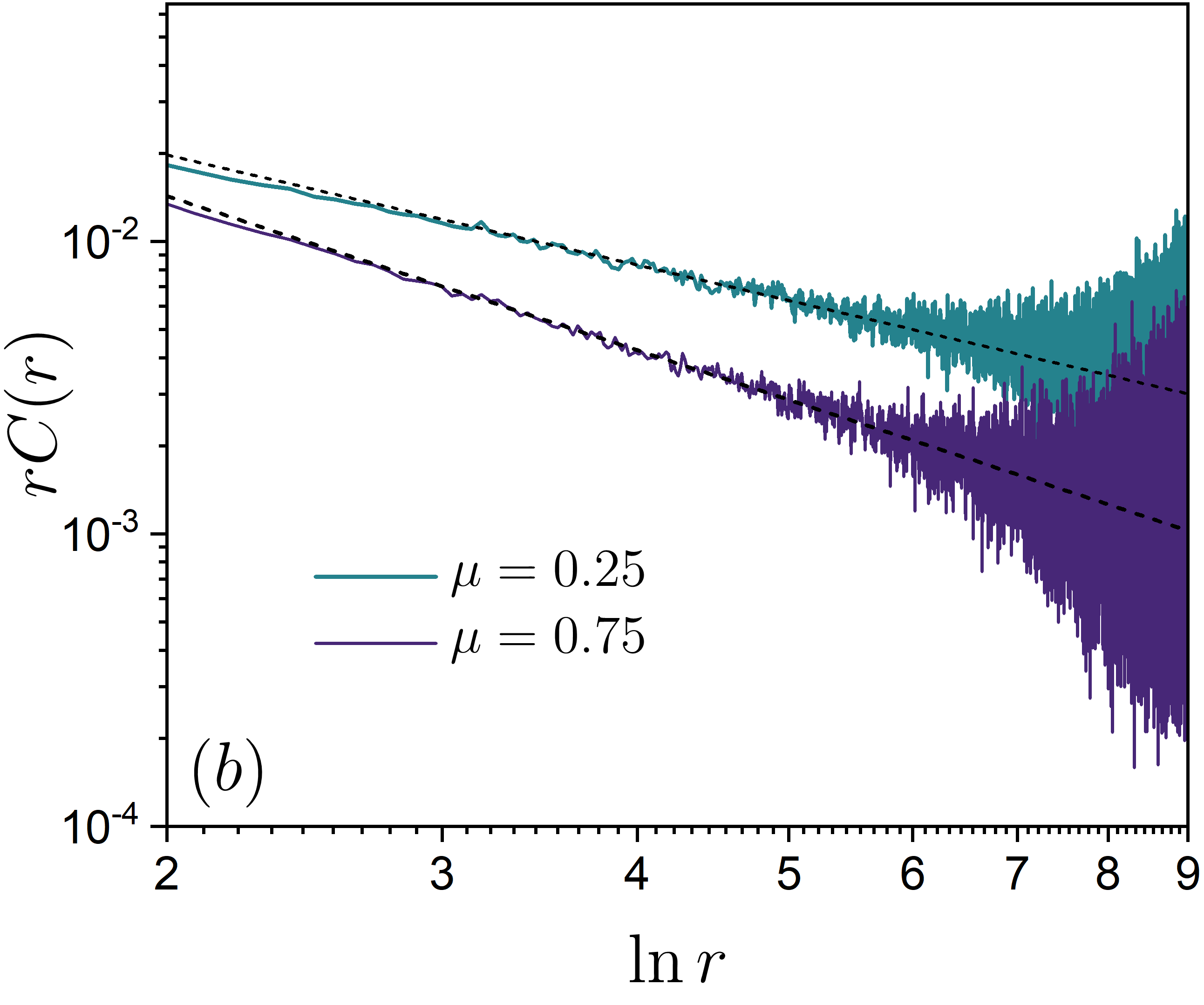}
 \caption{\label{figHOSRBM_mu}(a) Critical localization behavior of $\langle P_2\rangle$ for the generalized SRBM model Eq.~\eqref{defGSRBM} with $\beta=1$ and different $\mu$ values as indicated by labels. The black dashed lines are fits by $\langle P_2\rangle =  c_0 (\ln N)^{-\mu}+P_2^\infty$ with $c_0$ and $P_2^\infty$ two fitting parameters, confirming Eq.~\eqref{eq:P2GSRBM}. Disorder averaging ranges from $360,000$ realizations for $N=2^6$ to $18,000$ realizations for $N=2^{12}$. (b) Spatial decay for the correlation function  $C(r)$ in the generalized SRBM model with $\beta=1$ for different $\mu$ values as indicated by labels. The data are plotted as $rC(r)$ a function of $\ln r$ in log-log scale, the black dashed lines are fits by $rC(r)=A(\ln r)^{-(1+\mu)}$ with $A$ a fitting parameter, confirming Eq.~\eqref{eq:CorGSRBM}. Results have been averaged over $720$ realizations with $N=2^{14}$.}
 \end{figure*}

\section{Generalized SRBM ensembles and critical localization}\label{sec7}

In the context of random regular and Erdős-Rényi graphs, an analytical exploration has revealed that $\langle P_2\rangle\sim (\ln N)^{-1/2}+\langle P_2^{N=\infty}\rangle$, where $\langle P_2^{N=\infty} \rangle>0$ signifies a localized behavior in the thermodynamic limit \cite{PhysRevLett.67.2049,PhysRevLett.72.526,PhysRevB.99.024202,fyodorov1992novel,ADMirlin_1991}, which we called ``critical localization''. In this Section, we consider the generalized SRBM ensembles defined in Section \ref{secgenSRBM}, and show that they encapsulate this scenario. 

The behavior of eigenstate moments $\langle P_q \rangle$ in these ensembles can be also obtained through the perturbation theory assisted by weighted L\'evy sums illustrated in Sec.~\ref{sec3}. For these generalized ensembles, the analytical procedure is in the same manner by replacing Eq.~\eqref{eqsumv} and Eq.~\eqref{eqsumv2q} as
\begin{equation}
\begin{split}      \sum_{\vec{r}}\langle|V(\vec{r})|\rangle&\simeq 2b\langle|u_{\Vec{r}}|\rangle \int_{1+\epsilon}^{N}\frac{dr}{r(\ln r)^{1+\mu}}\\&\simeq  \text{Const} -\frac{2b\langle|u_{r}|\rangle}{\mu}(\ln N)^{-\mu},
\end{split}
\end{equation}
\begin{equation}
\begin{split}
\sum_{\vec{r}}\langle|V(\vec{r})|^{2q}\rangle&\simeq 2\langle |u_{\Vec{r}}|^{2q}\rangle\int_{1+\epsilon}^{N} \frac{dr}{[r(\ln r)^{1+\mu}]^{2q}}\\
      &\underrel{\ln N\rightarrow\infty}{\simeq}\frac{2\langle |u_{\Vec{r}}|^{2q}\rangle b^{2q}N^{1-2q}(\ln N)^{-2q(1+\mu)}}{1-2q}.
\end{split}
\end{equation}
Following same steps as those in Sec.~\ref{sec3}, the final results reads
\begin{equation}\small
\begin{split}
      \langle P_{q}\rangle
     &\underrel{\ln N\rightarrow\infty}{\simeq}1+\frac{2b^{2q}\Gamma(1/2+q)\Gamma(1/2-q)N^{1-2q}(\ln N)^{-2q(1+\mu)}}{\pi(1-2q)}
\end{split}
\end{equation}
for $q<1/2$ and
\begin{equation}\label{eq:P2GSRBM}
\begin{split}
      \langle P_{q}\rangle
     &\simeq P_q^{\infty}+\frac{\Gamma(q-1/2)}{\Gamma(q-1)}\frac{4b }{\mu\sqrt{\pi}}(\ln N)^{-\mu} \; ,
\end{split}
\end{equation}
for $q>1/2$, with $P_q^\infty>0$ indicative of a localized behavior. Numerical results presented in Fig.~\ref{figHOSRBM} (a) and Fig.~\ref{figHOSRBM_mu} (a) are well fitted by the analytical prediction Eq.~\eqref{eq:P2GSRBM}.

\begin{figure}
\includegraphics[width=0.48\textwidth]{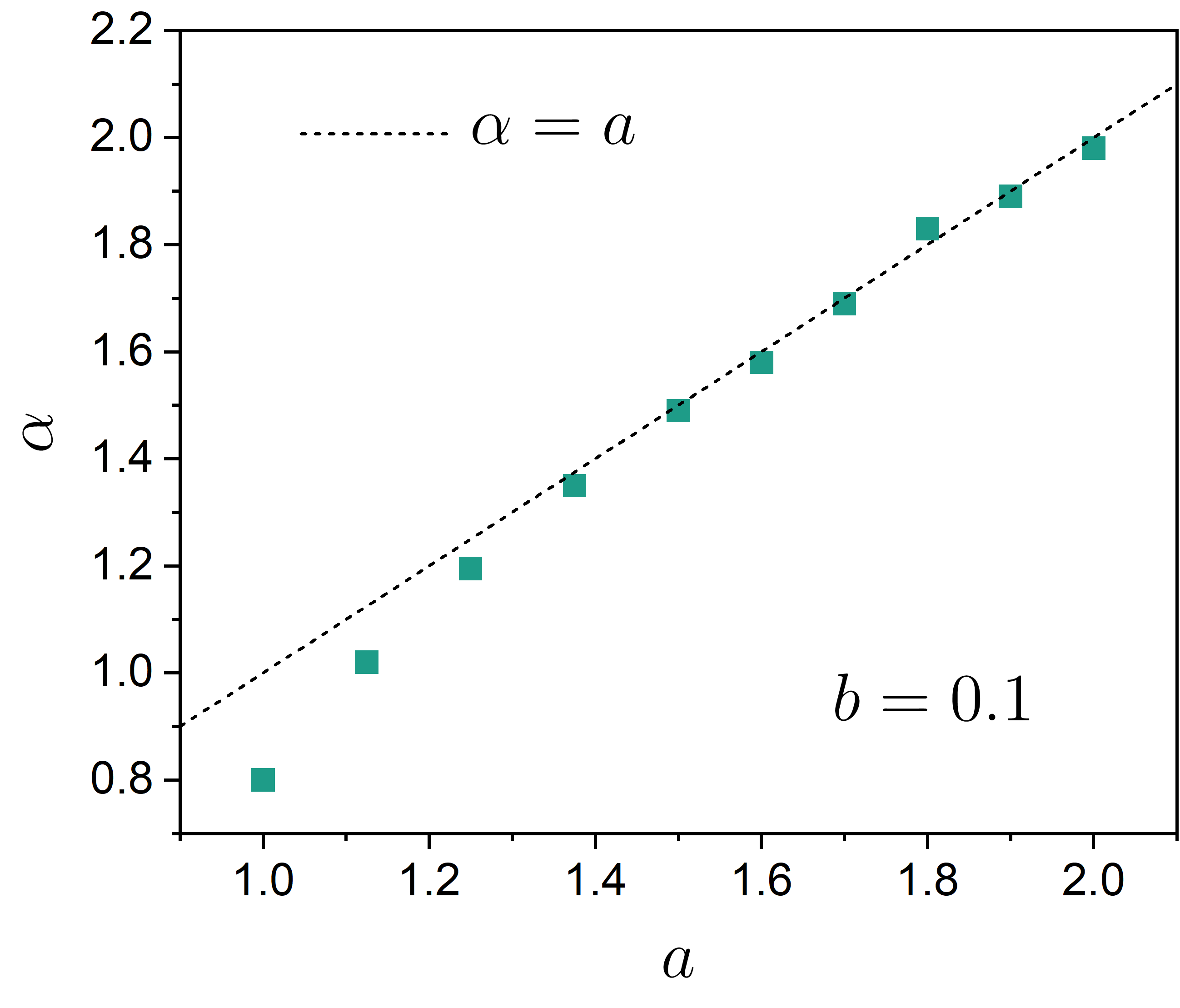}
 \caption{\label{figCorPRBM}The decay exponent $\alpha$ of the correlation function $C(r)$ for different $a\ge 1$ values of the PRBM model Eq.~\eqref{defPRBM} with $b=0.1$. The exponent $\alpha$ is obtained from a power-law fitting of the correlation function $ C(r)=A r^{-\alpha}$   with $A$ and $\alpha$ two fitting parameters, showing a convergence to $\alpha=a$. }
 \end{figure}

In addition, we have also studied the average correlation function $C(r)\equiv C^{11}(r)$ defined in Eq.~\eqref{eq:cor}. Here, we deduce the functional form of $C(r)$ for these generalized SRBM ensembles as
\begin{equation}\label{eq:CorGSRBM}
    C(r) \sim r^{-1} (\ln r)^{-(1+\mu)}.
\end{equation}
This expression, derived in a manner similar to the correlation function $C(r)$ for the SRBM ensemble in Section \ref{sec6}, combines a $r^{-1}$ factor stemming from the logarithmic support of wavefunctions, and a $(\ln r)^{-(1+\mu)}$ term indicating algebraic localization on this support. This contrasts with the multifractal behavior on a logarithmic support associated with previously considered log-multifractality. This scenario mirrors the localized phase of the PRBM model with $a>1$, where the eigenstates are algebraically localized and the correlation function behaves as $C(r)\sim r^{-a}$ (see Fig. \ref{figCorPRBM} and \cite{C.Yeung_1987,BORGONOVI1999317,CASATI1999293}), with the correspondence $r \rightarrow \ln r$ and the additional logarithmic support.

\begin{figure*}
\includegraphics[width=0.48\textwidth]{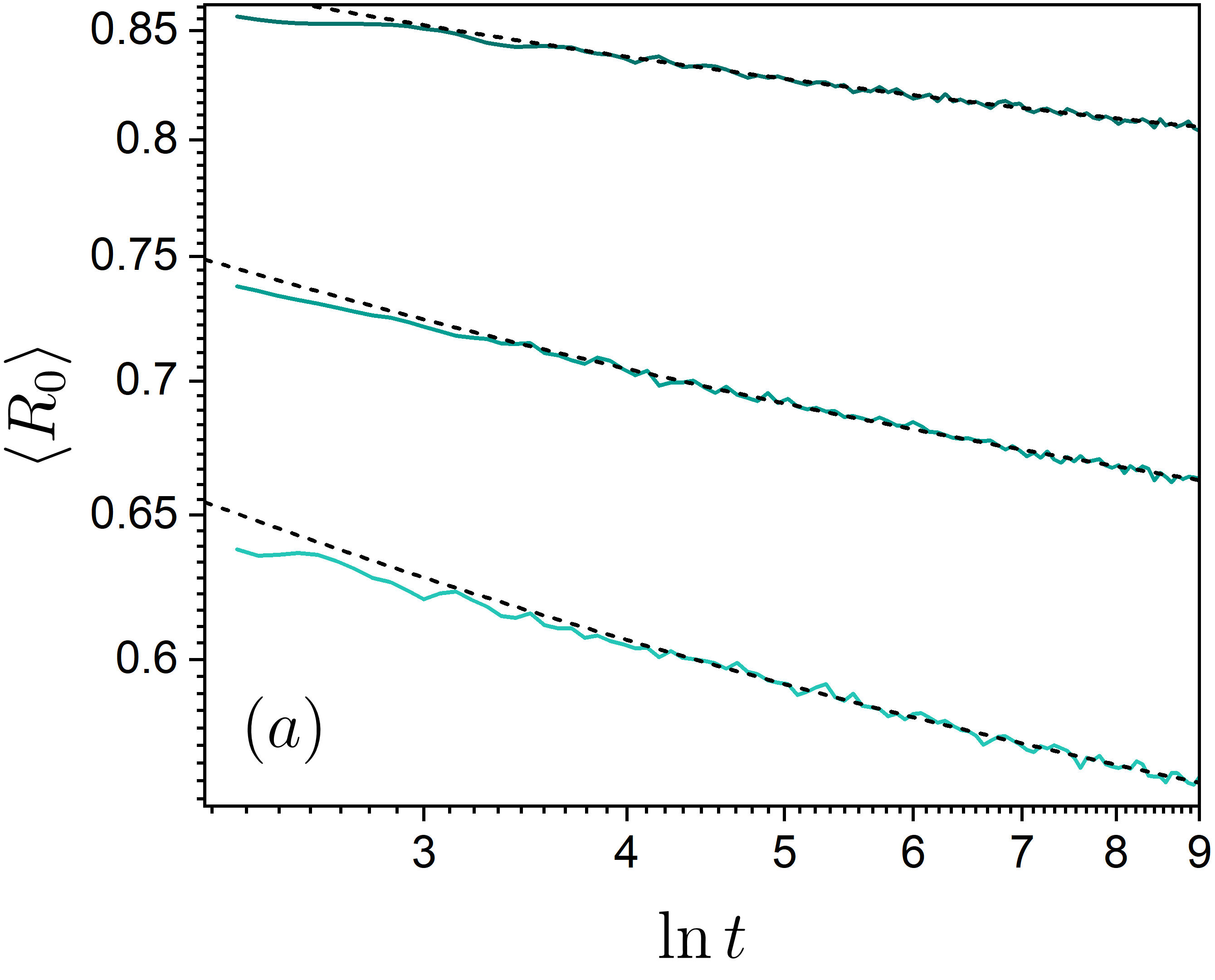}
\includegraphics[width=0.48\textwidth]{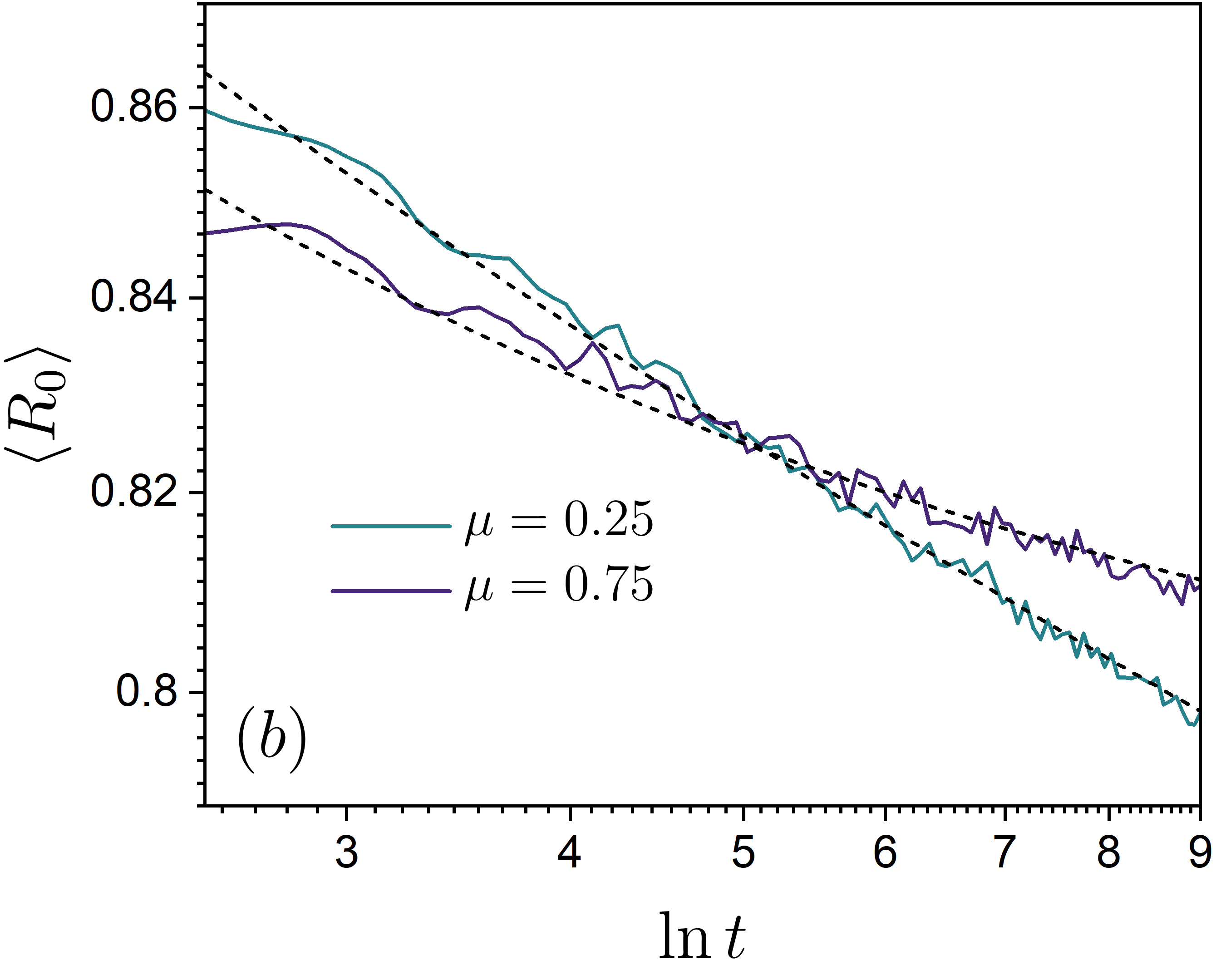}

    \caption{\label{figR0HO}  The return probability $\langle R_0 \rangle$ in the generalized SRBM model Eq.~\eqref{defGSRBM} with $\beta=1$. (a) $\mu=1/2$ for different $b$ values: $b=0.05,b=0.10,0.15$, from top to bottom. (b) Other $\mu$ values as indicated by labels. The black dashed lines are fits $\langle R_0 \rangle=R_0^\infty+A(\ln t)^{-\mu}$ with $R_0^\infty$ and $A$ two fitting parameters, see Eq.~\eqref{eq:R0GSRBM}. These results for $\mu=1/2$ are in perfect agreement with the critical dynamics predicted in random regular and Erdös-Rényi graphs of effective infinite dimensionality, see \cite{PhysRevB.34.6394, ZIRNBAUER1986375, PhysRevLett.72.526, PhysRevB.99.024202}. Results have been averaged over $14,400$ disorder configurations with the system size $N=2^{10}$.} 
\end{figure*}

Remarkably, this functional form coincides with predictions for the Anderson transition on random regular graphs (RRG) and the Bethe lattice when $\mu=1/2$. Numerical results in Fig.~\ref{figHOSRBM} (b) and Fig.~\ref{figHOSRBM_mu} (b) support this proposed functional form, suggesting a potential connection between the parameter $\mu$ and the detailed structure of various graphs with infinite-dimension (e.g., it has been observed that the case when $\mu\approx0.3$ in Eq.~\eqref{eq:CorGSRBM} reproduces the critical behavior of $C(r)$ in small-world networks with average connectivity $\approx2.12$, see \cite{PhysRevB.106.214202}). In addtion, numerical data for the dynamics of the return probability, presented in Fig.~\ref{figR0HO}, shows the decay behavior 
\begin{equation}\label{eq:R0GSRBM}
    \langle R_0\rangle\simeq R_0^{\infty}+A(\ln t)^{-\mu}
\end{equation} 
with $A$ a constant, which coincides with the predictions in Ref.~\cite{PhysRevB.99.024202} for the critical decay of the return probability in Anderson transition on RRG when $\mu=1/2$. 

\section{Conclusion}\label{sec8}

In conclusion, the introduction of the new random matrix models SRBM and SRUM provides a fundamental framework for understanding the critical behavior of the Anderson transition in infinite dimension. The critical behavior investigated in this work is distinct from conventional multifractality observed in finite dimensions \cite{castellani1986multifractal, PhysRevLett.102.106406,PhysRevA.80.043626,PhysRevB.84.134209,PhysRevB.95.094204,PhysRevLett.76.1687,PhysRevB.75.174203,doi:10.7566/JPSJ.83.084711,PhysRevLett.124.186601} and will be of importance for analytical comprehension of the Anderson transitions. This work indicates that the characteristic critical behavior is the pronounced, logarithmic, finite size effects at the Anderson transition point. Indeed, the previous lack of precise knowledge regarding this critical behavior has sparked significant debates concerning the nature of the delocalized phase, whether it is ergodic or not \cite{biroli2012difference, PhysRevLett.113.046806, PhysRevLett.117.156601, PhysRevB.94.220203,PhysRevB.96.214204,PhysRevLett.118.166801, KRAVTSOV2018148, PhysRevB.99.214202, PhysRevB.101.100201, Parisi_2020, PhysRevB.98.134205, TIKHONOV2021168525, biroli2018delocalization, PhysRevResearch.2.012020, PhysRevB.106.214202, PhysRevResearch.2.043346, 10.21468/SciPostPhys.15.2.045, pino2023correlated, pino2020scaling}, as well as the determination of critical exponents \cite{PhysRevLett.117.156601,PhysRevB.99.214202,pino2023correlated}. Furthermore, the parallels between the Anderson transition in infinite dimension and the enigmatic phenomenon of many-body localization underscore the importance of addressing finite-size effects and intricate dynamics. Despite a mathematical proof \cite{PhysRevLett.117.027201,https://doi.org/10.1002/andp.201600278}, experimental observations \cite{doi:10.1126/science.aaa7432,PhysRevLett.119.260401,doi:10.1126/science.aaf8834} and numerous numerical simulations \cite{sierant2024manybody}, the existence of many-body localization has been called into question due to the persistence of these subtle and poorly understood effects \cite{PhysRevB.95.155129,PhysRevB.105.174205,PhysRevB.106.L020202,leonard2022signatures,PhysRevB.100.104204,PhysRevE.102.062144,PhysRevB.102.064207,PhysRevB.103.024203,PhysRevE.104.054105,PhysRevLett.127.230603,ABANIN2021168415,PhysRevLett.124.186601,Panda_2019,PhysRevB.102.100202}.

Random matrix theory \cite{mehta2004random} has served as a powerful tool for offering universal insights in complex physical systems, independent of their microsopic details, including phenomena such as quantum chaos~\cite{PhysRevLett.52.1} and Anderson localization~\cite{PhysRevB.19.783,Efetov1983,PhysRevLett.67.2049,PhysRevLett.67.2405,PhysRevE.54.3221, PhysRevB.62.7920}. Notably, extensions of random matrix Wigner-Dyson ensembles like the power-law random banded matrix ensemble have been pivotal in elucidating multifractal properties at the Anderson transition in finite dimensions \cite{PhysRevE.54.3221, PhysRevB.62.7920}. Recent advancements, exemplified by the Rosenzweig-Porter ensemble and its variations, have provided crucial insights into multifractal phases akin to the many-body localized state \cite{Kravtsov_2015, 10.21468/SciPostPhys.6.1.014, PhysRevResearch.2.043346, von2019non, Truong_2016, PhysRevE.98.032139, Monthus_2017, Amini_2017, PhysRevB.103.104205, kravtsov2020localization, PhysRevResearch.2.043346, 10.21468/SciPostPhys.11.2.045,PhysRevB.103.104205}. However, a model for the strong multifractal critical behavior in infinite dimensions was absent.

We have thus filled in a gap through the SRBM and SRUM models. Our models feature a long-range decay of off-diagonal matrix elements, characterized by $(r \ln r)^{-1}$ where $r$ signifies the distance to the diagonal. This decay, distinct from the PRBM critical decay $r^{-1}$, incorporates a crucial logarithmic correction. Through our analysis, we demonstrate that this logarithmic term generates a strongly multifractal behavior, where multifractal dimensions $D_q$ exhibit positive values for $q<1/2$, while vanishing with system size for $q>1/2$. Remarkably, we unveil a logarithmic multifractality for $q>1/2$, where moments of wave function amplitudes scale as a power law of the logarithm of system size: $\langle P_q \rangle \sim (\ln N)^{-d_q}$. Additionally, the decay of the return probability of an expanding wave-packet initially peaked follows a power law of the logarithm of time, $\langle R_0(t) \rangle \sim (\ln t)^{-d_2}$, with the logarithmic multifractal dimensions $d_q$ whose expression we derive in the limit of small bandwidth $b\ll 1$. Our findings also encompass detailed descriptions of correlation functions and wave-packet dynamics. Conceptually, our model paints a picture of wavefunctions residing on a support comprising approximately $\ln N$ sites and exhibiting multifractal fluctuations on this support. This scenario echoes the behavior observed at the Anderson transition in random graphs of effective infinite dimension, where wavefunctions delocalize across a few rare branches consisting of $\ln N$ sites \cite{biroli2012difference, PhysRevLett.118.166801, PhysRevResearch.2.012020, PhysRevResearch.2.012020, biroli2023largedeviation}.

Nevertheless, our precise predictions regarding logarithmic multifractality stand in contrast with those for critical behavior on Cayley trees, random regular, and Erdös-Renyi graphs \cite{PhysRevB.34.6394, ZIRNBAUER1986375, PhysRevLett.72.526, PhysRevB.99.024202}. In these cases, $\langle P_2 \rangle \approx P_2^\infty + \text{cst} (\ln N)^{-1/2}$ converges to a constant with logarithmic finite-size corrections. This phenomenon, termed critical localization, deviates from the anticipated logarithmic multifractal behavior. Remarkably, we can extend our SRBM model to encompass such behavior by introducing a parameter $\mu >0$ and considering decays of off-diagonal elements of the random matrix model as $[r (\ln r)^{1+\mu}]^{-1}$. Our analysis demonstrates that this generalized model yields $\langle P_q \rangle \approx P_q^\infty + \text{cst} (\ln N)^{-\mu}$ for $q>1/2$ and $\langle R_0(t) \rangle \approx \langle R_0^{N=\infty}\rangle+\text{cst} (\ln t)^{-\mu}$, aligning perfectly for $\mu=1/2$ with results for Cayley trees, random regular, and Erdös-Renyi graphs \cite{PhysRevB.34.6394, ZIRNBAUER1986375, PhysRevLett.72.526, PhysRevB.99.024202}. Interestingly, while distinguishing between the behaviors of $\langle P_2(N) \rangle$ and $\langle R_0(t)\rangle $ in logarithmic multifractality and critical localization proves challenging, a clear distinction emerges in the behavior of the correlation function $r C(r)\sim (\ln r)^{-\alpha}$: $\alpha=1-d_2<1$ for logarithmic multifractality, while $\alpha = 1+\mu>1$ for critical localization.

Our results suggest the existence of a novel class of Anderson transitions in infinite dimensions with logarithmic multifractal critical behavior, distinct from those observed in Cayley trees, random regular, Erdös-Renyi, and smallworld graphs, all of which exhibit critical localization behavior \cite{PhysRevB.34.6394, ZIRNBAUER1986375, PhysRevLett.72.526, PhysRevB.99.024202,PhysRevB.106.214202}. Logarithmic multifractality can be interpreted as a logarithmic scale invariance which, to the best of our knowledge, remains unprecedented in the description of phase transitions. It denotes a novel form of strongly non-ergodic behavior, wherein states are delocalized yet multifractal on a support comprising approximately $\ln N$ sites, akin to a few branches on random graphs.

One of the most intriguing prospects stemming from this work involves integrating our models embodying strong multifractality with Rosenzweig-Porter type models of multifractal phases. Such an integration promises analytical insights into finite-size and finite-time effects that afflict non-ergodic phase transitions, including the elusive many-body localization transition.

\acknowledgements
This study was supported by
research funding Grants No.~ANR-18-CE30-0017 and ANR-19-CE30-0013, and by the Singapore
Ministry of Education Academic Research Fund Tier I (WBS
No.~R-144-000-437-114). We thank Calcul en Midi-Pyrénées
(CALMIP) and the National Supercomputing Centre (NSCC) of Singapore for computational resources and assistance. W.~Chen is supported by the President's Graduate Fellowship at National University of Singapore and the Merlion Ph.D. Scholarship awarded by the French Embassy in Singapore.


\appendix
\section{Decay of the amplitudes of unitary matrix elements $|U_{ij}|$}\label{appendixA}
In this Appendix, we evaluate the decay behavior of the amplitudes of unitary matrix elements $|U_{ij}|$ through perturbative arguments. Starting from 
\begin{equation}
\label{appeqVij}
\begin{split}
W_{ij}&=\sum_{k=1}^{N}F_{ik}V(2\pi k/N)F_{kj}^{-1} \\
&=\sum_{k=1}^{N}\frac{1}{N}e^{2i\pi (i-j)k/N}V(\frac{2\pi k}{N}),
\end{split}
\end{equation}
\begin{widetext}
one can evaluate the summation approximately as an integral,
\begin{equation}
\begin{split}
-W_{ij}=-\frac{1}{2\pi}\int_0^{2\pi}V(q)e^{{\rm i}rq}dq=-\frac{1}{2\pi}\int_0^{2\pi} \ln\left[-\ln(a|\sin\frac{q}{2}|)\right] e^{{\rm i}rq}dq,
\end{split}    
\end{equation}
with $r\equiv i-j$. To evaluate this integral, introduce the auxiliary integral
\begin{equation}
\label{defIbeta}
\begin{split}
     I(\beta,x)&=\frac{1}{2\pi}\int_0^{2\pi} \left[-\ln(a|\sin\frac{q}{2}|)\right]^\beta e^{{\rm i}xq}dq.
\end{split}
\end{equation}
Notice that the integrand is singular at $q=\pi$, to discriminate the singularity when $q\rightarrow\pi$, we separate the integral as
\begin{equation}
\begin{split}
  I(\beta,x)&=\frac{1}{2\pi}\int_0^{\pi} \left[-\ln(a|\sin\frac{q}{2}|)\right]^\beta e^{{\rm i}xq}dq+\frac{1}{2\pi}\int_\pi^{2\pi} \left[-\ln(a|\sin\frac{q}{2}|)\right]^\beta e^{{\rm i}xq}dq\\
  &=\frac{1}{2\pi}\int_0^{\pi} \left[-\ln(a|\sin\frac{q}{2}|)\right]^\beta e^{{\rm i}xq}dq+c.c.\underrel{q\rightarrow0}{\simeq}\frac{1}{2\pi}\int_0^{\pi} \left[-\ln(\frac{aq}{2})\right]^\beta e^{{\rm i}xq}dq+c.c.\\
  &=\frac{1}{a\pi}\int_0^{\frac{a\pi}{2}} \left(-{\ln t}\right)^\beta e^{{\rm i}\frac{2x}{a}t}dt+c.c.
\end{split}    
\end{equation}
To proceed, we adapt Theorem 1 in \cite{WONG1978173}. We have
\begin{equation}
\begin{split}
     J(\alpha,\beta,x)&=\int_0^c q^{\alpha-1}(-\ln q)^\beta e^{{\rm i}xq}dq \underrel{x\rightarrow+\infty}{\sim}\frac{e^{{\rm i}\alpha\pi/2}}{x^\alpha}\sum_{r=0}^{\infty}c_r(\alpha,\beta)(\ln x)^{\beta-r}+e^{{\rm i}xc}\sum_{s=0}^{\infty}(-1)^s g_s(\alpha,\beta)(ix)^{-s-1},
\end{split}
\end{equation}
where $g_s(\alpha,\beta)$ is the $s$th derivative of $q^{\alpha-1}(-\ln q)^\beta$ at $q=c$, and
\begin{equation}
\begin{split}
    c_r(\alpha,\beta)=(-1)^r\binom{\beta}{r}\sum_{k=0}^{r}\binom{r}{k}\Gamma^{(k)}(\alpha)\left(\frac{\pi i}{2}\right)^{r-k}.
\end{split}
\end{equation} 

Therefore,
\begin{equation}
\label{eq16}
\begin{split}
     \int_0^{\frac{a\pi}{2}}\left(-{\ln t}\right)^\beta e^{{\rm i}\frac{2x}{a}t}dt=J(1,\beta,\frac{2x}{a})
     &\underrel{x\rightarrow+\infty}{\sim}\frac{ia}{2x}\left[c_0(\beta)[\ln (\frac{2x}{a})]^{\beta}+c_1(\beta)[\ln (\frac{2x}{a})]^{\beta-1}+c_2(\beta)[\ln (\frac{2x}{a})]^{\beta-2}+\dots\right]\\
     &+e^{{\rm i}\pi x}\left[g_0(\beta)(\frac{2ix}{a})^{-1}+g_1(\beta)(\frac{2x}{a})^{-2}+\dots\right],
\end{split}
\end{equation}
with
\begin{equation}
\begin{split}
c_0(\beta)=1,\quad c_1(\beta)=-\beta(\frac{\pi i}{2}-\gamma)&,\quad c_2(\beta)=\frac{\beta(\beta-1)}{2}\left(-\frac{\pi^2}{12}-i\gamma\pi+\gamma^2\right),\\
g_0(\beta)=[-\ln(\frac{a\pi}{2}) ]^\beta, &\quad g_1(\beta)=-\frac{2\beta[-\ln(\frac{a\pi}{2}) ]^{\beta-1}}{a\pi},
\end{split}
\end{equation}
where $\gamma$ is the Euler's constant. By adding Eq.~\eqref{eq16} with its complex conjugate, we obtain
\begin{equation}\label{eq18}
\begin{split}
   I(\beta,x)\simeq&\frac{1}{2\pi x}\left\{\beta\pi[\ln (\frac{2x}{a})]^{\beta-1}+\beta(\beta-1)\gamma\pi[\ln (\frac{2x}{a})]^{\beta-2}+\dots\right\}+\frac{2e^{{\rm i}\pi x}}{a\pi}g_1(\beta)(\frac{2x}{a})^{-2}+\dots\\
   &\underrel{x\rightarrow+\infty}{\simeq}\frac{\beta}{2 x}[\ln (\frac{2x}{a})]^{\beta-1}\underrel{\beta\rightarrow0}{\simeq}\frac{\beta}{2 x}[\ln (\frac{2x}{a})]^{-1}\left[ 1+\beta\ln \ln (\frac{2x}{a}) +\dots \right].
\end{split}
\end{equation}
\end{widetext}
Note that from its definition \eqref{defIbeta} we have
\begin{equation}
\begin{split}
    \lim_{\beta\rightarrow0} I(\beta,x)&\simeq\frac{1}{2\pi}\int_0^{2\pi}\{ 1+\beta\ln\left[-\ln(a|\sin\frac{q}{2}|)\right]\}e^{{\rm i}xq}dq\\
    &=\frac{1}{2\pi}\int_0^{2\pi}\beta\ln\left[-\ln(a|\sin\frac{q}{2}|)\right]e^{{\rm i}xq}dq,
\end{split}
\end{equation}
i.e., $-\beta W_{ij}\simeq \lim_{\beta\rightarrow0} I(\beta,|i-j|)$. Comparing with terms at first order in $\beta$ in Eq.~\eqref{eq18}, we get
\begin{equation}
   |W_{ij}|\simeq\frac{1}{2|i-j|}[\ln (\frac{2|i-j|}{a})]^{-1}
\end{equation}
and hence
\begin{equation}
\label{uijapp}
|U_{ij}|\simeq K|W_{ij}|\simeq\frac{K}{2|i-j|\ln|i-j|}
\end{equation}
for $|i-j|\gg1$.


%
%

\section{Calculation of the return probability for SRMB in the case $\beta=2$}
\label{returnprobbeta2}
Starting from  Eq.~(14) of \cite{Kravtsov_2011} we get
\begin{equation}
   \langle R_0^{(1)} \rangle=\frac{2\sqrt{\pi}}{Nt}\sum_{i\neq j}^{N}\sum_{k=1}^{\infty}\frac{(-2b_{ij}t^2)^k}{(k-1)!}\frac{k}{2k-1}.
\end{equation}
Replace the double sum $\sum_{i\neq j}^N$ as 2$\sum_{x=1}^{N}(N-x)$,
\begin{equation}
\begin{split}
    \langle R_0^{(1)} \rangle&=\frac{4\sqrt{\pi}}{Nt}\sum_{x=1}^{N}(N-x)\sum_{k=1}^{\infty}\frac{(-2b_{x}t^2)^k}{(k-1)!}\frac{k}{2k-1}\\
    &\underrel{N\rightarrow\infty}{\simeq}\frac{4\sqrt{\pi}}{t}\int_{1}^{\infty}dx\sum_{k=1}^{\infty}\frac{(-2b_{x}t^2)^k}{(k-1)!}\frac{k}{2k-1}.\\
\end{split}
\end{equation}
Note that
\begin{equation}
    \sum_{k=1}^{\infty}\frac{(-2b_{x}t^2)^k}{(k-1)!}\frac{k}{2k-1}=\sqrt{2b_{x}t^2}\int_{0}^{\sqrt{2b_{x}t^2}}dy(y^2-1)e^{-y^2},
\end{equation}
hence
\begin{equation}
\begin{split}
     \langle R_0^{(1)} \rangle&=\frac{4\sqrt{\pi}}{t}\int_{1}^{\infty}dx\sqrt{2b_{x}t^2}\int_{0}^{\sqrt{2b_{x}t^2}}dy(y^2-1)e^{-y^2}\\
     &=4\sqrt{\pi}\int_{1}^{\infty}dx\frac{b}{x\ln x}\int_{0}^{\frac{bt}{x\ln x}}dy(y^2-1)e^{-y^2}.
\end{split}
\end{equation}
To proceed, we change the variable $y\rightarrow y^{\prime}=(x\ln x)y$ and arrive at
\begin{equation}\small
 \langle R_0^{(1)} \rangle=4\sqrt{\pi}\int_{1}^{\infty}dx\frac{b}{(x\ln x)^2}\int_{0}^{bt}dy^{\prime}[(\frac{y^{\prime}}{x\ln x})^2-1]e^{-(\frac{y^{\prime}}{x\ln x})^2},
 \end{equation}
hence
\begin{equation}
 \frac{\partial \langle R_0^{(1)} \rangle}{\partial t}=4\sqrt{\pi}\int_{1}^{\infty}dx\frac{b^2}{(x\ln x)^2}[(\frac{bt}{x\ln x})^2-1]e^{-(\frac{bt}{x\ln x})^2}.
\end{equation}
Changing again the variable $x\rightarrow u=\frac{x\ln x}{bt}$ with $\ln x=\mathcal W(btu)$, we arrive at 
\begin{equation}
   \frac{\partial \langle R_0^{(1)} \rangle}{\partial t}=4\sqrt{\pi}\int_{0}^{\infty}du\frac{bt}{\mathcal W(btu)+1}\frac{1}{(ut)^2}(\frac{1}{u^2}-1)e^{-\frac{1}{u^2}}.
\end{equation}
Therefore if the return probability scales as $\langle R_0 \rangle\sim (\ln t)^{-d_2}$ then
\begin{equation}
\begin{split}
    -d_2&=t\ln t\frac{\partial \langle R_0^{(1)} \rangle}{\partial t}\\
    &=4b\sqrt{\pi}\int_{0}^{\infty}du\frac{\ln t}{\mathcal W(btu)+1}\frac{1}{u^2}(\frac{1}{u^2}-1)e^{-\frac{1}{u^2}}.
\end{split}
\end{equation}
In the limit $t\rightarrow\infty$, $\mathcal W(x)\underrel{x\rightarrow\infty}{\sim}\ln x$, and thus
\begin{equation}
    \underrel{t\rightarrow\infty}{\lim}\frac{\ln t}{\mathcal W(btu)+1}\simeq \underrel{t\rightarrow\infty}{\lim}\frac{\ln t}{\ln t+\ln b+\ln u+1}\simeq 1.
\end{equation}
Finally
\begin{equation}
d_2\simeq4b\sqrt{\pi}\int_{0}^{\infty}du\frac{1}{u^2}(\frac{1}{u^2}-1)e^{-\frac{1}{u^2}}=-\pi b,
\end{equation}
which indeed corresponds to the value found at Eq.~\eqref{eq:dq_}.
\bibliography{modified_apssamp}

\end{document}